\documentclass[11pt]{article}
\usepackage{latexsym}
\usepackage{graphicx}
\usepackage{authblk}
\graphicspath{{./}{figures}}
\usepackage{tikz}          
\usetikzlibrary{arrows.meta} 
\usetikzlibrary{arrows}
\usetikzlibrary{automata}
\usetikzlibrary{positioning, calc, shapes.multipart, shadows, backgrounds, fit}
\usepackage{enumitem}
\newlist{enumdescript}{description}{1}
\setlist[enumdescript]{labelwidth=6em, leftmargin=!}
\usepackage{array}
\usepackage{xcolor}
\usepackage{epstopdf}
\usepackage{amssymb}
\usepackage{amsthm}
\usepackage{amsmath,enumerate,rotate}
\usepackage{hyperref}
\newcommand{\R}{\mathbb R}

\def\be#1\ee{\begin{equation}#1\end{equation}}

\newtheorem{proposition}{\bf Proposition}[section]

\newtheorem{remark}{\bf Remark}[section]

\newcommand{\bq}{\begin{equation}}
\newcommand{\eq}{\end{equation}}

\def\bqa{\begin{eqnarray}}
\def\eqa{\end{eqnarray}}

\newcommand{\bd}{\begin{displaymath}}
\newcommand{\ed}{\end{displaymath}}
\newcommand{\ba}{\begin{eqnarray}}
\newcommand{\ea}{\end{eqnarray}}

\def\R{\mathbb{R}}

\begin{document}

\title{How opinions shape epidemics: a graphon-based kinetic approach}
\author[1]{Abu Safyan Ali}
\author[1]{Elisa Calzola}
\author[1]{Giacomo Dimarco}
\author[1,2]{Lorenzo Pareschi}
\author[3]{Thomas Rey}
\affil[1]{Department of Mathematics and Computer Science \& Center for Modeling, Computing and Statistics (CMCS), University of Ferrara, via Machiavelli 30, 44121 Ferrara, Italy}
\affil[2]{Maxwell Institute for Mathematical Sciences and Department of Mathematics, Heriot-Watt University, Edinburgh, UK}
\affil[3]{Universit\'e C\^ote d’Azur, CNRS, LJAD, Parc Valrose, F-06108 Nice, France}

\maketitle

\begin{abstract}
	Understanding the mutual influence between social behavior and physical health is crucial for designing effective epidemic mitigation strategies. Individual interactions drive the evolution of opinions, which in turn shape how infectious diseases are perceived and consequently how they spread within a population, for instance through the adoption or rejection of preventive measures. At the same time, the distribution and dynamics of physical contacts play a fundamental role in determining transmission patterns. To this end, we develop a mathematical framework to analyze the coupled dynamics of opinion formation, disease transmission, and physical contacts by employing graphon-based networks, which capture heterogeneous and large-scale connectivity patterns typical of realistic social structures. The epidemic compartmental model further incorporates a kinetic description of microscopic level physical contacts, allowing for a consistent multiscale representation of interaction patterns. Starting from a microscopic description governed by interpersonal compromise and intrinsic self-thinking processes, we derive a kinetic compartmental epidemic model on graphons via a mean-field limit. This formulation allows us to investigate the joint evolution of the disease state and the opinion distribution, with a particular focus on the role of social networks and physical contacts.
	Numerical experiments demonstrate that the graphon-kinetic approach provides a comprehensive representation of the coupled opinion-epidemic dynamics, revealing new possibilities for controlling disease spread by shaping population opinion patterns.
\end{abstract}

\noindent
{\bf Keywords}: Opinion dynamics, epidemic modelling, coupled social-epidemic dynamics, graphon networks.\\

\tableofcontents

\section{Introduction}\label{sec:intro}
As the number of COVID-19 cases in 2020 spiked, it became evident that diminishing the physical contacts as well as wearing masks and avoiding social gatherings were crucial safety precautions to safeguard public health, particularly in the absence of medications or vaccinations \cite{Albi2024,bellomo2022predicting, bertaglia2021spatial, Perthame,zanella2021data}. On the other hand, it is clear that the effectiveness of lockdown and preventive measures is strongly tied to individual opinions regarding protective behaviors, which themselves depend on personal situational awareness. Indeed, following recent studies \cite{ Dezecache2020,Durham2011, Tunccgencc2021social}, social effects and opinion formation phenomena play a significant role in adherence to safety standards, since individuals respond differently to rising case numbers, and societal norms often shift when shared perceptions emerge. 

In this context, opinion formation models have been extensively studied over the past decades, reflecting a broad and growing interest in the behavior of large-scale interacting agent systems across diverse disciplines \cite{Barre2017, bicego25,Ciallella2021, Fornasier2011, Motsch2014}. Classical models rely on the assumption that individual opinions evolve through binary interactions and external influences such as social media. Most of these models focus on individual-level dynamics, which collectively lead to large and complex systems \cite{Degroot1974reaching, Goddard2021, Hegselmann2002opinion, Iacomini03042023, Jabin2014}. In many such settings, microscopic interaction rules give rise to complex patterns and emergent collective behavior. At the same time, it has been shown in the last years \cite{ Albi2014,bonandin24,Bondesan2024, During2015, Franceschi2023,  Pareschi2019,Pareschi2017, Toscani2006, Zanella2023} that kinetic theory is able to provide a very powerful framework for understanding the driving mechanisms and properties of opinion dynamics starting from simple microscopic interaction and through suitable scaling limits. 

On the other hand, network theory is crucial for understanding complex dynamics in many fields, including epidemiology \cite{Danon2011, lloyd2007network}. Real-world social contact networks, characterized by heterogeneous interaction strengths, clustering, and degree distributions, shape how information and opinions spread. Through the use of such methods, opinion formation models explain how social interactions generate consensus, polarization, or clustered opinions \cite{Albi2024,Albi2017,   Das2014modeling,Fagioli2024, li2020effect}. Network topology strongly influences these dynamics: highly connected individuals can drive rapid opinion shifts, while tightly knit groups may resist external influence. In addition, the increasing availability of data from social platforms has advanced the study of social influence on voting behavior and vaccination campaigns \cite{ Albi2024,Albi2017}, informing research on opinion control \cite{Albi2016}, polarization \cite{Amelkin2017polar,Lee2014}, and influence strategies \cite{Toscani2018opinion}.

The above arguments suggest that coupling epidemic dynamics with network structure and opinion formation is essential to adequately capture mutual influence effects. In addition, the role and estimation of the distribution of contacts between individuals are widely recognized as key factors in pathogen transmission \cite{French,Perthame,Dol}. Indeed, classical epidemic models assume homogeneous mixing among individuals \cite{Beckley2013modeling, Kendall1956deterministic,Kermack1927contribution}, whereas real epidemics arise from repeated interactions shaped by both contact-related disease features and social behavior \cite{Dimarco2022optimal,Loy2021viral, Marca2022sir}. Contact-network-based compartmental models extend this perspective by linking transmission to transitions between health states through an underlying network topology \cite{Pastor2002immunization, Salathe2010dynamics, Sun2017epidemic}. For instance, small-world networks can slow inter-cluster spread despite fast local outbreaks, thereby allowing more time for intervention \cite{Iotti2017infection, Saif2024sir,Wang2003complex}. Social behavior further shapes epidemic trajectories, as opinions on vaccination, distancing, and disease severity directly influence protective actions. In this context, the large size of modern networks, often comprising millions of nodes, motivates the study of their asymptotic topological properties. Graph theory provides an effective framework for describing such large-scale structures and their emergent collective behaviors \cite{Caron2023, Van2024random}. Graphons represent large networks as continuous objects, replacing the discrete adjacency matrix with a continuum connectivity function. Building on this perspective, recent works \cite{Bayraktar2023graphon, Nurisso2024network} have investigated the asymptotic behavior of network dynamics through graphon-based kinetic and mean-field equations. Recent work has also explored the interplay between opinion formation and epidemic spreading in large-scale networks; for instance, \cite{bondesan2026} proposes a graphon-based kinetic framework for the joint evolution of opinions and SEIR epidemic dynamics on heterogeneous networks, and \cite{naldi26} presents a related graphon-based kinetic model of opinion-driven epidemic dynamics modulated by graphons. While both approaches investigate the coupling between social behavior and disease transmission, the focus of our work is different, since here we introduce a multiscale kinetic formulation that explicitly accounts for the distribution and dynamics of heterogeneous physical contacts, which interact with both the opinion formation process and the epidemic spread. This additional layer allows us to capture the behavioral adaptation of social contacts and its feedback on the epidemic dynamics within the graphon-based network structure.

In this paper, we develop a mathematical framework to analyze in detail the interplay between opinion formation and disease dynamics, accounting for the role of physical contacts within social network structures. 
Opinion dynamics are first described at the microscopic level and then, in the continuum limit, through graphons acting as interaction kernels associated with shared or dependent latent variables. Physical contacts are also modeled starting from a microscopic description of individual interactions and, through a mean-field limit, are incorporated at the kinetic level in terms of a probability distribution of the number of contacts. This setting provides a realistic representation of behavioral responses, where individuals adopt protective actions as a consequence of their opinions. More precisely, the novelty of our approach lies in the unified coupling of graphon-based opinion dynamics, a kinetic description of heterogeneous physical contacts, and a kinetic SEIR framework, allowing social influence, contact heterogeneity, and epidemic transmission to be modeled within a single multiscale structure.

The rest of the paper is organized as follows. Section \ref{SEIR-model} introduces the coupled compartmental model incorporating physical contacts and opinion exchange over a social network, driven by binary interactions and indirect influence through graphon-based connectivity. In the same section, we derive a Fokker--Planck asymptotic approximation for opinion dynamics on social networks, describing the temporal evolution of the joint distribution of opinions and graphon-encoded connections. Section \ref{real-data-garphon} discusses graphon estimation from real network data, illustrating how connectivity patterns influence opinion dynamics in complex communities. Section \ref{results} presents numerical simulations highlighting the impact of model parameters and highly connected individuals, both on the physical and on the social side, on disease outcomes, as well as the role of influential agents in shaping public opinion and affecting epidemic trajectories. Section~\ref{sec:conclusion} is devoted to conclusions and outlines directions for future research. Finally, in Appendix~\ref{appendix} we analyze the evolution of the macroscopic observables of the mean-field epidemic model.

\section{A coupled model of opinion-driven epidemic dynamics with graphon based connectivity structure and physical interactions}\label{SEIR-model}
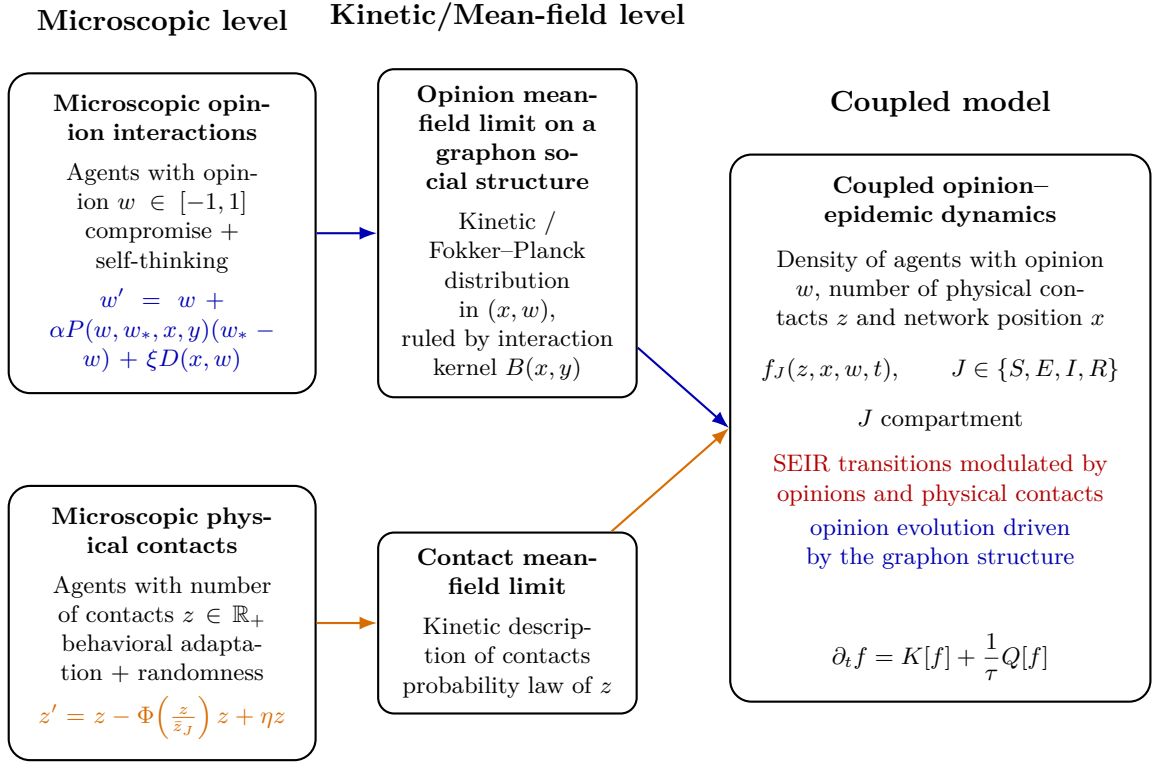
\begin{figure}[t!]
	\centering
	\begin{tikzpicture}[
		>=Latex,
		node distance=1.2cm and 1.4cm,
		every node/.style={font=\footnotesize},
		box/.style={
			draw,
			rounded corners=8pt,
			thick,
			align=center,
			minimum height=2.0cm,
			text width=3.5cm,
			inner sep=8pt
		},
		smallbox/.style={
			draw,
			rounded corners=6pt,
			thick,
			align=center,
			minimum height=1.4cm,
			text width=3cm,
			inner sep=6pt
		},
		flow/.style={->, thick},
		feedback/.style={->, thick, dashed},
		opinion/.style={blue!70!black},
		epidemic/.style={red!70!black},
		contact/.style={orange!85!black},
		graphon/.style={cyan!60!black}
		]
		
		\node[box] (microop) {
			\textbf{Microscopic opinion interactions}\\[4pt]
			Agents with opinion $w \in [-1,1]$\\
			compromise + self-thinking\\[4pt]
			{\color{blue!70!black}
				$w' = w + \alpha P(w,w_\ast,x,y)(w_\ast-w)+\xi D(x,w)$}
		};
		
		\node[box, below=of microop] (microcon) {
			\textbf{Microscopic physical contacts}\\[4pt]
			Agents with number of contacts $z \in \mathbb{R}_+$\\
			behavioral adaptation + randomness\\[4pt]
			{\color{orange!85!black}
				$z' = z - \Phi\!\left(\frac{z}{\bar z_J}\right) z + \eta z$}
		};
		
		\node[smallbox, right=.8cm of microop] (graphon) {
			\textbf{Opinion mean-field limit on a\\graphon social structure}\\[4pt]
			Kinetic / Fokker--Planck\\
			distribution in $(x,w)$,\\
			ruled by interaction kernel $B(x,y)$
		};

		\node[smallbox, right=.8cm of microcon] (fpcon) {
			\textbf{Contact mean-field limit}\\[4pt]
			Kinetic description of contacts\\
			probability law of $z$
		};
		
		\node (C) at ($(graphon.center)!0.5!(fpcon.center)+(2cm,0)$) {};
		\node[box, right=.8cm of C, text width=5cm, minimum height=5.8cm] (coupled) {
			\textbf{Coupled opinion--epidemic dynamics}\\[6pt]
			Density of agents with opinion $w$, number of physical contacts $z$ and network position $x$
			\[
			f_J(z,x,w,t), \qquad J \in \{S,E,I,R\}
			\]
			$J$ compartment\\[6pt]
			{\color{red!70!black}
				SEIR transitions modulated by opinions and physical contacts}\\[2pt]
			{\color{blue!70!black}
				opinion evolution driven by the graphon structure}\\[6pt]
			\[
			\partial_t f = K[f] + \frac{1}{\tau}Q[f]
			\]
		};
		
		\draw[flow, opinion] (microop) -- node[above, sloped] {} (graphon);
		\draw[flow, opinion] (graphon) -- (coupled.west);
		
		\draw[flow, contact] (microcon) -- node[below, sloped] {} (fpcon);
		\draw[flow, contact] (fpcon) -- (coupled.west);
		
		
		\node[above=0.35cm of microop, font=\normalsize\bfseries] {Microscopic level};
		\node[above=0.35cm of graphon, font=\normalsize\bfseries] {Kinetic/Mean-field level};
		\node[above=0.35cm of coupled, font=\normalsize\bfseries] {Coupled model};
		
	\end{tikzpicture}
	\caption{Schematic representation of the proposed multiscale opinion--epidemic model. Opinion interactions are described at the microscopic level and, in the continuum limit, through a graphon-based kinetic/Fokker--Planck framework. Physical contacts are also modeled microscopically and then represented macroscopically through a probability distribution of the number of contacts. Both ingredients are coupled in the compartmental SEIR dynamics through the joint density $f_J(z,x,w,t)$.}
	\label{fig:diagr}
\end{figure}

We propose a compartmental model that integrates the main factors driving epidemic dynamics, namely public perception of preventive measures and the influence of the daily number of physical contacts on the incidence rate, so as to capture the complex interplay between opinion formation, social interactions, and disease transmission.
In what follows, we draw inspiration from \cite{ albi2025impact,Zanella2023} to include consensus formation mechanisms within the epidemic framework, and from \cite{Perthame} to account for the role of physical contacts. A schematic representation of the proposed model, together with the corresponding scalings, is provided in Figure \ref{fig:diagr}.

%
%
%
%
%
%
%

\subsection{Graphon based social connectivity structure}\label{SEIR-model_sub1}
\begin{figure}[h]
\centering
\begin{tikzpicture}
	\node (img1) at (0,0) {
		\includegraphics[width=3.5cm,height=3.5cm]{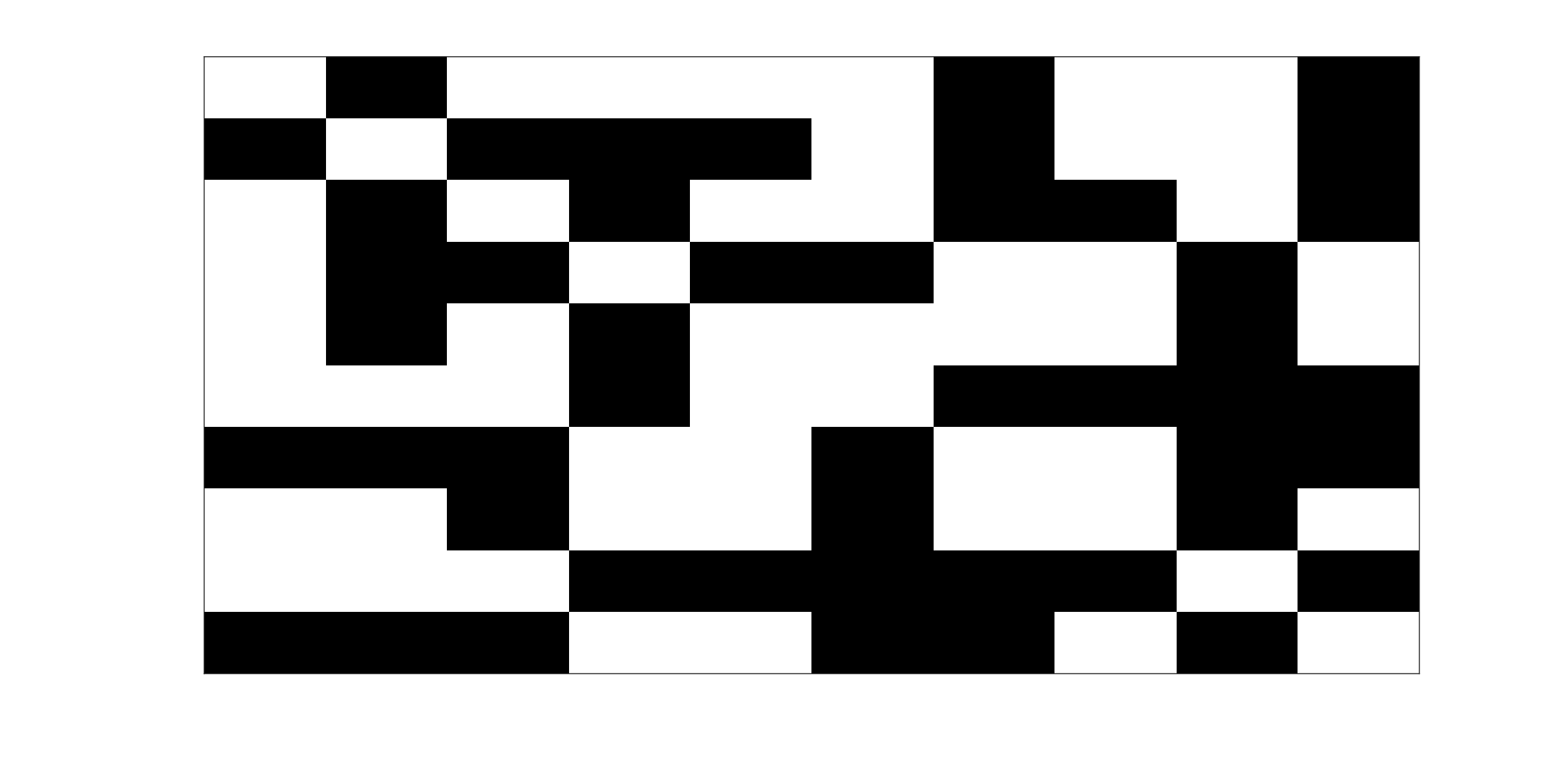}
	};
	
	\node (img2) at (5,0) {
		\includegraphics[width=3.5cm,height=3.5cm]{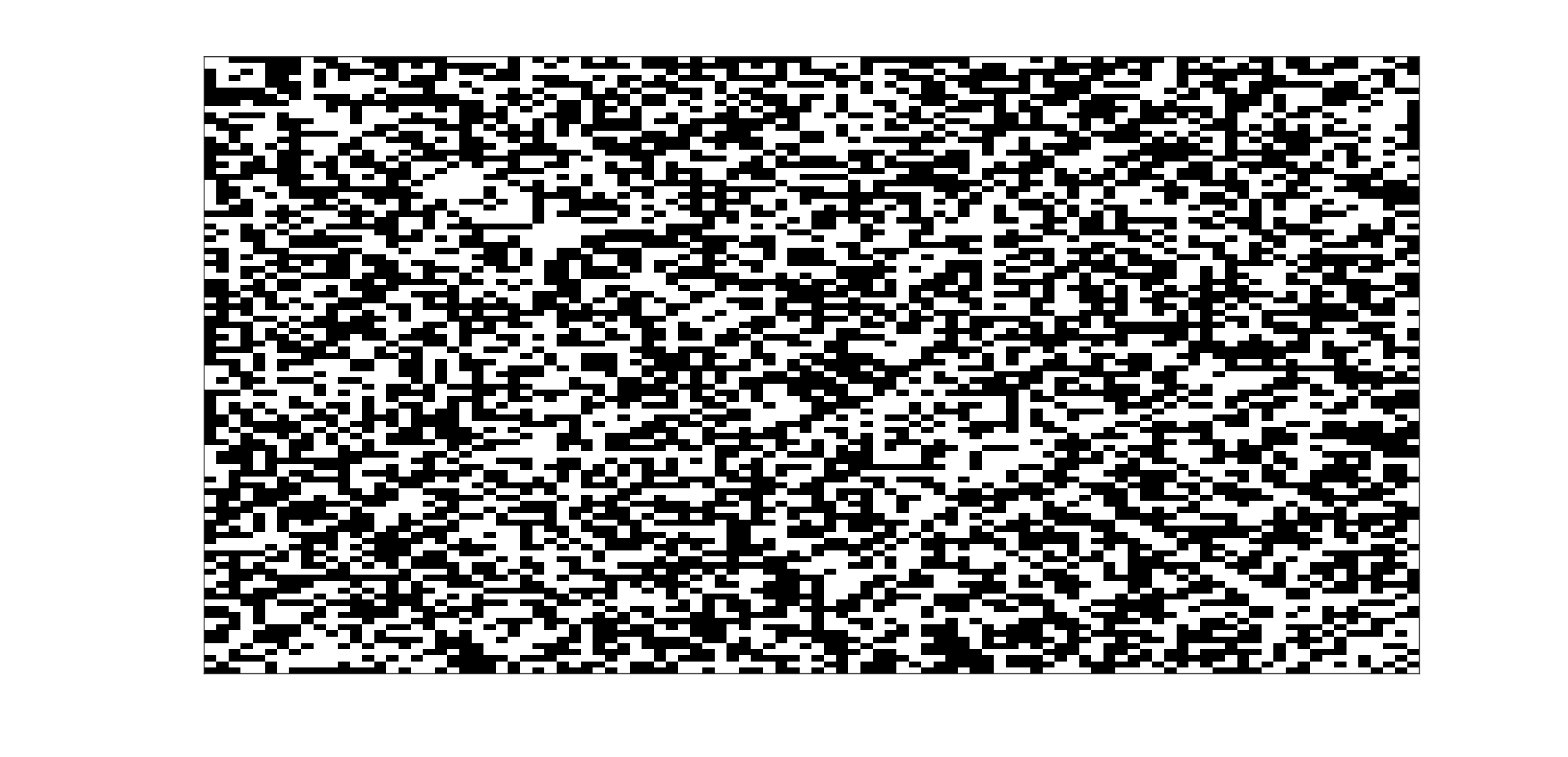}
	};
	
	\node (img3) at (10,0) {
		\includegraphics[width=3.5cm,height=3.5cm]{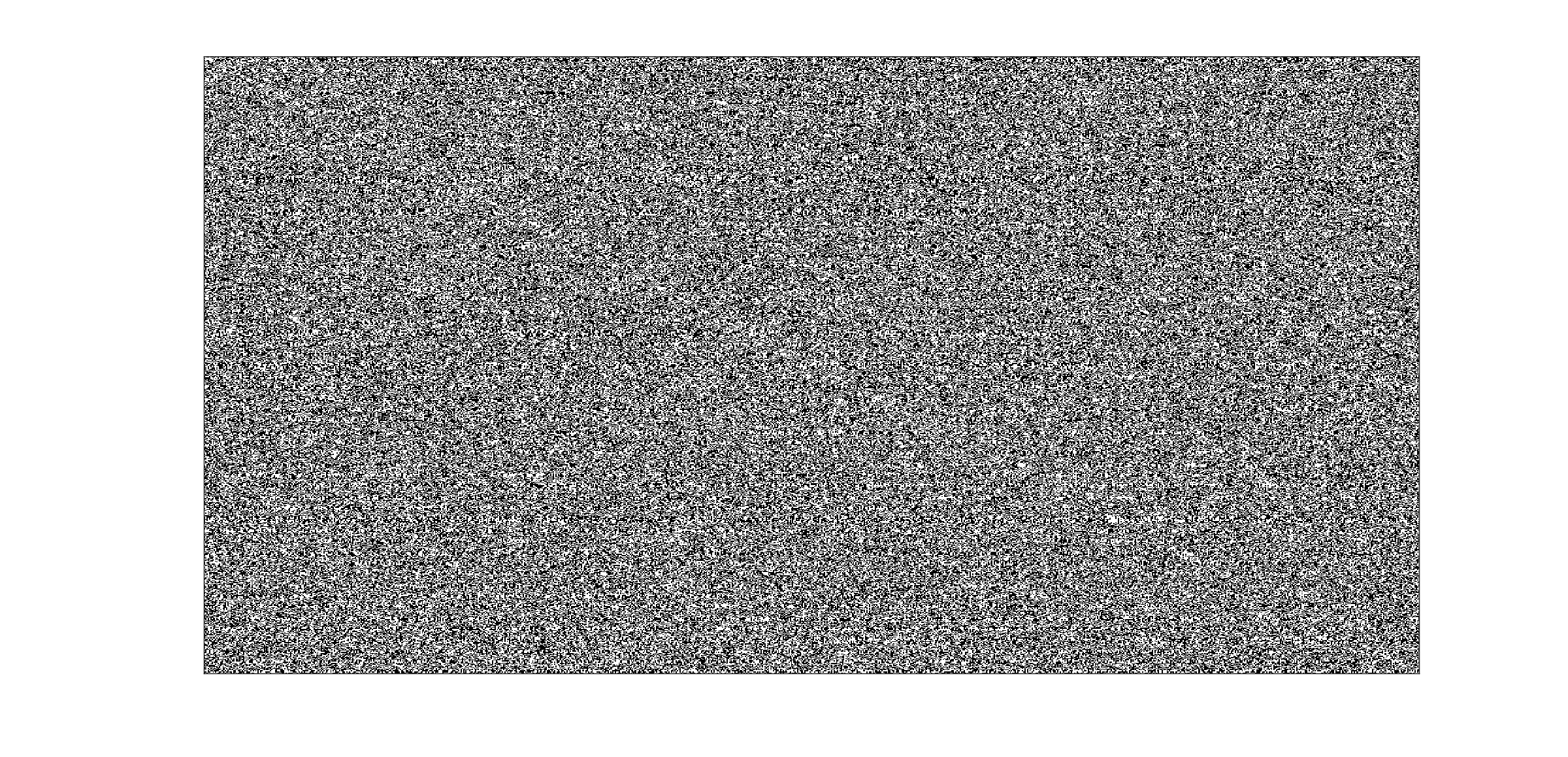}
	};
	
	\draw[-{Stealth[scale=1.5]}, thick] (img1.east) -- ++(0.55cm,0) -- (img2.west);
	\draw[-{Stealth[scale=1.5]}, thick] (img2.east) -- ++(0.55cm,0) -- (img3.west);
\end{tikzpicture}
\caption{Pixel representations of graphs with $n=10, 100, 1000$.}
\label{fig:graphon}
\end{figure}

One of our aims is to examine patterns and interactions among individuals through the lens of social network structures. In this context, graph theory provides a well-established framework for modeling such connections. As datasets increase in size, graphons arise as a continuous counterpart of finite graphs, enabling the analysis of large-scale interaction patterns; see, for instance, \cite{ Coppini2022note, During2024breaking,glasscock2015graphon, lovasz2006limits}.  
We begin by introducing an unweighted graph $\mathcal{G}$, defined as a pair $(\mathcal{V}, \mathcal{E})$, where $\mathcal{V}(\mathcal{G})$ denotes the set of vertices (or nodes) and $\mathcal{E}(\mathcal{G})$ denotes the set of edges connecting pairs of distinct nodes, all with identical weight. The associated adjacency matrix $\mathcal{A}(\mathcal{G}) \in \{0,1\}^{n \times n}$ is defined entrywise by
\begin{equation}\nonumber
\mathcal{A}_{ij}(\mathcal{G}) =
\begin{cases}
	1 & \text{if } (i, j) \in \mathcal{E}(\mathcal{G}), \\[4pt]
	0 & \text{otherwise.}
\end{cases}
\end{equation}
These matrices are commonly visualized through pixel representations, namely discretizations of the unit square $[0,1]\times[0,1]$ into an $n\times n$ grid of squares of size $\frac{1}{n}\times\frac{1}{n}$, where $n$ is the number of nodes. According to the adjacency matrix, the pixel located at $\bigl(\frac{i-1}{n}, \frac{j-1}{n}\bigr)$ is defined as
\begin{equation}\nonumber
\text{Pixel}(x,y)=
\begin{cases}
	\text{black} & \text{if } x\in\bigl[\tfrac{i-1}{n},\tfrac{i}{n}\bigr),~y\in\bigl[\tfrac{j-1}{n},\tfrac{j}{n}\bigr) \text{ and } \mathcal{A}_{ij}(\mathcal{G})= 1,\\[4pt]
	\text{white} & \text{otherwise.}
\end{cases}
\end{equation}
\begin{figure}[h!]
\centering
\begin{tikzpicture}
	\node (matrix) at (0,0) {
		$\begin{pmatrix}
			0 & 1 & 0 & 1 & 0 & 1 \\
			1 & 1 & 1 & 0 & 0 & 0 \\
			0 & 1 & 0 & 1 & 0 & 1 \\
			1 & 0 & 1 & 0 & 1 & 0 \\
			0 & 0 & 0 & 1 & 1 & 1 \\
			1 & 0 & 1 & 0 & 1 & 0 
		\end{pmatrix}$
	};
	
	\node (image) at (6,0) {
		\includegraphics[width=5cm,height=3.5cm]{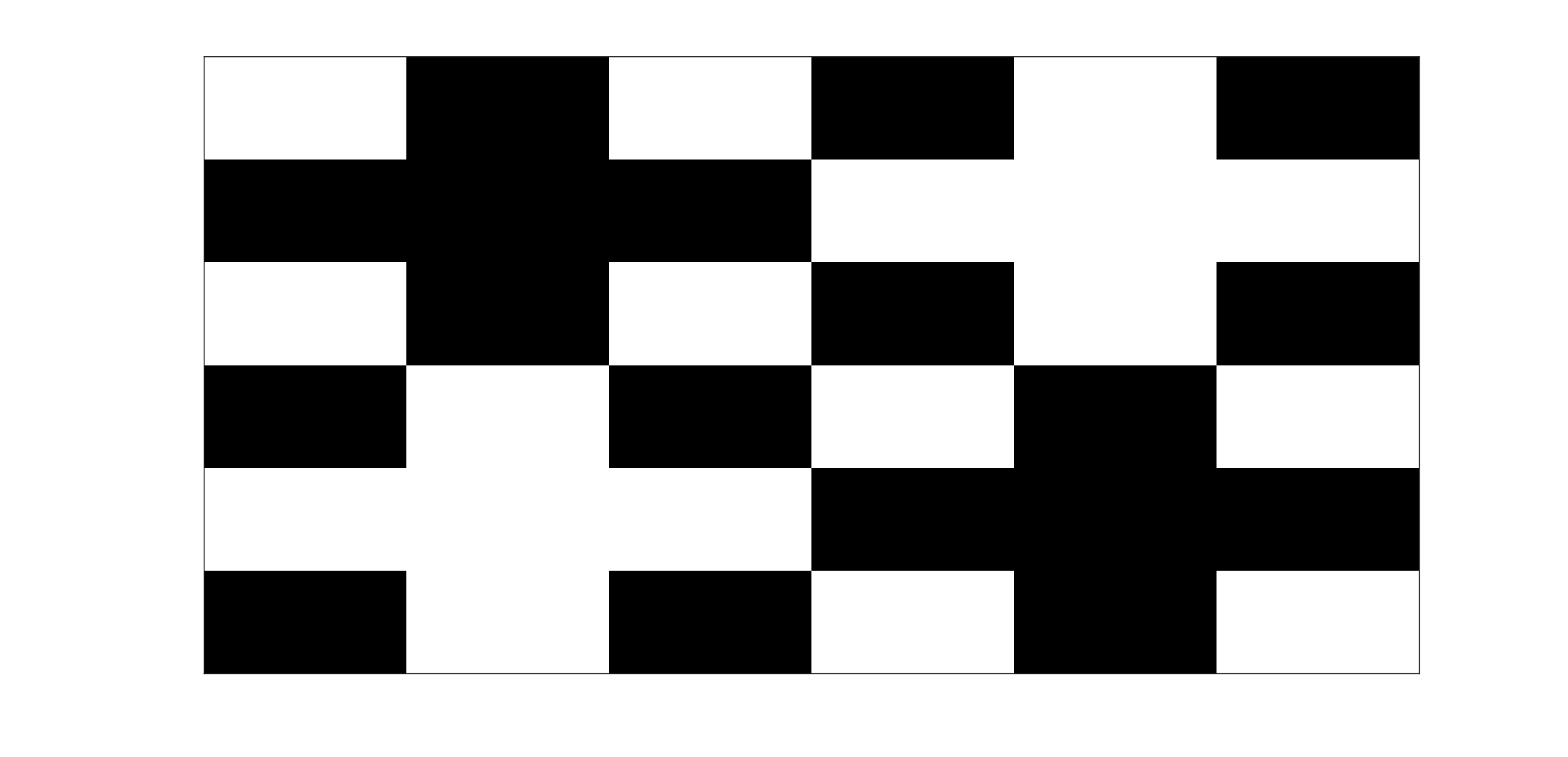}
	};
	
	\draw[-{Stealth[scale=1.5]}, thick] (matrix.east) -- (image.west);
\end{tikzpicture}
\caption{$6 \times 6$ adjacency matrix and corresponding pixel visualization.}
\end{figure}
In this framework, the notion of a graphon formalizes the convergence of sequences of graphs toward a smooth, continuous limit. More precisely, a graphon represents the limit object of a sequence of large, dense graphs (see Figure \ref{fig:graphon} for an illustration).
A graphon $\mathcal{B}$ is defined as a bounded, measurable, and symmetric function $\mathcal{B} : \varPsi \times \varPsi \to [0,1]$, with $\varPsi \subset [0,1]$. To interpret the space of graphons as the completion of the space of finite graphs and to make convergence precise, one introduces a distance $D(\mathcal{B}, \mathcal{C})$ between two graphons $\mathcal{B}$ and $\mathcal{C}$ (see \cite{glasscock2015graphon}) given by
\[
D(\mathcal{B}, \mathcal{C})
= \inf_{\varphi, \psi} \sup_{S, T} \left|
\int_{S \times T} \mathcal{B}(\varphi(x), \varphi(y))
-\mathcal{C}(\psi(x), \psi(y)) \, dx \, dy
\right|,
\]
where the infimum is taken over all relabelings $\varphi$ and $\psi$, and the supremum is taken over all measurable subsets $S,T \subset [0,1]$, up to weak isomorphism. This distance quantifies the maximal discrepancy between the integrals of the two graphons over measurable rectangles, minimized over all possible relabelings. Under these definitions, one can show that every graphon arises as the $D$-limit of a sequence of finite graphs.

We assume that $x \in \varPsi = [0,1]$ represents the position of individuals in the network (i.e., their connectivity) over a given graphon $\mathcal{B}$, while $w \in \Phi = [-1,1]$ encodes the individual opinion regarding protection measures.
In particular, $w = 1$ corresponds to full adherence to protective behaviors, whereas $w = -1$ represents strong opposition to such measures (including social distancing, mask wearing, public health guidelines, or vaccination), potentially driven by misinformation or ideological beliefs.
In the following, we describe the role of the physical contacts into the epidemic dynamics.

\subsection{Social contact dynamics}\label{SEIR-model_sub2}
The formation of the distribution of social contacts can be described by accounting for typical human behavioral patterns, and in particular the tendency, in the absence of epidemics, to seek opportunities for socialization (see \cite{Perthame} for details). As observed in \cite{French}, individuals tend to stabilize around a characteristic number of daily contacts, which depends on the social habits of a given country. We denote this reference value by $\bar{z}_M$, interpreted as the mean number of daily contacts in the population under investigation. It is worth noting that, in the presence of an epidemic, the characteristic mean number of contacts $\bar{z}_M$ may reasonably vary in time, even in the absence of external interventions such as lockdowns, due to the perceived risk associated with social interactions. Accordingly, even when not explicitly stated, we will assume $\bar{z}_M = \bar{z}_M(t)$.

To describe the disease dynamics, in the following, we adopt a kinetic compartmental approach based on the Susceptible-Exposed-Infected-Recovered (SEIR) model \cite{hethcote2000mathematics} taking into account the physical contacts, coupled with the evolution of individuals’ opinion distributions over a graphon-based social structure. Although other compartmental epidemic models could be considered depending on the desired level of detail, the SEIR paradigm provides a suitable framework for our purposes. We then assume that individuals belonging to different classes may exhibit different mean numbers of social contacts, infected for instance more likely have a lower level of average contacts. Accordingly, the microscopic update of the number of contacts for an individual in the class $\mathcal J\in\mathcal C=\{S,E,I,R\}$ is modeled as
\begin{equation}\label{coll}
	z_J^* = z - \Phi^\varepsilon\!\left(\frac{z}{\bar z_J}\right)z + \eta_\varepsilon z.
\end{equation}
In the above equation, in a single interaction, the number of contacts $z$ may change due to two distinct mechanisms, both proportional to $z$. The first contribution, governed by the coefficient $\Phi^\varepsilon(\cdot)$, may take both positive and negative values and represents the predictable variation of social contacts driven by individual behavioral adaptation. The second term accounts for random fluctuations and models the intrinsic unpredictability of the process.
We assume that the random variable $\eta_\varepsilon$ has zero mean and variance of order $\varepsilon>0$, namely
\[
\langle \eta_\varepsilon \rangle = 0, 
\qquad 
\langle \eta_\varepsilon^2 \rangle = \varepsilon \lambda,
\qquad \lambda > 0.
\]
Inspired by \cite{Perthame}, we choose $\Phi^\varepsilon$ as a function of the normalized variable $s = z/\bar z_J$, defined by
\begin{equation}\label{vd}
	\Phi_\delta^\varepsilon(s)
	=
	\mu \,
	\frac{\exp\!\left(\varepsilon \frac{s^\delta - 1}{\delta}\right)-1}
	{\exp\!\left(\varepsilon \frac{s^\delta - 1}{\delta}\right)+1},
	\qquad s \ge 0,
\end{equation}
where $\mu$ denotes the maximal variation of $z$ attainable in a single interaction, $0<\delta \le 1$ characterizes the intensity of the behavioral response, and $\varepsilon>0$ is the interaction intensity parameter. Therefore, the regime $\varepsilon \ll 1$ corresponds to small variations of the expected change $\langle z_J^* - z\rangle$, and both the adaptive and random effects are consistently scaled with the interaction intensity $\varepsilon$. We emphasize that the function $\Phi^\varepsilon$ plays the role of the so-called value function in the prospect theory of Kahneman and Tversky \cite{TK}. Inspired by this theory, it incorporates the mathematical features of expected human behavior. In particular, the main hypothesis is that, relative to the mean value $\bar{z}_J$, it is perceived as easier to increase the value of $z$, that is, individuals tend to seek larger networks than to decrease it.

Once the elementary interaction \eqref{coll} is defined, the sole time evolution of the distribution of the number $z$ of social contacts can be described by resorting to a kinetic collision-like approach, see \cite{pareschi2013interacting}. More precisely, let $f_J(z,t)$ denote the density of individuals in class $J \in \mathcal{C}=\{S,E,I,R\}$ having $z \in \mathbb{R}_+$ social contacts at time $t$ (retrospectively of the other quantities at the moment). Then the effect of the interaction can be expressed in weak form as follows: for any smooth test function $\varphi(z)$ (representing an observable quantity), we have
\begin{equation}
	\label{kin-w}
	\dfrac{d}{dt}\int_{\mathbb{R}_+}\varphi(z) f_J(z,t)\,dz
	=
	\Bigg\langle 
	\int_{\mathbb{R}_+} B(z)\,\bigl(\varphi(z_J^*)-\varphi(z)\bigr)\, f_J(z,t)\,dz
	\Bigg\rangle .
\end{equation}
Here, the expectation operator $\langle \cdot \rangle$ accounts for the randomness introduced by the variable $\eta_\varepsilon$ in the microscopic update rule \eqref{coll}, whereas the function $B(z)$ represents the interaction frequency. The right-hand side of \eqref{kin-w} measures, at time $t>0$, the variation of the density in class $J$ due to individuals whose number of contacts changes from $z$ to $z_J^*$ as a result of the interaction.
Following the approach proposed in \cite{Perthame}, we specify the interaction kernel $B(z)$ by assuming that the frequency of changes in the number of social contacts depends on $z$ itself through the power law
\[
B(z) = z^{-b},
\]
for some constant $b>0$. This choice assigns a lower interaction frequency to individuals who already have a large number of contacts, while attributing a higher interaction frequency to individuals with a small number of contacts as natural.
The parameter $b$ can be selected by observing that, for small values of $z$, the mean individual rate of variation predicted by the value function behaves as $z^{\delta-1}$. Therefore, choosing $b=\delta$ yields a collective rate of variation that is independent of $\delta$, whereas $\delta$ still retains its role in characterizing the individual behavioral response encoded in the value function. 

We now assume that the time scale governing the evolution of social contacts is sufficiently fast so that, with respect to the time scale of the epidemic and opinion dynamics, the contact distribution can be regarded as quasi-stationary, we only admit that the average number $\bar{z}_M$ may change depending on the epidemic evolution. Under this assumption, it is meaningful to study the stationary solution of \eqref{kin-w}. To investigate the long-time behavior, we introduce the time scaling
\[
\tau = \varepsilon t, 
\qquad 
f_{J,\varepsilon}(z,\tau) = f_J\!\left(z,\frac{\tau}{\varepsilon}\right),
\]
which separates the time scale of the epidemic and opinion dynamics from the (faster) time scale of the contact dynamics (even if the dependence on these quantities is not explicitly given at the moment). Then, by standard arguments in kinetic theory (see \cite{pareschi2013interacting}), one can derive the corresponding Fokker--Planck equation from \eqref{kin-w}
\begin{equation}\label{fp_contacts}
	\frac{\partial}{\partial \tau} f_J(z,\tau)
	=
	\frac{\mu}{2\delta}\frac{\partial}{\partial z}
	\left[
	z^{1-\delta}
	\left(
	\left(\frac{z}{\bar z_J}\right)^{\delta}-1
	\right)
	f_J(z,\tau)
	\right]
	+
	\frac{\lambda}{2}\frac{\partial^2}{\partial z^2}
	\left(
	z^{2-\delta} f_J(z,\tau)
	\right).
\end{equation}
The equilibrium distribution of this surrogate Fokker--Planck model \eqref{fp_contacts} can be computed explicitly. Introducing the ratio $\nu = \mu/\lambda$, the stationary states are given by
\begin{equation}\label{equilibrio}
	f_J^\infty(z)
	=
	C_J(\bar z_J,\delta,\nu)\,
	z^{\nu/\delta+\delta-2}
	\exp\left\{
	-\frac{\nu}{\delta^2}
	\left(\frac{z}{\bar z_J}\right)^\delta
	\right\},
\end{equation}
where $C_J>0$ is a normalization constant. The steady state \eqref{equilibrio} can be equivalently rewritten as a generalized Gamma probability density $f_\infty(z;\theta,\chi,\delta)$, namely
\begin{equation}\label{equili}
	f_{J,\infty}(z;\theta,\chi,\delta)
	=
	\frac{\delta}{\theta^\chi}\,
	\frac{1}{\Gamma\!\left(\chi/\delta\right)}
	\,z^{\chi-1}
	\exp\left\{
	-\left(\frac{z}{\theta}\right)^\delta
	\right\},
\end{equation}
characterized by the shape parameter $\chi>0$, the scale parameter $\theta>0$, and the exponent $\delta>0$. In the present setting, these parameters are given by
\begin{equation}\label{para}
	\chi = \frac{\nu}{\delta}+\delta-1,
	\qquad
	\theta = \bar z_J\left(\frac{\delta^2}{\nu}\right)^{1/\delta}.
\end{equation}

\subsection{A model of opinion-driven epidemic dynamics with graphon based connectivity structure}\label{SEIR-model_sub3}
\begin{figure}[h!] \centering \begin{tabular}{ccc} \includegraphics[height=3.6cm]{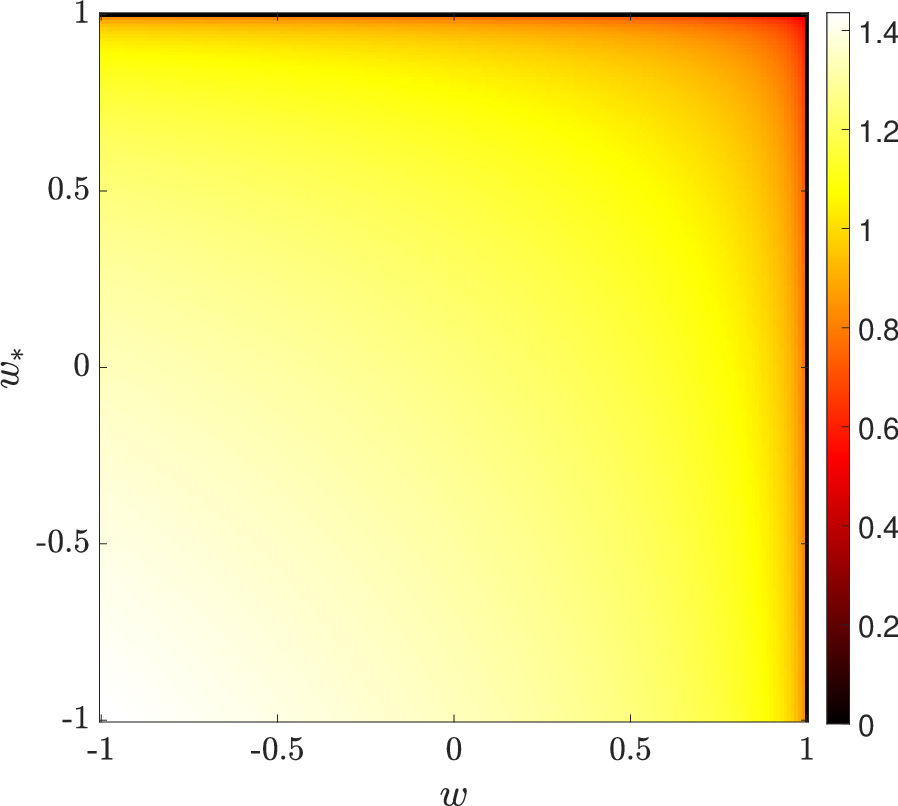} & \includegraphics[height=3.6cm]{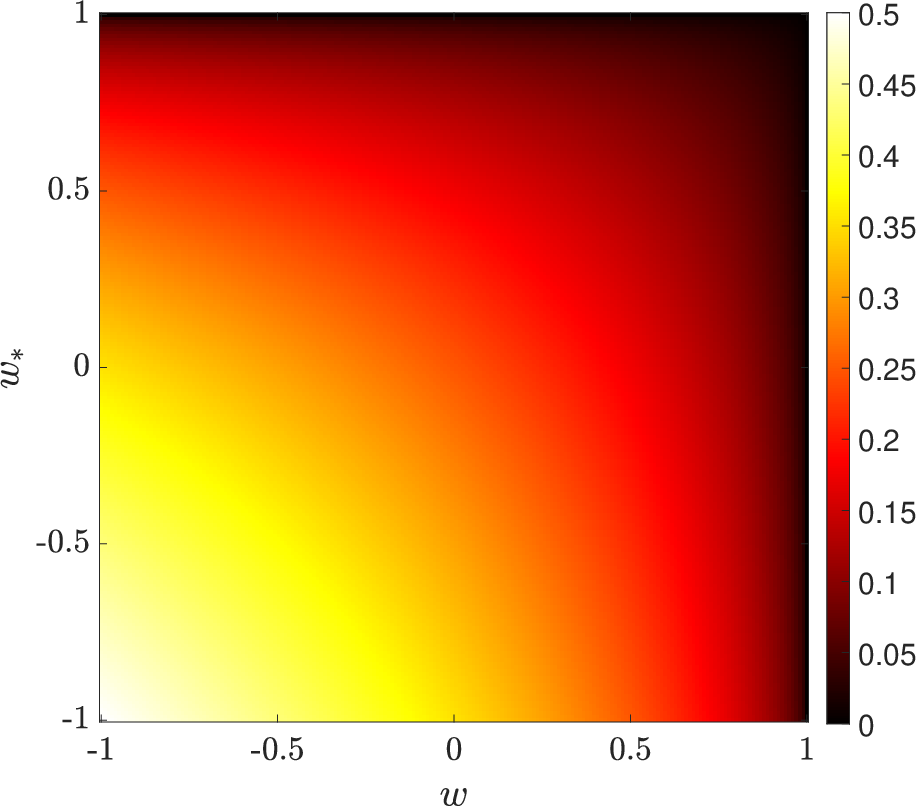} & \includegraphics[height=3.6cm]{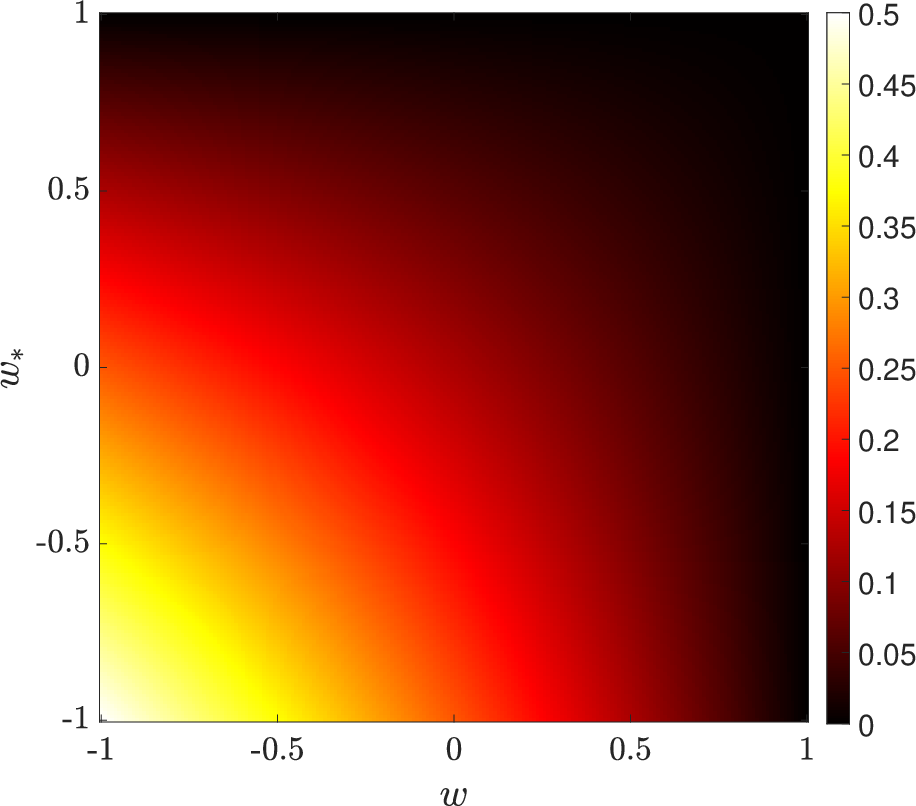} \\
		$\eta = 0.1$ & $\eta = 0.5$ & $\eta = 1$ \end{tabular} \caption{Function $	\kappa(w, w_*)$ in \eqref{eq:kappa} for $\eta = 0.1$ (left), $\eta = 0.5$ (middle) and $\eta = 1$ (right). For all cases, $\beta = 0.5$.} \label{fig:disease_transmission} \end{figure}

We introduce now the complete model. Let $f_{\mathcal J}(z,x,w,t)$, $\mathcal J\in\mathcal C=\{S,E,I,R\}$, denote the density of agents at time $t\ge 0$ with opinion $w$, network position $x$ (as defined by the graphon $\mathcal B$) and number of physical contact $z$, belonging to compartment $\mathcal J$. The total density of agents is then
\[
\sum_{\mathcal J\in\mathcal C} f_{\mathcal J}(z,x,w,t)=f(z,x,w,t), 
\qquad 
\int_{\R^+\times\varPsi\times\varPhi} f_{\mathcal J}(z,x,w,t)\,dz\,dx\,dw =1,
\qquad \forall t\ge 0.
\]
In this setting, opinion dynamics and graphon-driven social interactions are integrated with the SEIR structure through the kinetic system
\begin{equation}
	\partial_t \mathbf{f}(z,x,w,t)
	= \boldsymbol{\mathcal K}(z,x,w,\mathbf{f}(z,x,w,t))
	+ \frac{1}{\tau}\,\boldsymbol{\mathcal Q}(\mathbf{f}(z,x,w,t)),
	\label{general_model}
\end{equation}
where $\mathbf f=(f_{\mathcal J})_{\mathcal J\in\mathcal C}$.  
The vector $\boldsymbol{\mathcal K}$ describes disease transmission and compartmental transitions, with each component depending on the individual opinion $w$ on the number of physical connections $z$ and implicitly through the opinion dynamics on the social network position $x$.  
The operator $\boldsymbol{\mathcal Q}$ is a integral-type term encoding opinion variation due to social interactions mediated by the graphon structure $\mathcal B(x,y)$.  
The parameter $\tau>0$ represents the relative time scale of opinion dynamics with respect to the epidemic evolution. Under this framework, the SEIR model within a graphon structure is governed by the following set of kinetic equations:
\begin{equation}
	\label{eq:SEIR}
	\left\{
	\begin{aligned}
		\partial_t f_S &= - \mathcal K(f_I,f_S)(z,x,w,t)
		+ \frac{1}{\tau}\sum_{\mathcal J\in\mathcal C}\mathcal Q(f_S,f_{\mathcal J})(z,x,w,t), \\[4pt]
		\partial_t f_E &= \mathcal K(f_I,f_S)(z,x,w,t) - \sigma_E f_E
		+ \frac{1}{\tau}\sum_{\mathcal J\in\mathcal C}\mathcal Q(f_E,f_{\mathcal J})(z,x,w,t), \\[4pt]
		\partial_t f_I &= \sigma_E f_E - \gamma f_I
		+ \frac{1}{\tau}\sum_{\mathcal J\in\mathcal C}\mathcal Q(f_I,f_{\mathcal J})(z,x,w,t), \\[4pt]
		\partial_t f_R &= \gamma f_I
		+ \frac{1}{\tau}\sum_{\mathcal J\in\mathcal C}\mathcal Q(f_R,f_{\mathcal J})(z,x,w,t).
	\end{aligned}
	\right.
\end{equation}
Here, $1/\sigma_E$ denotes the average latent period and $1/\gamma$ the mean infectious period. The operator $\mathcal Q(\cdot,\cdot)$ models opinion interactions and will be specified later. The disease transmission is characterized by the local incidence rate
\begin{equation}
	\mathcal K(f_I,f_S)(z,x,w,t)
	= f_S(z,x,w,t)
	\int_{\R^+\times\varPsi\times\varPhi}\kappa(w,w_*;z,z_*)\,f_I(z_*,x_*,w_*,t)\,dz_*\,dx_*\,dw_*,
	\label{eq:incidenc_rate}
\end{equation}
where $\kappa$ represents the rate of infection modulated by opinions, physical contacts and, indirectly, by network connectivity. A possible choice is
\begin{equation}
	\kappa(w,w_*;z,z_*)
	= \frac{\beta}{4^\eta}(1-w)^\eta(1-w_*)^\eta z^{\alpha}z^{\alpha_*}_*,
	\label{eq:kappa}
\end{equation}
with $\beta$ the baseline transmission rate, $\eta>0$ the parameter encoding the effect of protective behavior while $\alpha$, $\alpha_*$ positive constants which cause the incidence rate to be dependent on the product of the number of contacts of susceptible and infected people. With the above choice, individuals with negative opinions ($w\approx -1$) exhibit higher susceptibility, especially when interacting with similarly non-compliant individuals. Conversely, when either $w\approx 1$ or $w_*\approx 1$, the transmission probability decreases substantially. A visualization of \eqref{eq:kappa} in the case $z=z_*=1$, $\beta = 0.5$ is provided in Figure~\ref{fig:disease_transmission}. Note that the kernel \eqref{eq:kappa} does not explicitly depend on network positions $(x,y)$; however, the opinion dynamics driven by the graphon $\mathcal B(x,y)$ influences the distribution of opinions over time, thereby indirectly affecting epidemic spread as specified later. In particular, highly connected nodes (influencers) and the connectivity patterns encoded by $\mathcal B$ play a central role in shaping collective behavior. 

\subsection{Opinion formation under graphon structures} \label{opinion-model}
We now focus on the opinion dynamics. We consider a binary interaction system in which two agents with opinions $w, w_*$ and network positions $x, y$ (on the graphon $\mathcal{B}$) update their post-interaction opinions according to
\begin{equation}
	\begin{aligned}
		w' &= w + \alpha_{\mathcal{JH}}\,\mathit{P}(w,w_*,x,y)\,(w_* - w)
		+ \zeta_{\mathcal{HJ}}\,\mathcal D(x,w),\\
		w_*' &= w_* + \alpha_{\mathcal{JH}}\,\mathit{P}(w_*,w,y,x)\,(w - w_*)
		+ \zeta_{\mathcal{JH}}\,\mathcal D(y,w_*),
	\end{aligned}
	\label{binaryinteraction}
\end{equation}
where $\alpha_{\mathcal{JH}} \in [0,\tfrac{1}{2}]$ is the compromise parameter and $\mathit{P} \in [0,1]$ modulates the interaction strength based on the agents’ positions on the graphon. The function $\mathcal D(x,w)$ represents local diffusion (self-thinking or access to information), while the random variables $\zeta_{\mathcal{HJ}}, \zeta_{\mathcal{JH}}$ are centered, independent, and identically distributed:
\[
\langle \zeta_{\mathcal{HJ}} \rangle = \langle \zeta_{\mathcal{JH}} \rangle = 0, 
\qquad
\langle \zeta_{\mathcal{HJ}}^{2} \rangle = \langle \zeta_{\mathcal{JH}}^{2} \rangle = \sigma^{2} > 0.
\]
A common choice is $\mathcal D(x,w)$ vanishing at the boundary of $\Phi = [-1,1]$, ensuring that opinions remain admissible. The update rule thus combines a deterministic compromise term with a stochastic fluctuation modulated by $\mathcal D$. We now discuss different models for the interaction operator.

\medskip
\noindent\textbf{Opinion-based interaction.}
If the interaction kernel depends only on opinion similarity, one may use non-increasing functions such as $\mathit{P}(|w|)$ \cite{Toscani2006} or $\mathit{P}(|w-w_*|)$, the latter corresponding to the bounded-confidence framework \cite{Hegselmann2002opinion}. A standard choice is
\begin{equation}
	P(w,w_*) = \chi(|w-w_*| \le \Delta), \qquad w,w_* \in [-1,1],
	\label{bounded_confidence}
\end{equation}
where $\Delta \in [0,2]$ is the confidence threshold \cite{pareschi2013interacting}. In this case, agents interact only when their opinions are sufficiently close, producing non-local effects such as clustering or fragmentation. This formulation neglects network heterogeneity, as interactions depend solely on opinion proximity.

\medskip
\noindent\textbf{Network-based interaction.}
If instead the interaction kernel depends only on network positions, a common choice \cite{During2024breaking} is
\begin{equation}
	\mathit{P}(x,y)=\exp\!\Big(-\gamma\,\tfrac{d_i(x)}{d_i(y)}\Big),
	\qquad
	\mathit{P}(x,y)=\Big(1+\tfrac{d_i(x)}{d_i(y)}\Big)^{-\gamma},
	\qquad \gamma>0,
	\label{graphon_cases}
\end{equation}
where the in-degree function is defined as
\[
d_i(x)=\int_{\varPsi} \mathcal{B}(x,y)\,dy.
\]
The ratio $\tfrac{d_i(x)}{d_i(y)}$ determines how connectivity shapes influence: highly connected agents tend to be more resistant to external persuasion, while weakly connected agents are more susceptible. The exponential and power-law forms differ in how sharply influence decays.

\medskip
\noindent\textbf{Coupled opinion--network interaction.}
Generalizing and following partly \cite{albi2025impact}, a more realistic kernel depends on both opinion similarity and network connectivity:
\begin{equation}
	P(w,w_*,x,y)=\chi\big(|w-w_*|\le\Delta(x,y)\big)\,\mathcal{B}(x,y),
	\label{opinion_network}
\end{equation}
where $\Delta(x,y)\in[0,2]$ is a position-dependent confidence threshold and $\mathcal{B}(x,y)\in[0,1]$ encodes the connection intensity. This formulation couples social structure and opinion proximity, yielding richer dynamics capable of capturing both behavioral similarity and network effects.

\medskip
In general, since the opinion variable $w$ is defined on the bounded domain $[-1,1]$, it is natural to restrict interactions to those that preserve this domain. The following result provides a sufficient condition ensuring this property (see \cite{Toscani2018opinion,Albi2024,During2024breaking} for details).
\begin{proposition}
	Assume that for any $x \in [0,1]$, $\mathcal D(x,\pm1)=0$. Then, for pre-interaction values $w,w_* \in [-1,1]$, the binary interaction \eqref{binaryinteraction} preserves the bounds, i.e., $w', w_*' \in [-1,1]$, provided that
	\begin{equation}
		0 < \mathit{P}(w, w_*, x, y) \leq 1, \quad 0 < \alpha_{\mathcal{JH}} \leq \tfrac{1}{2}, \quad 
		|\zeta_{\mathcal{JH}}| \leq (1 - \lambda^*) d,
		\label{prop_1}
	\end{equation}
	where
	\begin{equation}
		\lambda^* = \min_{\substack{w,w_* \in [-1,1] \\ x,y \in [0,1]}} 
		\alpha_{\mathcal{JH}}\, \mathit{P}(w, w_*, x, y), 
		\qquad
		d = \min_{\substack{w \in [-1,1] \\ x \in [0,1]}}
		\left\{
		\frac{1 - |w|}{\mathcal D(x,w)} : \mathcal D(x,w) \neq 0
		\right\}.
		\label{prop_2}
	\end{equation}
\end{proposition}
Moreover, under the same assumptions, if the interaction function $\mathit{P}$ is symmetric, the mean opinion is preserved during interactions, while in the absence of stochastic terms the mean energy is dissipated, indicating convergence toward a compromise.

We now turn to a continuous description of the microscopic binary interaction \eqref{binaryinteraction}, given by a Boltzmann-type equation. For simplicity, in the remainder of this section we neglect the dependence on the number of physical contacts. We denote by $f_{\mathcal J}(x,w,t)$ the joint probability density of opinions and network positions, while $\mathcal Q_{\mathcal J}(f_{\mathcal J},f_{\mathcal H})(x,w,t)$ represents the collision operator accounting for interactions between distributions.
The resulting kinetic equation reads
\begin{equation}
	\partial_t f_{\mathcal J}(x,w,t)
	=
	\frac{1}{\tau}
	\sum_{\mathcal H \in \mathcal C}
	\mathcal Q_{\mathcal J}(f_{\mathcal J},f_{\mathcal H})(x,w,t),
	\label{collisionoperater1}
\end{equation}
where $\tau>0$ denotes the interaction frequency. The weak formulation of \eqref{collisionoperater1}, tested against a smooth function
$\varphi(x,w)$, is given by
\begin{equation}
	\begin{aligned}
		&\frac{\mathrm{d}}{\mathrm{d}t}
		\int_{\varPsi \times \varPhi}
		f_{\mathcal J}(x,w,t)\,\varphi(x,w)\,
		\mathrm{d}x\, \mathrm{d}w
		=
		\int_{\varPsi \times \varPhi}
		\mathcal Q_{\mathcal J}(f_{\mathcal J},f_{\mathcal H})(x,w,t)\,
		\varphi(x,w)\,
		\mathrm{d}x\, \mathrm{d}w=
		\\[6pt]
		&=
		\frac{1}{2}
		\int_{\varPsi^2 \times \varPhi^2}
		\mathcal B(x,y)
		\Big\langle
		\varphi(x,w') + \varphi(y,w_*')
		-
		\varphi(x,w) - \varphi(y,w_*)
		\Big\rangle
		\\[3pt]
		&\qquad\qquad\times
		f_{\mathcal J}(x,w,t)\,
		f_{\mathcal H}(y,w_*,t)\,
		\mathrm{d}x\, \mathrm{d}y\, \mathrm{d}w\, \mathrm{d}w_* .
	\end{aligned}
	\label{weekform}
\end{equation}
In \eqref{weekform}, the operator $\langle \cdot \rangle$ denotes the expectation
with respect to the random variables
$\zeta_{\mathcal{JH}}$ and $\zeta_{\mathcal{HJ}}$.
The parameter $\tau$ acts as a scaling factor reflecting the fact that opinion
formation and disease diffusion may occur on different time scales: smaller values
of $\tau$ correspond to faster opinion dynamics relative to the epidemic process.
Finally, we emphasize that the interaction operator
$\mathcal Q_{\mathcal J}(\cdot,\cdot)$ does not depend explicitly on the underlying
network realization. Indeed, the social structure is assumed to be stationary in
time, while the graphon $\mathcal B(\cdot,\cdot)$ characterizes the connectivity
pattern of the system by modulating the interaction intensity between agents
located at positions $x,y \in [0,1]$ in the social space.

The qualitative behavior of the kinetic model \eqref{collisionoperater1} is analytically challenging due to its intrinsic complexity. To investigate its long-time dynamics, we consider a Fokker--Planck-type approximation obtained through the so-called quasi-invariant opinion limit \cite{Toscani2006}. 
This procedure relies on a suitable rescaling of the interaction parameters, leading to a regime of frequent but weak interactions, in analogy with the grazing collision limit for classical Boltzmann equations. From a modeling perspective, this corresponds to a situation in which opinions evolve through a large number of small interaction steps. In this regime, the kinetic description converges to a mean-field Fokker--Planck equation \cite{pareschi2013interacting}, which retains the essential microscopic interaction mechanisms while allowing for a more tractable analysis of equilibria and asymptotic behavior. To derive this limit model, we introduce, as before, a small parameter $\varepsilon > 0$ and consider the quasi-invariant scaling
\begin{equation}
	\alpha_{\mathcal{JH}} \mapsto \varepsilon \alpha_{\mathcal{JH}},
	\qquad
	\sigma_{\mathcal{JH}}^2 \mapsto \varepsilon \sigma_{\mathcal{JH}}^2.
	\label{eq:scaling}
\end{equation}
The rescaled random variables $\zeta_\varepsilon^{\mathcal{JH}}$ and
$\zeta_\varepsilon^{\mathcal{HJ}}$ are assumed independent with
\begin{equation}
	\begin{aligned}
		\langle \zeta_\varepsilon^{\mathcal{JH}} \rangle
		= \langle \zeta_\varepsilon^{\mathcal{HJ}} \rangle = 0, \quad 
		\langle (\zeta_\varepsilon^{\mathcal{JH}})^2 \rangle
		= \langle (\zeta_\varepsilon^{\mathcal{HJ}})^2 \rangle
		= \varepsilon \sigma_{\mathcal{JH}}^2, \quad 
		\langle (\zeta_\varepsilon^{\mathcal{JH}})^3 \rangle
		= \langle (\zeta_\varepsilon^{\mathcal{HJ}})^3 \rangle < \infty .
	\end{aligned}
	\label{eq:scaling_rv}
\end{equation}
Introducing the slow time scale $t_\varepsilon = \varepsilon t$, the weak form
of \eqref{collisionoperater1} becomes
\begin{equation}
	\frac{\mathrm{d}}{\mathrm{d}t_\varepsilon}
	\int_{\varPsi \times \varPhi}
	f_{\mathcal J,\varepsilon}(x,w,t_\varepsilon)\,\varphi(x,w)\,
	\mathrm{d}x\,\mathrm{d}w
	=
	\frac{1}{\varepsilon \tau}
	\sum_{\mathcal H\in\mathcal C}
	\mathcal Q_{\mathcal J}(f_{\mathcal J,\varepsilon},f_{\mathcal H,\varepsilon}).
	\label{eq:rescaled_weak}
\end{equation}
To derive the grazing limit, we perform a Taylor expansion of
$\varphi(x,w')$ around $w$:
\[
\begin{aligned}
	\langle \varphi(x,w') - \varphi(x,w) \rangle
	= \,&
	\partial_w \varphi(x,w)\,\langle w'-w\rangle
	+ \frac12 \partial_w^2 \varphi(x,w)\,\langle (w'-w)^2\rangle \\
	&
	+ \frac16 \partial_w^3 \varphi(x,\tilde w)\,\langle (w'-w)^3\rangle,
\end{aligned}
\]
for some $\tilde w$ between $w$ and $w'$. Using the scaling
\eqref{eq:scaling}--\eqref{eq:scaling_rv} and the interaction rule
\eqref{opinion-model}, we obtain
\begin{equation}
	\begin{aligned}
		\langle \varphi(x,w') - \varphi(x,w) \rangle
		&=
		\varepsilon \alpha_{\mathcal{JH}}
		\partial_w \varphi(x,w)\,
		\mathit P(w,w_\ast,x,y)(w_\ast-w)
		\\
		&\quad
		+ \frac{\varepsilon}{2}
		\partial_w^2 \varphi(x,w)\,
		\Big(
		\sigma_{\mathcal{JH}}^2 \mathcal D^2(x,w)
		+ \varepsilon \alpha_{\mathcal{JH}}^2
		\mathit P^2(w,w_\ast,x,y)(w_\ast-w)^2
		\Big)
		\\
		&\quad
		+ \varepsilon^2 \mathcal R_\varepsilon(\varphi),
	\end{aligned}
	\label{eq:taylor}
\end{equation}
where the remainder $\mathcal R_\varepsilon(\varphi)\to 0$ as
$\varepsilon\to0$. Substituting \eqref{eq:taylor} into \eqref{eq:rescaled_weak} and letting
$\varepsilon\to0$, we formally obtain the limiting Fokker--Planck operator
\begin{equation}
	\begin{aligned}
		\overline{\mathcal Q}_{\mathcal J}(f_{\mathcal J},f_{\mathcal H})
		&=
		- \alpha_{\mathcal{JH}}\,
		\partial_w\!\left(
		\mathcal P[f_{\mathcal H}](x,w,t)\, f_{\mathcal J}(x,w,t)
		\right)
		\\
		&\quad
		+ \frac{\sigma_{\mathcal{JH}}^2}{2}\,
		\partial_w^2\!\left(
		\mathcal L[f_{\mathcal H}](x,t)\,
		\mathcal D^2(x,w)\, f_{\mathcal J}(x,w,t)
		\right),
	\end{aligned}
	\label{eq:FP_operator}
\end{equation}
where
\begin{equation}
	\begin{aligned}
		\mathcal P[f_{\mathcal H}](x,w,t)
		&=
		\int_{\varPsi\times\varPhi}
		\mathcal B(x,y)\,
		\mathit P(w,w_\ast,x,y)(w_\ast-w)\,
		f_{\mathcal H}(y,w_\ast,t)\,
		\mathrm{d}y\,\mathrm{d}w_\ast,
		\\
		\mathcal L[f_{\mathcal H}](x,t)
		&=
		\int_{\varPsi\times\varPhi}
		\mathcal B(x,y)\,
		f_{\mathcal H}(y,w_\ast,t)\,
		\mathrm{d}y\,\mathrm{d}w_\ast.
	\end{aligned}
\end{equation}
The resulting Fokker--Planck model associated with
\eqref{collisionoperater1} reads
\begin{equation}
	\partial_t f_{\mathcal J}(x,w,t)
	=
	\frac{1}{\tau}
	\sum_{\mathcal H\in\mathcal C}
	\overline{\mathcal Q}_{\mathcal J}(f_{\mathcal J},f_{\mathcal H})(x,w,t),
	\label{collisionoperater3}
\end{equation}
supplemented with zero-flux boundary conditions at $w=\pm1$.

Under this quasi-invariant limit, the original Boltzmann-type operator
$\mathcal Q_{\mathcal J}$ in \eqref{collisionoperater1} is replaced by the reduced-complexity Fokker--Planck operator $\overline{\mathcal Q}_{\mathcal J}$ in \eqref{eq:FP_operator}. This asymptotic provides a mean-field description of the microscopic interaction dynamics, where the drift term encodes compromise effects driven by social interactions, while the diffusion term captures stochastic fluctuations associated with individual variability and information uncertainty.
In particular, the resulting model highlights how the graphon structure modulates both transport and diffusion mechanisms, yielding a nonlocal drift--diffusion equation whose coefficients depend on the underlying network topology.

We can now couple this description with the epidemic dynamics. Restoring the dependence on the number of physical contacts $z$, we obtain the following mean-field SEIR system:
\begin{equation}
	\left\{
	\begin{aligned}
		\partial_t f_S
		&= -\mathcal K(f_I,f_S)(z,x,w,t)
		+ \frac{1}{\tau}\sum_{\mathcal J\in\mathcal C}
		\overline{\mathcal Q}\!\left(f_S,f_{\mathcal J}\right),\\
		\partial_t f_E
		&= \mathcal K(f_I,f_S)(z,x,w,t) - \sigma_E\,f_E
		+ \frac{1}{\tau}\sum_{\mathcal J\in\mathcal C}
		\overline{\mathcal Q}\!\left(f_E,f_{\mathcal J}\right),\\
		\partial_t f_I
		&= \sigma_E\,f_E - \gamma\,f_I
		+ \frac{1}{\tau}\sum_{\mathcal J\in\mathcal C}
		\overline{\mathcal Q}\!\left(f_I,f_{\mathcal J}\right),\\
		\partial_t f_R
		&= \gamma\,f_I
		+ \frac{1}{\tau}\sum_{\mathcal J\in\mathcal C}
		\overline{\mathcal Q}\!\left(f_R,f_{\mathcal J}\right).
	\end{aligned}
	\right.
	\label{eq:SEIR_updated}
\end{equation}
System \eqref{eq:SEIR_updated} describes the epidemic evolution while consistently incorporating socio-behavioral effects through the opinion variable $w \in [-1,1]$, the number of contacts $z$, and the network position $x \in [0,1]$ encoded by the graphon $\mathcal B(x,y)$. In this framework, the spread of the disease is not only driven by biological parameters, but is dynamically modulated by the distribution of opinions and the underlying social structure, allowing for a more realistic representation of feedback mechanisms between behavior and epidemic propagation.

\begin{remark}
	Starting from the epidemic system \eqref{eq:SEIR_updated}, it is possible to derive a macroscopic description of the dynamics in terms of aggregate quantities. In particular, by introducing suitable moments of the distribution functions, one can formally obtain a system of ordinary differential equations of compartmental type, analogous to classical models such as SIR or SEIR \cite{Beckley2013modeling,Kendall1956deterministic,Kermack1927contribution}. The relevant macroscopic quantities can be recovered as following: the total masses are given by
	\[
	\rho_{\mathcal J}(t)=\int_{\R^+\times\varPsi\times\varPhi} f_{\mathcal J}(z,x,w,t)\,dz\,dx\,dw,
	\qquad \mathcal J\in\mathcal C,
	\]
	the corresponding mean opinions by
	\[
	m_{\mathcal J,w}(t)=\frac{1}{\rho_{\mathcal J}(t)}
	\int_{\R^+\times\varPsi\times\varPhi} w\, f_{\mathcal J}(z,x,w,t)\,dz\,dx\,dw,
	\qquad \mathcal J\in\mathcal C,
	\]
	while the mean number of physical contacts as
	\[
	m_{\mathcal J,z}(t)=\frac{1}{\rho_{\mathcal J}(t)}
	\int_{\R^+\times\varPsi\times\varPhi} z\, f_{\mathcal J}(z,x,w,t)\,dz\,dx\,dw,
	\qquad \mathcal J\in\mathcal C.
	\]
	The mixed moment between opinions and connectivity are defined as
	\[
	m_{\mathcal J,zw}(t)=\frac{1}{\rho_{\mathcal J}(t)}
	\int_{\R^+\times\varPsi\times\varPhi} zw\, f_{\mathcal J}(z,x,w,t)\,dz\,dx\,dw,
	\qquad \mathcal J\in\mathcal C.
	\]
	A key difference, however, is that in our framework the effective parameters of the macroscopic model (e.g., transmission and recovery rates) are not prescribed \emph{a priori} on an empirical basis, but rather emerge from the underlying microscopic interaction mechanisms, including opinion dynamics, network structure, and contact patterns. In this sense, classical compartmental models can be interpreted as reduced or limiting cases of the more general kinetic description proposed here. A detailed derivation of the corresponding macroscopic system and its relation to standard compartmental models is provided in Appendix~\ref{appendix}.
\end{remark}

\section{Graphon estimation from observed network data} \label{real-data-garphon}
The main difficulty in network modeling is to forecast node interaction
probabilities accurately without imposing overly restrictive functional forms.
Over the last decade, machine learning and statistics have advanced
significantly through the development of statistical models for network data \cite{ Airoldi2008mixed, Hoff2002latent,Nowicki2001estimation, Soufiani2012graphlet,Wasserman2006all}. Nevertheless, fitting
a given parametric family becomes increasingly challenging as network complexity
grows. The approach we use is the so-called nonparametric exchangeable graph method \cite{Chan2014consistent}. In the following we briefly describe the methodology.
Let $G=(G_{ij})_{i,j\in\mathbb N}$ be an infinite random adjacency array with
entries $G_{ij}\in\{0,1\}$. The array $G$ is (jointly) exchangeable if
\[
(G_{ij}) {:=}\; (G_{\theta(i)\,\theta(j)})
\qquad\text{for every permutation }\theta.
\]
By the Aldous--Hoover theorem, $G$ is exchangeable if and only if there exists a
random measurable function $\mathcal F:[0,1]^3\to\{0,1\}$ such that
\[
(G_{ij}) {:=}\; \bigl(\mathcal F(U_i,U_j,U_{ij})\bigr),
\]
where $(U_i)_{i\in\mathbb N}$ are i.i.d. uniformly distributed in $[0,1]$ latent
variables and $(U_{ij})_{i,j\in\mathbb N}$ are i.i.d. uniformly distributed in $[0,1]$
edge-specific random variables, independent of $(U_i)_{i\in\mathbb N}$. In this setting, graphons correspond to a canonical and widely used special case. In particular, a graphon is a symmetric measurable function
$\mathcal B:[0,1]^2\to[0,1]$ such that
\[
\mathcal F(U_i,U_j,U_{ij})
=
\mathbf 1_{\{\,U_{ij}\le \mathcal B(U_i,U_j)\,\}},
\]
i.e., conditional on the network positions $U_i$ and $U_j$, the edge $(i,j)$ is
present with probability $\mathcal B(U_i,U_j)$. Thus, a sample from a graphon can be
generated in two stages:
\begin{enumdescript}
	\item \textbf{Stage 1:} Assign to each node a latent (network) position
	\[
	U_i \stackrel{\text{i.i.d.}}{\sim} \mathrm{Uniform}[0,1].
	\]
	\item \textbf{Stage 2:} For each pair of nodes $i,j$, generate an edge according to
	\[
	G_{ij}\mid U_i,U_j \sim \mathrm{Bernoulli}\bigl(\mathcal B(U_i,U_j)\bigr).
	\]
\end{enumdescript}

\begin{figure}[htp]
	\centering
	\begin{tabular}{cccc}
		\includegraphics[height=3cm]{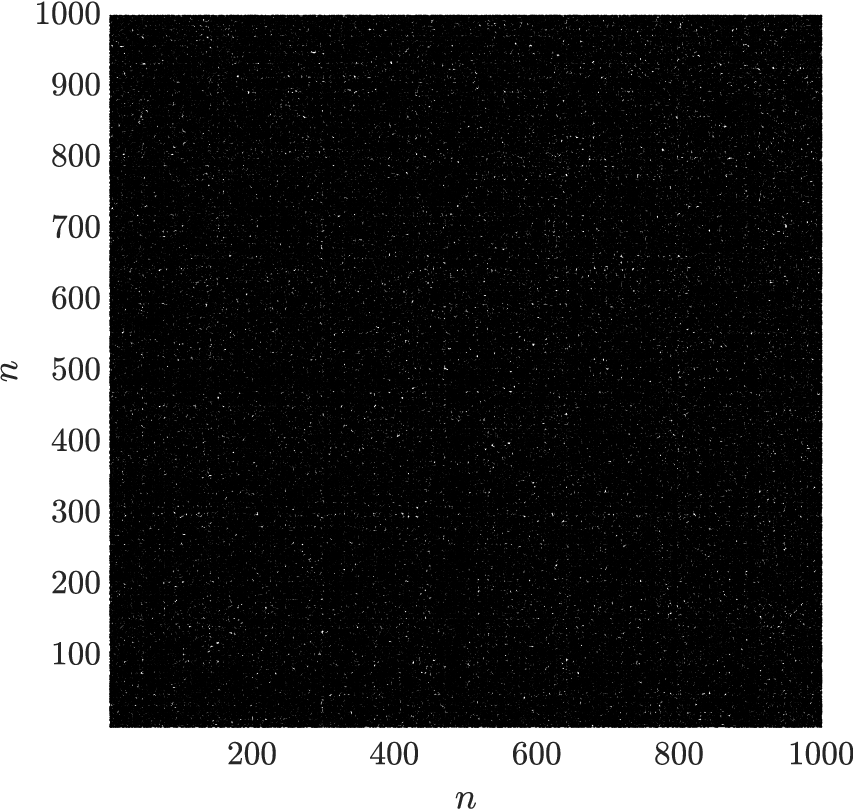} &
		\includegraphics[height=3cm]{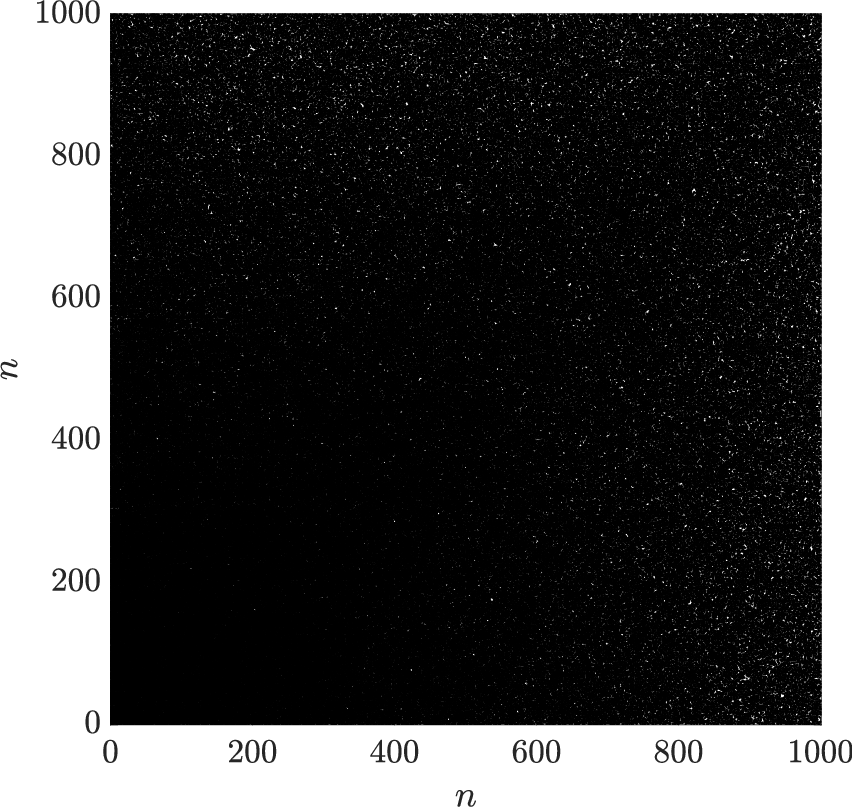} &
		\includegraphics[height=3cm]{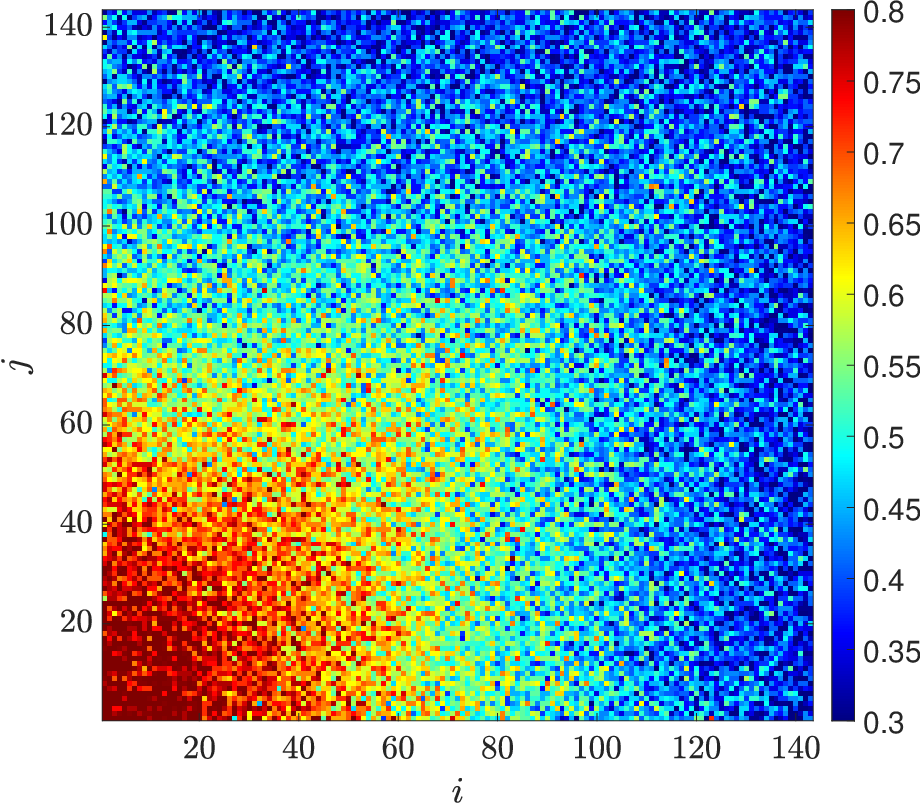} &
		\includegraphics[height=3cm]{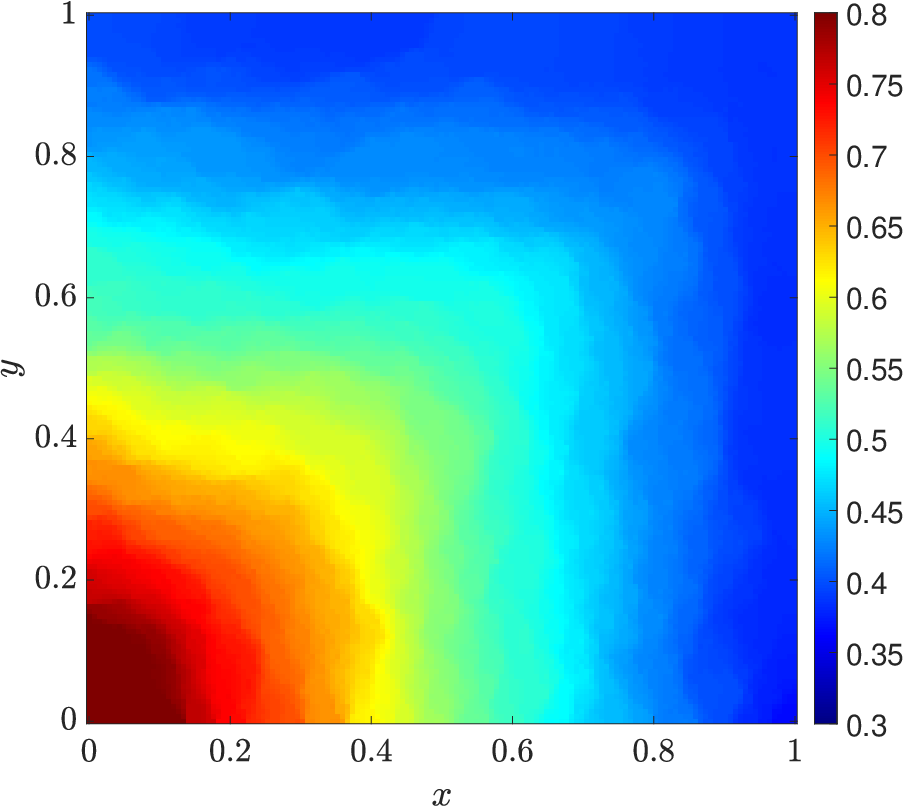} 
		\\[0.2cm]
		\includegraphics[height=3cm]{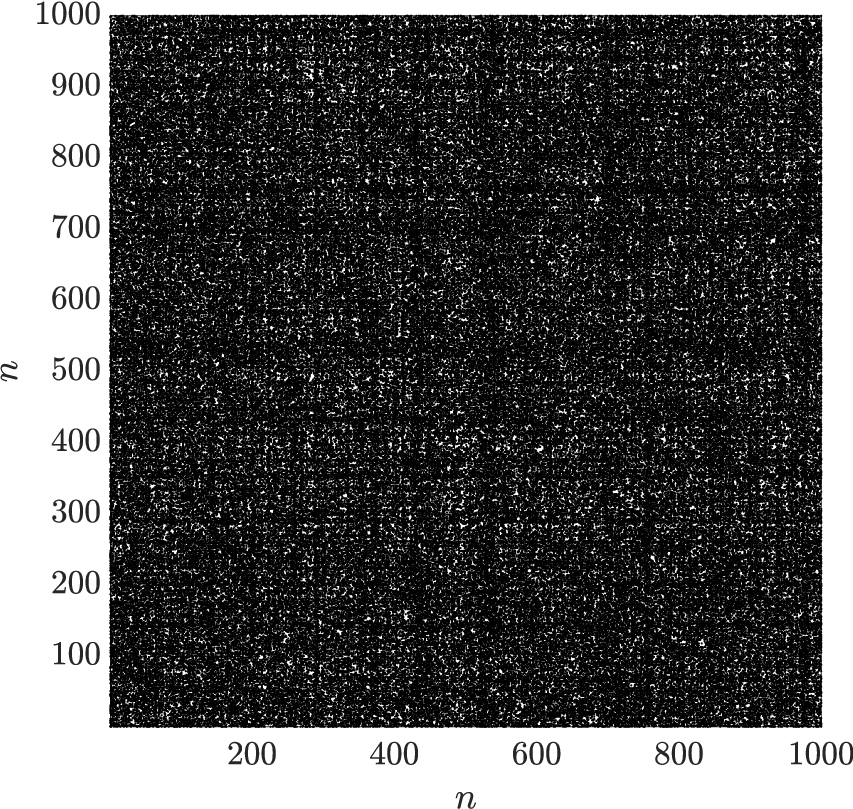} &
		\includegraphics[height=3cm]{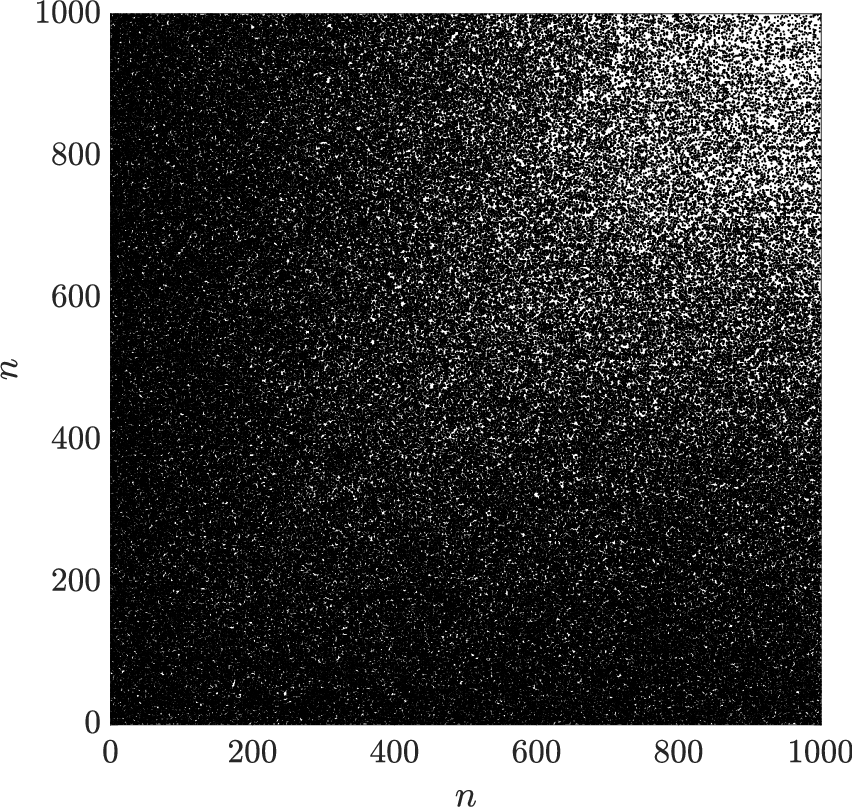} &
		\includegraphics[height=3cm]{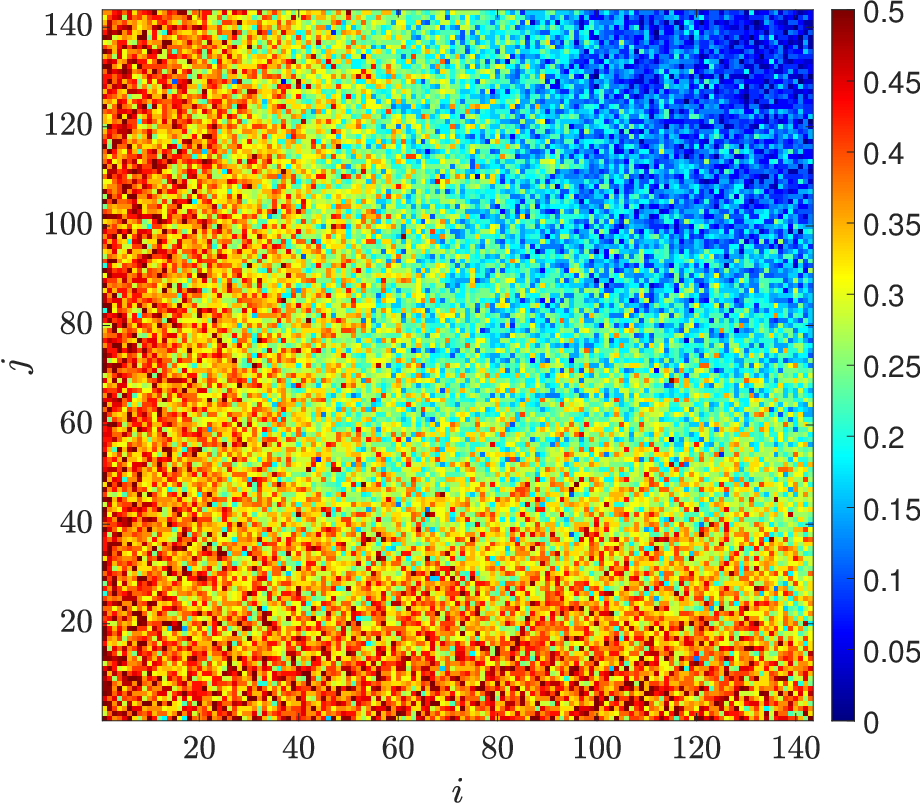} &
		\includegraphics[height=3cm]{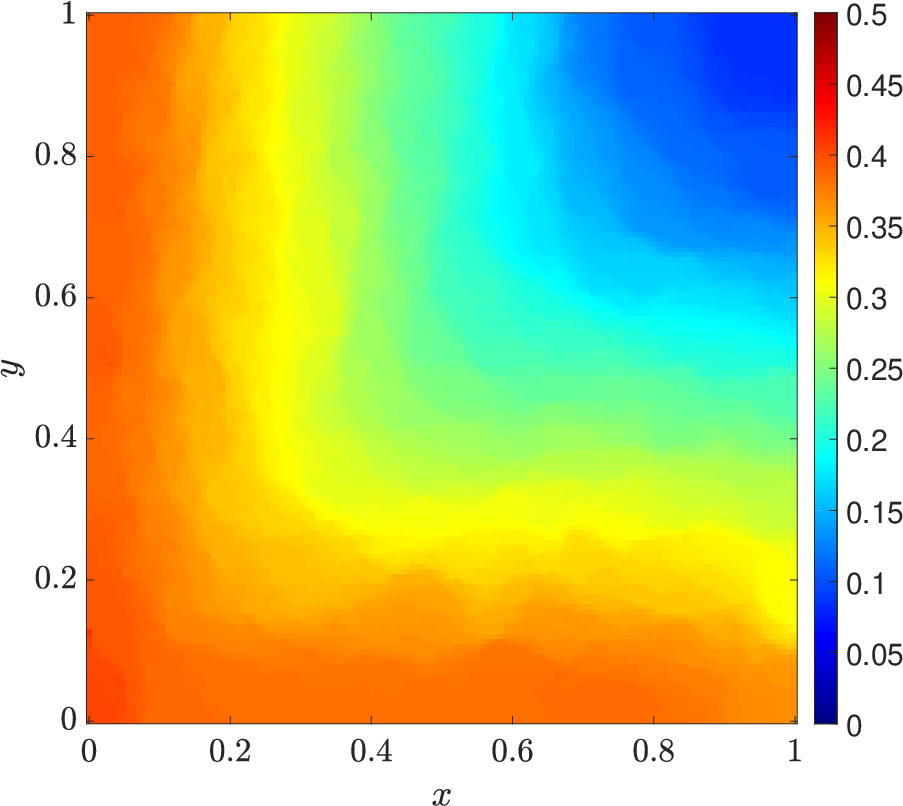} \\
		Observed graph & Sorted graph & Local histogram & Estimated graphon \\
		$G \in \{0,1\}^{\mathfrak{n}\times \mathfrak{n}}$ &
		$\widehat{\mathcal{A}} \in \{0,1\}^{\mathfrak{n} \times \mathfrak{n}}$ &
		$\widehat{\mathcal{H}} \in [0,1]^{\mathrm{k} \times \mathrm{k}}$ &
		$\widehat{\mathcal{B}}^{tv} \in [0,1]^{\mathrm{k} \times \mathrm{k}}$ \\
	\end{tabular}
	\caption{Illustration of the SAS algorithm for two different scenarios.}
	\label{fig:SAS_method}
\end{figure}

We now adopt the exchangeable graph model framework to estimate the graphon from observed network data. Specifically, we rely on the Sorting-And-Smoothing (SAS) algorithm \cite{Chan2014consistent}, a histogram-based approach that consists of sorting empirical node degrees, rearranging the adjacency matrix accordingly, and constructing a histogram representation, which is subsequently smoothed via total variation minimization. The SAS algorithm targets the canonical graphon by sorting nodes according to their empirical degrees, which converge to the canonical degree distribution as the network size increases.
The main steps of the Sorting-And-Smoothing procedure are illustrated in Fig.~\ref{fig:SAS_method}. In the first step, the rows and columns of the observed graph
$ G \in \{0,1\}^{\mathfrak{n}\times \mathfrak{n}} $
are reordered based on empirical degrees, yielding the sorted graph
$ \widehat{\mathcal{A}} \in \{0,1\}^{\mathfrak{n}\times \mathfrak{n}} $.
In the second step, a local histogram
$ \widehat{\mathcal{H}} \in [0,1]^{k \times k} $
is computed from $ \widehat{\mathcal{A}} $, and the final graphon estimate
$ \widehat{\mathcal{B}}^{\mathrm{tv}} \in [0,1]^{k \times k} $
is obtained by applying total variation minimization to smooth the histogram.
We detail the mathematical formulation of the Sorting-And-Smoothing (SAS) algorithm, following the schematic illustration in Fig.~\ref{fig:SAS_method}. The sorting step aims to reorder the observed graph $ G $ such that the corresponding empirical degrees form a monotonically non-decreasing sequence. Specifically, the empirical degree of node $ i $ is defined as
\[
d_i = \sum_{j=1}^{n} G_{ij}.
\]
We then introduce a permutation $ \widehat{\theta} $ such that
\[
d_{\widehat{\theta}(1)} \le d_{\widehat{\theta}(2)} \le \dots \le d_{\widehat{\theta}(n)},
\]
and define the sorted graph as
\[
\widehat{\mathcal{A}}_{ij} = G_{\widehat{\theta}(i)\,\widehat{\theta}(j)}.
\]
It is important to note that, depending on the latent node configuration, the true permutation that determines the canonical graphon may differ from the empirical permutation $ \widehat{\theta} $ obtained from degree sorting. We therefore denote by $ \theta $ the oracle permutation that orders the latent variables $ \mathcal{U}_i $ such that
\[
\mathcal{U}_{\theta(1)} \le \mathcal{U}_{\theta(2)} \le \dots \le \mathcal{U}_{\theta(n)}.
\]
The corresponding oracle-ordered graph is given by
\[
\mathcal{A}_{ij} = G_{\theta(i)\,\theta(j)}.
\]
In the smoothing step, following the canonical representation, graphon estimation reduces to identifying a smooth surface that best approximates the sorted adjacency matrix $ \widehat{\mathcal{A}} $. To this end, we employ a simplified stochastic block model approximation, in which the continuous graphon is represented by a piecewise-constant function. Specifically, the histogram estimator is defined as
\[
\widehat{\mathcal{H}}_{ij}
= \frac{1}{h^2}
\sum_{i_1=1}^{h}
\sum_{j_1=1}^{h}
\widehat{\mathcal{A}}_{i h + i_1,\; j h + j_1},
\]
and, analogously, the oracle histogram is given by
\[
\mathcal{H}_{ij}
= \frac{1}{h^2}
\sum_{i_1=1}^{h}
\sum_{j_1=1}^{h}
\mathcal{A}_{i h + i_1,\; j h + j_1},
\]
where $ h > 0 $ denotes the block size. These expressions show that $ \widehat{\mathcal{H}}_{ij} $ and $ \mathcal{H}_{ij} $ correspond to histogram representations of $ \widehat{\mathcal{A}} $ and $ \mathcal{A} $, respectively. Since the function is constant within each block, the dimensionality of the estimation problem is significantly reduced: instead of estimating $ \mathfrak{n}^2 $ parameters, only $ k^2 $ parameters are required, where the number of blocks is $ k = \lfloor \mathfrak{n} / h \rfloor $. 

Although the network histogram estimator is consistent, incorporating a total variation (TV) regularization step can significantly improve the error decay rate. This refinement relies on a sparsity assumption on the underlying graphon $ \mathcal{B} $. In particular, we assume that graphons are gradient-sparse, in analogy with natural images. Under this assumption, the discretized graphon $ \mathcal{B} $, represented on a $ k \times k $ grid, is expected to exhibit bounded total variation.
The total variation of $ \mathcal{B} $ is defined as
\[
\|\mathcal{B}\|_{\mathrm{TV}}
= \sum_{i=1}^{k} \sum_{j=1}^{k}
\sqrt{
	\left( \frac{\partial \mathcal{B}}{\partial x} \right)_{ij}^{2}
	+
	\left( \frac{\partial \mathcal{B}}{\partial y} \right)_{ij}^{2}
},
\]
where $ \frac{\partial \mathcal{B}}{\partial x} $ and
$ \frac{\partial \mathcal{B}}{\partial y} $ denote the horizontal and vertical finite differences of the discretized graphon, respectively.
By exploiting this total variation prior, the smoothing stage can be formulated as the following constrained optimization problem:
\[
\widehat{\mathcal{B}}^{\mathrm{tv}}
= \arg \min_{\widehat{\mathcal{R}}}
\|\widehat{\mathcal{R}}\|_{\mathrm{TV}},
\quad
\text{subject to }
\|\widehat{\mathcal{R}} - \widehat{\mathcal{H}}\|_{2}
\leq \varepsilon,
\]
where $ \varepsilon > 0 $ controls the fidelity between the TV-regularized solution
$ \widehat{\mathcal{R}} $ and the histogram estimator
$ \widehat{\mathcal{H}} $, and
$ \|\cdot\|_{2} $ denotes the Frobenius norm. the SAS algorithm's total computational complexity is $ O(\mathfrak{n}~ \text{log}~ \mathfrak{n} + \mathrm{k}^2 ~\text{log}~ \mathrm{k}^2)$  multiplications and $O(\mathfrak{n}^2)$ additions while  the magnitude of the final estimate $\widehat{\mathcal{B}}^{\text{est}}$ is $\mathfrak{n} \times \mathfrak{n}$.

\begin{figure}[htp]
	\centering
	\begin{tabular}{cccc}
		\includegraphics[height=3cm]{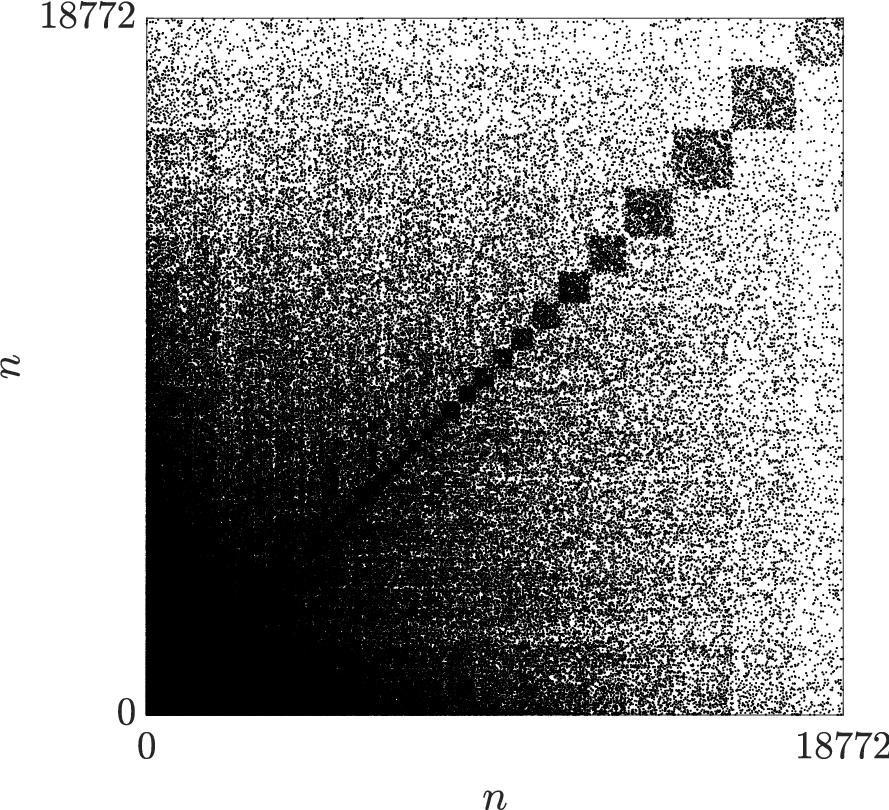}&\includegraphics[height=3cm]{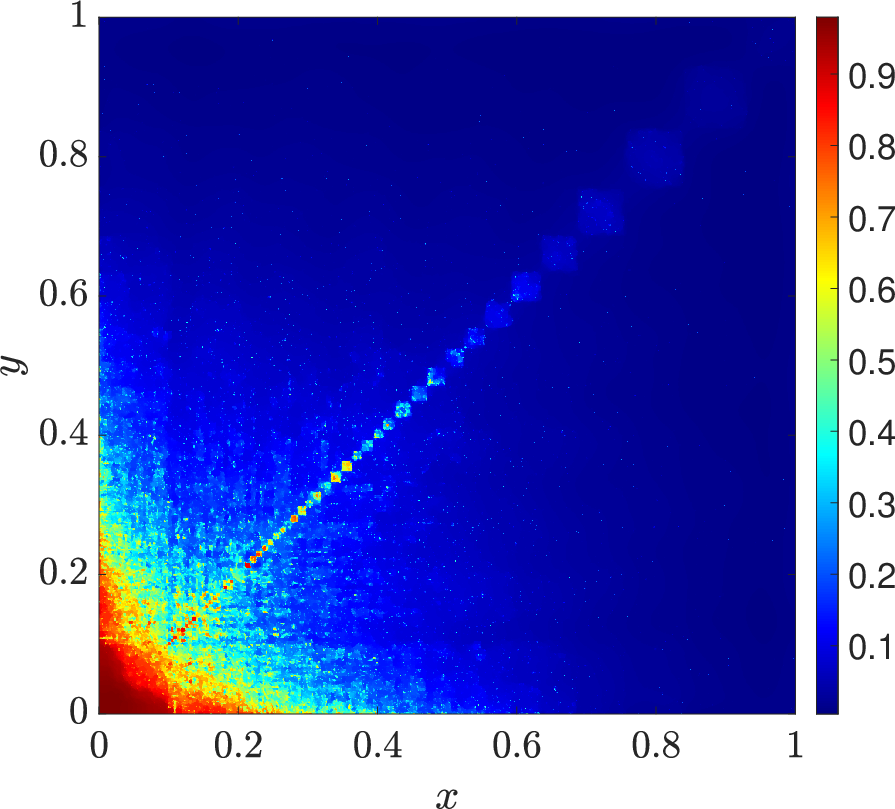} &
		\includegraphics[height=3cm]{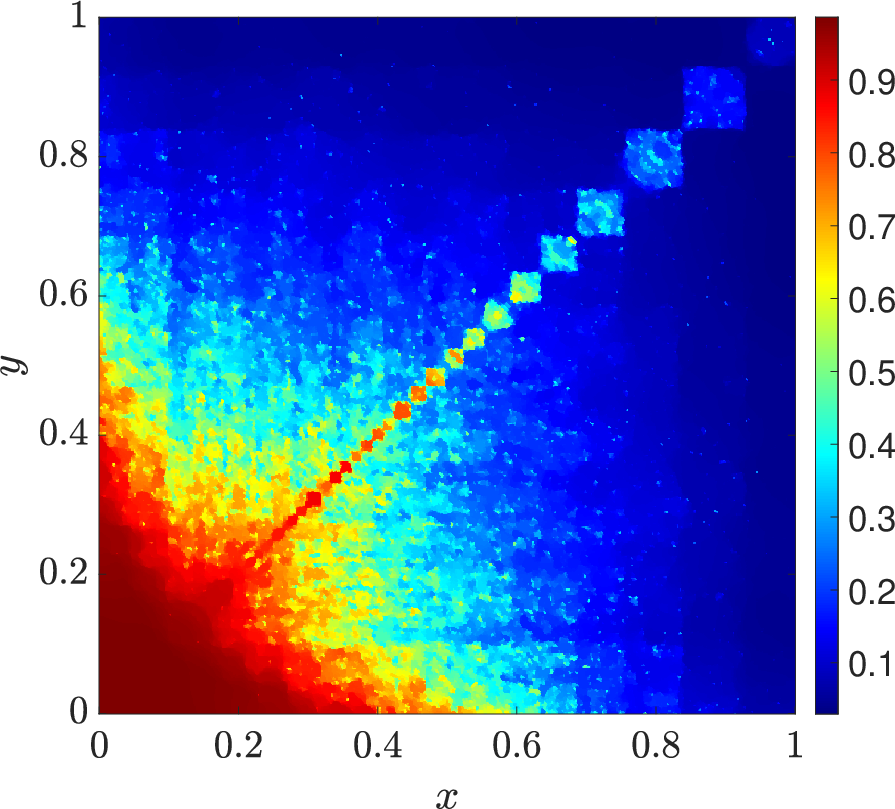} &
		\includegraphics[height=3cm]{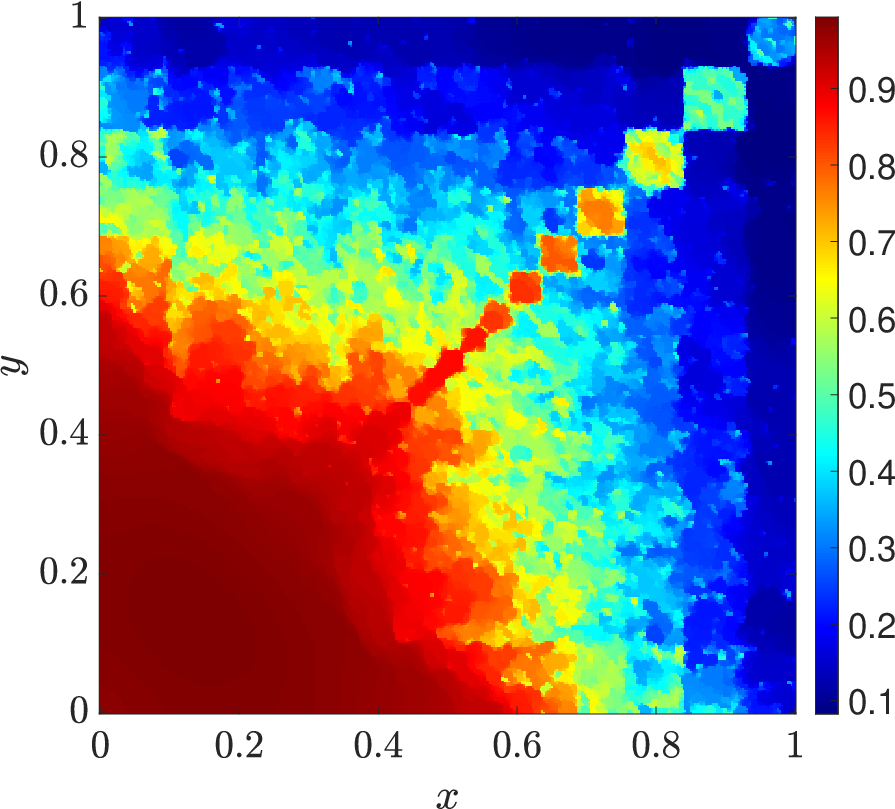} \\
		Sorted graph &
		$10$ bins &
		$25$ bins &
		$50$ bins \\
		\includegraphics[height=3cm]{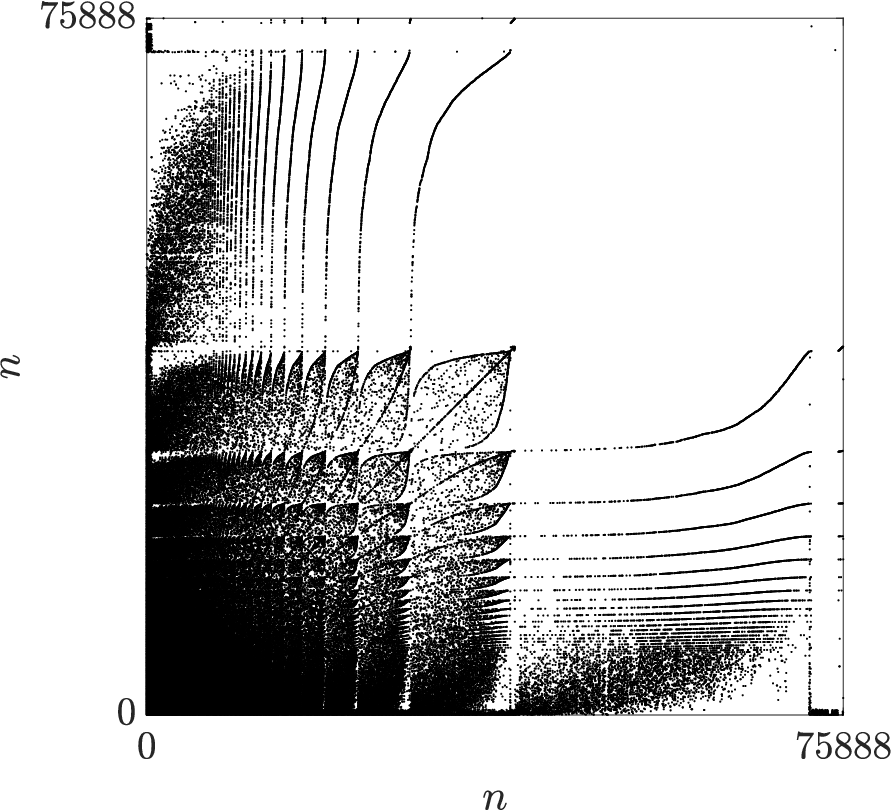}& \includegraphics[height=3cm]{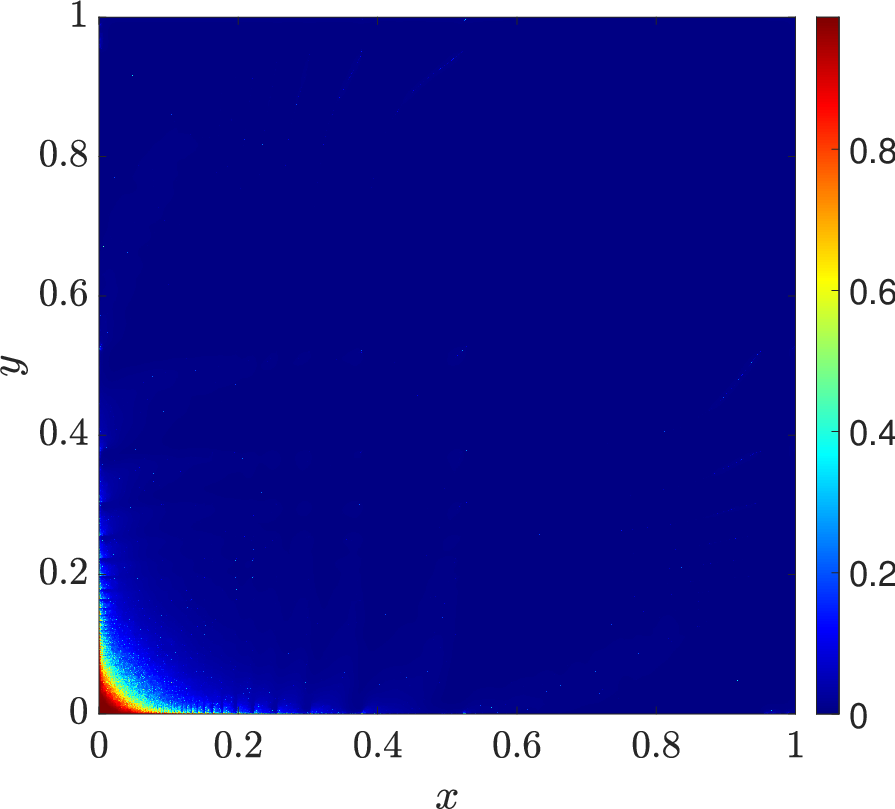} &
		\includegraphics[height=3cm]{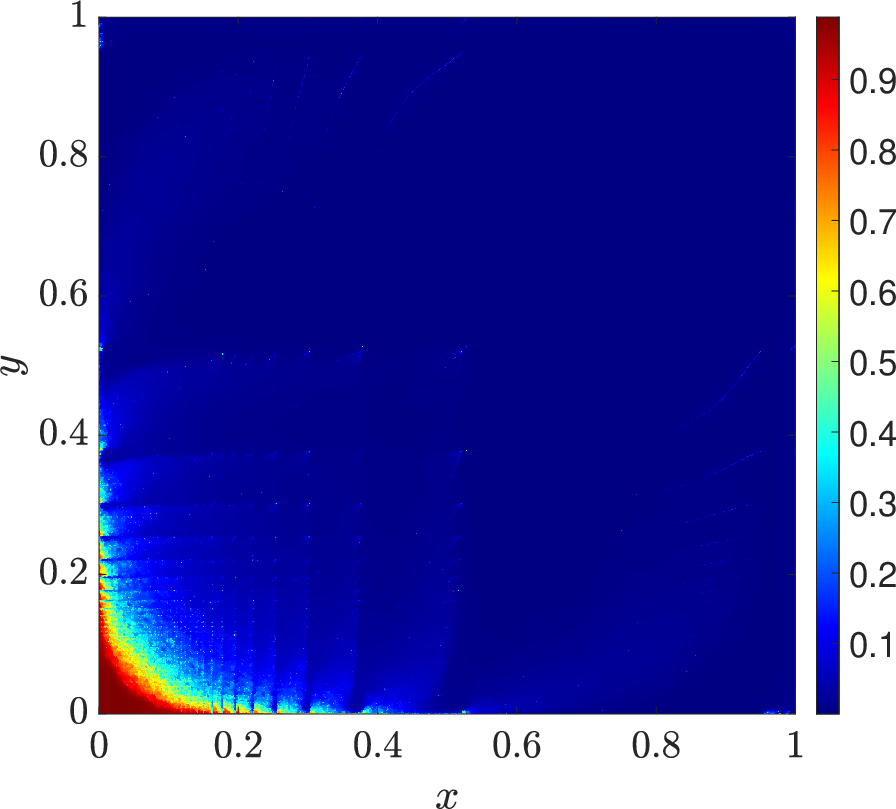} &
		\includegraphics[height=3cm]{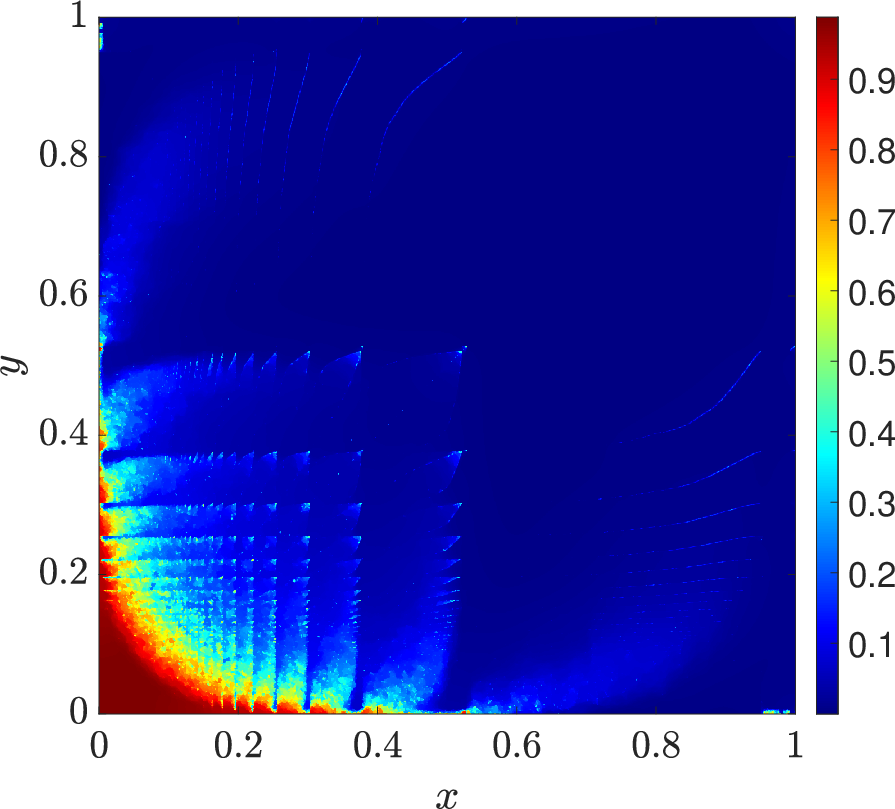} \\
		Sorted graph &
		$10$ bins &
		$25$ bins &
		$50$ bins \\
	\end{tabular}
	\caption{Estimated graphons for real networks. First row: Collaboration
		network of arXiv astro physics (ca-AstroPh) $n = 1.8 \times
		10^4$. Second row: who-trusts-whom network of Epinions.com (soc-
		Epinions1) $n = 7.5 \times 10^4$.}
	\label{fig:RealData}
\end{figure}

In the following, we use the above described algorithm within two real-world network datasets: the collaboration network of arXiv astrophysics (ca-AstroPh) and the who-trusts-whom network from Epinions.com (soc-Epinions-1). Both datasets are obtained from the Stanford Large Network Dataset Collection (SNAP)\footnote{\url{http://www.cise.ufl.edu/research/sparse/matrices/SNAP/}}.
The soc-Epinions-1 dataset corresponds to a directed (asymmetric) binary graph with approximately $7.5 \times 10^{4}$ nodes and $5.1 \times 10^{5}$ edges, whereas the ca-AstroPh dataset represents an undirected (symmetric) binary graph with about $1.8 \times 10^{4}$ nodes and $3.9 \times 10^{5}$ edges. To emulate a raw-data scenario in which node labels are intrinsically unordered, we randomly permute the rows and columns of the adjacency matrices for both networks.
Figure~\ref{fig:RealData} displays the sorted graphs and the visualizations of the two approximated graphons using different bin sizes in order to efficiently highlight their community structures. The estimated graphon for the ca-AstroPh network reveals tightly connected collaborations and relatively small communities. In contrast, the soc-Epinions-1 network exhibits recurring interaction patterns driven by influential nodes, which is reflected in the repeated structures observed in the corresponding graphon.

\section{Numerical Tests}\label{results}
In this section, we present several numerical experiments aimed at illustrating the qualitative behavior of the proposed model. To this end, we adopt an operator-splitting strategy that decouples the opinion dynamics from the epidemiological evolution.

Specifically, we first introduce the opinion update step
\[
f_\ell^\ast = \mathcal{I}_{\Delta t}(f_\ell^n),
\]
defined as
\begin{equation}
	\label{eq:opinion_step}
	\begin{cases}
		\partial_t f_\ell^\ast =
		\displaystyle \frac{1}{\tau}
		\sum_{\mathcal H \in \mathcal{C}}
		\overline{\mathcal{Q}}_{\ell}
		\bigl(f_\ell^\ast, f_{\mathcal H}^\ast\bigr)(x,w,t), \\[8pt]
		f_\ell^\ast(x,w,0) = f_\ell^n(x,w),
		\qquad \forall \ell \in \mathcal{C},
	\end{cases}
\end{equation}
followed by the epidemiological step
\[
f_{\mathcal{J}}^{\ast\ast} = \mathcal{E}_{\Delta t}(f_{\mathcal{J}}^\ast),
\]
which reads
\begin{equation}
	\label{eq:epidemic_step}
	\begin{cases}
		\partial_t f_S^{\ast\ast}
		= \mathcal{K}(f_S^{\ast\ast}, f_I^{\ast\ast})(x,w,t), \\[6pt]
		\partial_t f_E^{\ast\ast}
		= \mathcal{K}(f_S^{\ast\ast}, f_I^{\ast\ast})(x,w,t)
		- \sigma_E f_E^{\ast\ast}, \\[6pt]
		\partial_t f_I^{\ast\ast}
		= \sigma_E f_E^{\ast\ast} - \gamma f_I^{\ast\ast}, \\[6pt]
		\partial_t f_R^{\ast\ast}
		= \gamma f_I^{\ast\ast}, \\[8pt]
		f_\ell^{\ast\ast}(x,w,0)
		= f_{\mathcal{J}}^\ast(x,w,\Delta t).
	\end{cases}
\end{equation}
The solution at time $t^{n+1}$ is then obtained by combining the two evolution steps. In particular, a first-order operator-splitting scheme reads
\begin{equation}\nonumber
	f_{\mathcal{J}}^{n+1}(x,w)
	= \mathcal{E}_{\Delta t}
	\bigl(
	\mathcal{I}_{\Delta t}(f_{\mathcal{J}}^n(x,w))
	\bigr),
\end{equation}
where $\mathcal{I}_{\Delta t}$ and $\mathcal{E}_{\Delta t}$ denote the opinion consensus and epidemiological evolution operators, respectively.

The epidemiological system \eqref{eq:epidemic_step} is integrated in time using a fourth-order Runge--Kutta (RK4) scheme. For the opinion consensus step \eqref{eq:opinion_step}, we employ a first-order semi-implicit, structure-preserving numerical method \cite{Pareschi2018structure,Bartel2024structure} applied to
\[
\partial_t f_{\mathcal{J}}(x,w,t)
=
- \alpha_{\mathcal{JH}} \, \partial_w
\bigl(
\mathcal{P}[f](x,w,t)\, f(x,w,t)
\bigr)
+ \frac{\sigma_{\mathcal{JH}}^2}{2}
\, \partial_w^2
\Bigl(
\mathcal{H}[f](x,t)\,
\mathcal D^2(x,w)\,
f(x,w,t)
\Bigr),
\]
which ensures the preservation of steady states and maintains the structural properties of the underlying Fokker--Planck equation.

\subsection{On the opinion evolution on graphon networks}
\begin{figure}[h!] 
	\centering  \includegraphics[width=.3\textwidth]{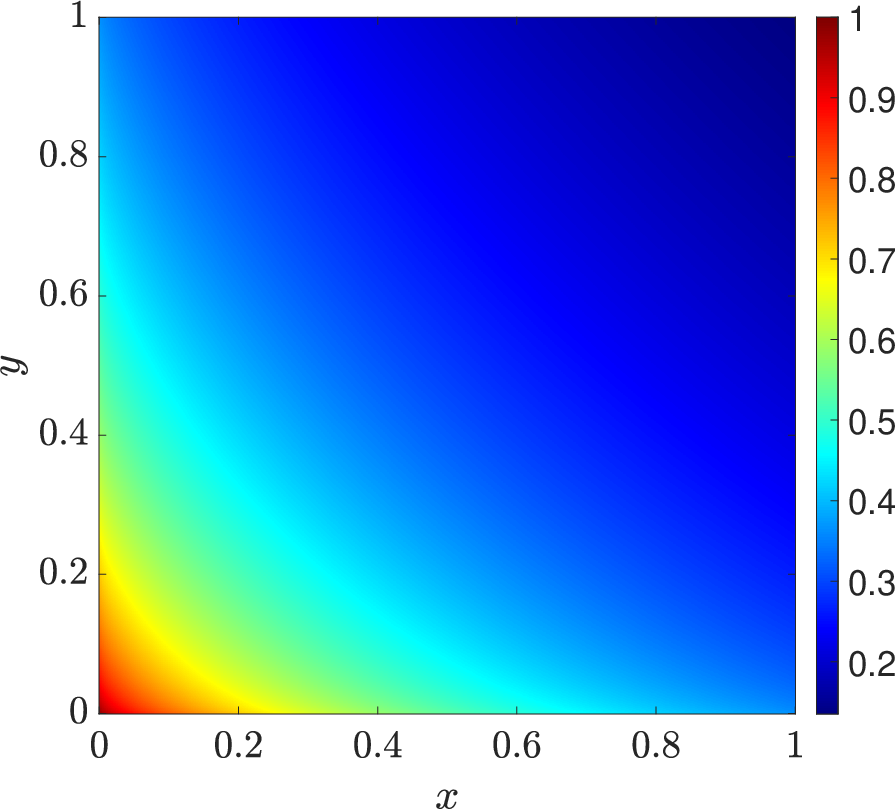}
	\quad \includegraphics[width=.3\textwidth]{Graphon_ca-Astroph50_new.png} 
	\quad \includegraphics[width=.3\textwidth]{Graphon_soc-Epinion50_new.png} 
	\caption{Left: plot of the separable graphon in \eqref{eq:separableg}. Center: plot of the real collaboration network of arXiv astro physics ($\mathcal{B}_A(x,y)$), reconstructed using $50$ bins. Right: plot of the real who-trusts-whom network ($\mathcal{B}_E(x,y)$), reconstructed using $50$ bins.} \label{fig:graphon_plots} \end{figure}

In this section, we investigate the evolution of the opinion density on a graphon network. We restrict to the case of a single population and neglect the role of physical contacts, focusing exclusively on the opinion dynamics. We consider the separable $L^\infty$ graphon
\begin{equation}\label{eq:separableg}
	\mathcal{B}_1 (x,y) = \exp\left(-\left(x^{0.7} + y^{0.7}\right)\right),
\end{equation}
as well as the two empirical graphons introduced in Section~\ref{real-data-garphon}, reconstructed using $50$ bins. A visual comparison of these three network structures is reported in Figure~\ref{fig:graphon_plots}. 

In this context, we aim to mimic a scenario in which the agents are initially divided into two groups. The first group consists of highly connected individuals, namely those with
\[
d_i(x) \geq \kappa \max(d_i(x)),
\]
while the remaining individuals form the second group. Due to the different shapes of the connectivity distributions, we set $\kappa = 0.7$ for the separable graphon \eqref{eq:separableg}, $\kappa = 0.88$ for the arXiv astroPh collaboration network (denoted by $\mathcal{B}_A(x,y)$), and $\kappa = 0.6$ for the who-trusts-whom network (denoted by $\mathcal{B}_E(x,y)$). The initial distribution is given by
\[
f(x,w,0) = K
\begin{cases}
	\exp\left\{-\frac{(w-m_1)^2}{2\sigma_1^2}\right\} d_i(x), & \text{if } d_i(x) \geq \kappa \max(d_i(x)), \\[6pt]
	\exp\left\{-\frac{(w-m_2)^2}{2\sigma_2^2}\right\} d_i(x), & \text{otherwise},
\end{cases}
\]
with $m_1=-0.5$, $m_2=0.5$, $\sigma_1=\sigma_2=0.2$, and $K$ a normalization constant such that
\[
\int_{-1}^1 \int_0^1 f(x,w,0)\,dx\,dw = 1.
\]
This setup reflects a situation in which a smaller fraction of highly connected individuals holds an opinion centered around the negative value $-0.5$, while the majority of the population is centered around the positive value $0.5$. We set $\mathcal D(w) = 1 - w^2$ and $\sigma_{\mathcal{JH}} = 10^{-1}$. Our goal is to show that, for suitable choices of the interaction kernel $\mathit P$, highly connected individuals can steer the rest of the population toward their opinion, while remaining relatively unaffected themselves. To this end, we consider three different interaction kernels:
\begin{equation}\label{eq:kernels}
	\begin{aligned}
		P_1(w,w_*,x,y) &= \chi\left(|w-w_*| < 1\right) H(x,y),\\
		P_2(w,w_*,x,y) &= \chi\left(|w-w_*| < \tfrac{3}{2} H(x,y)\right),\\
		P_3(w,w_*,x,y) &= \chi\left(|w-w_*| < \tfrac{3}{2} H(x,y)\right) H(x,y),
	\end{aligned}
\end{equation}
where
\[
H(x,y) = \frac{d_o(y)^\alpha}{d_i(x)^\alpha + d_o(y)^\alpha}, 
\qquad \alpha = 6.
\]

\begin{figure}[h] 
	\centering  
	\begin{tabular}{ccc}
		\includegraphics[height=4cm]{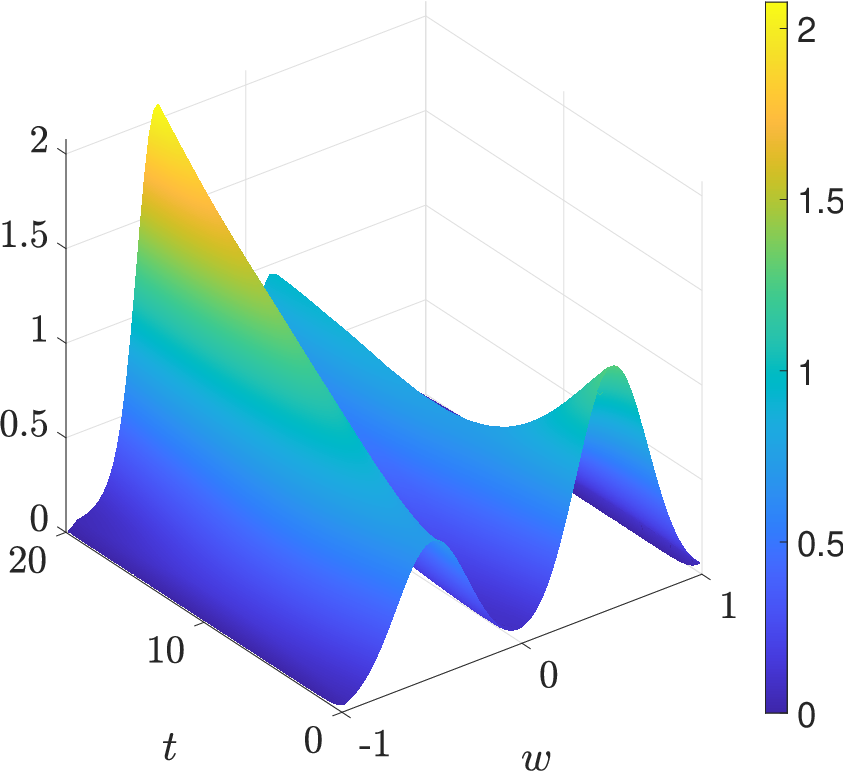} & \includegraphics[height=4cm]{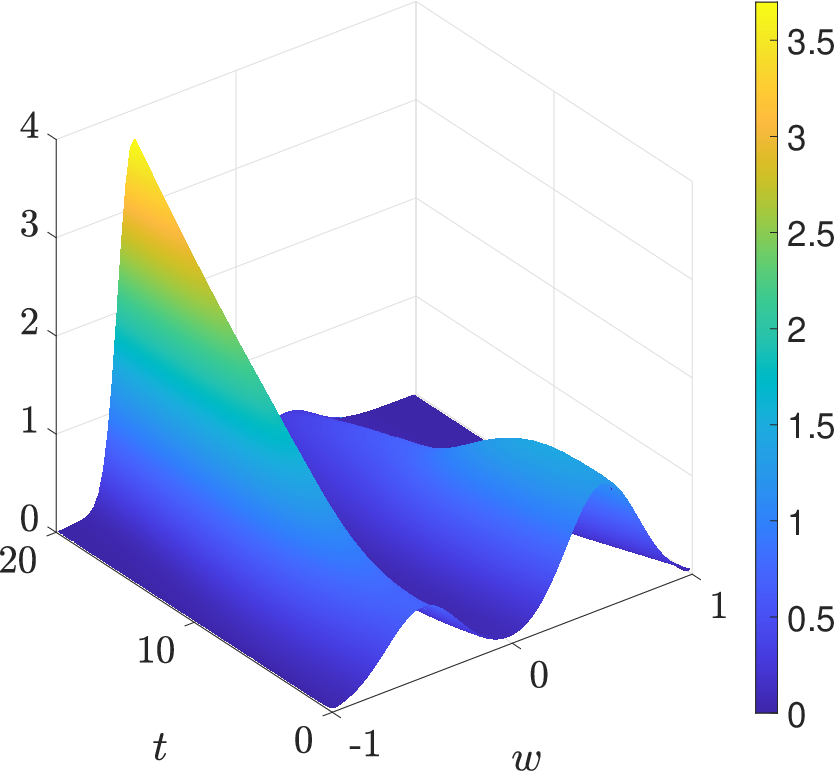}&
		\includegraphics[height=4cm]{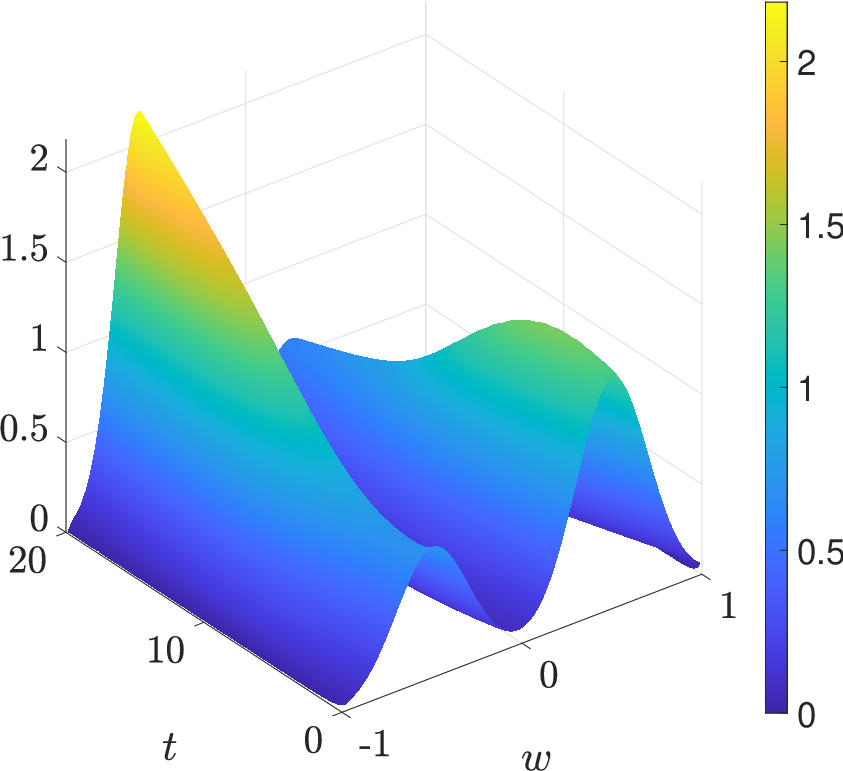} \\
		\includegraphics[height=4cm]{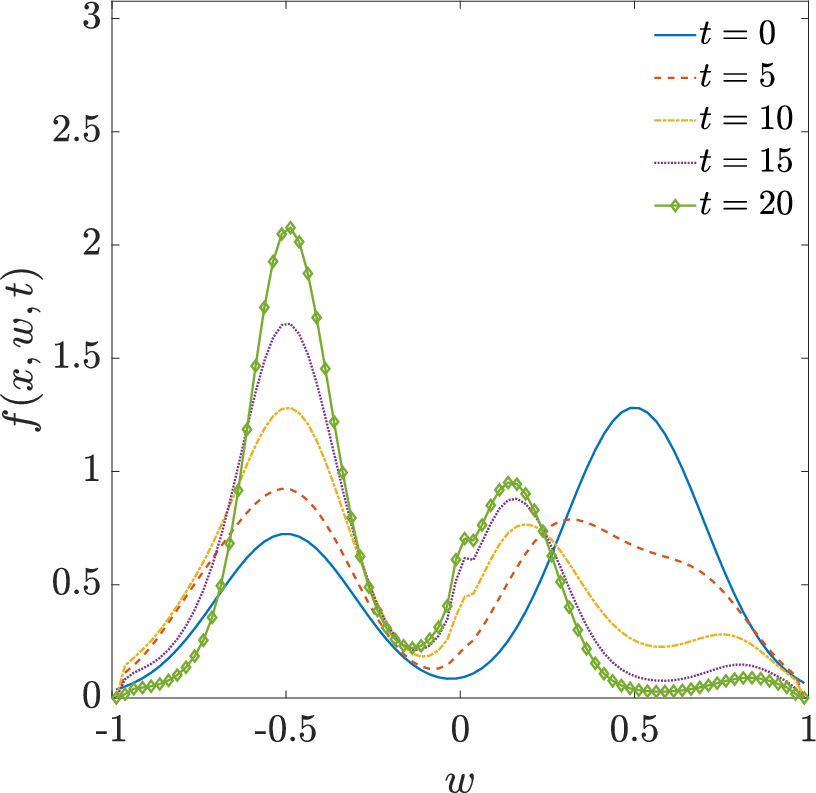} &
		\includegraphics[height=4cm]{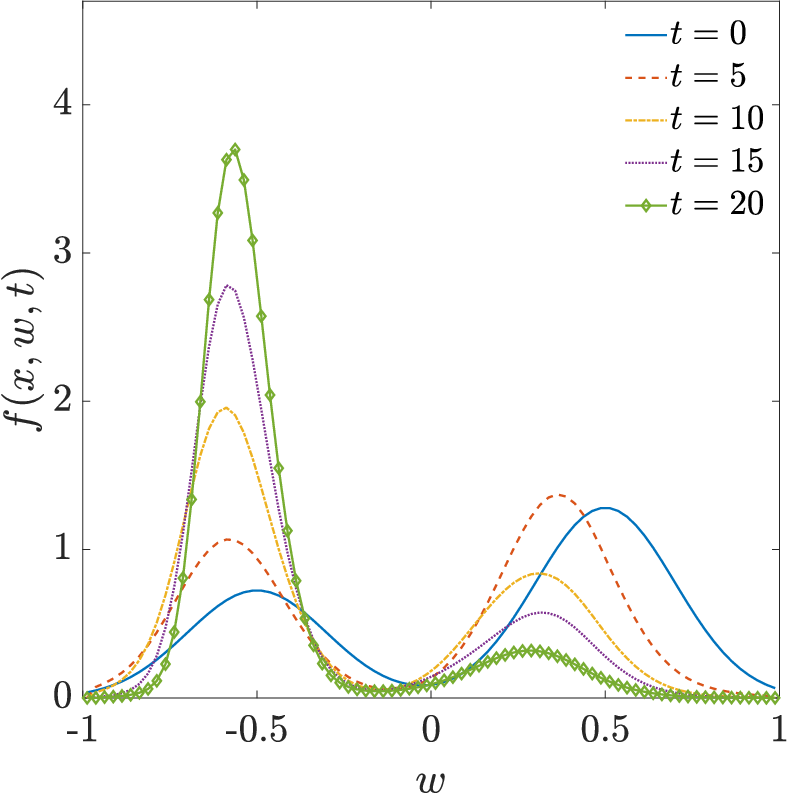} &
		\includegraphics[height=4cm]{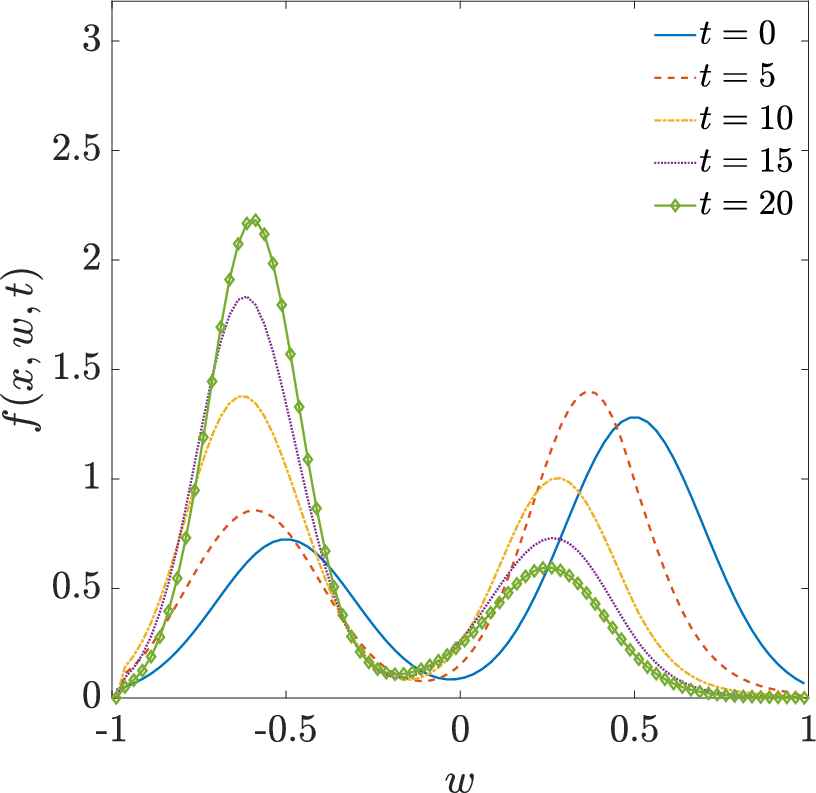} 
	\end{tabular}
	\caption{Separable graphon \eqref{eq:separableg}. Surface plot (top), different time frames (bottom) for the evolution of the opinions in time. First column: $P_1(w,w_*,x,y)$ in \eqref{eq:kernels}. Second column: $P_2(w,w_*,x,y)$ in \eqref{eq:kernels}. Third column: $P_3(w,w_*,x,y)$ in \eqref{eq:kernels}} \label{fig:separable} 
\end{figure}

\begin{figure}[h] 
	\centering  
	\begin{tabular}{ccc}
		\includegraphics[height=4cm]{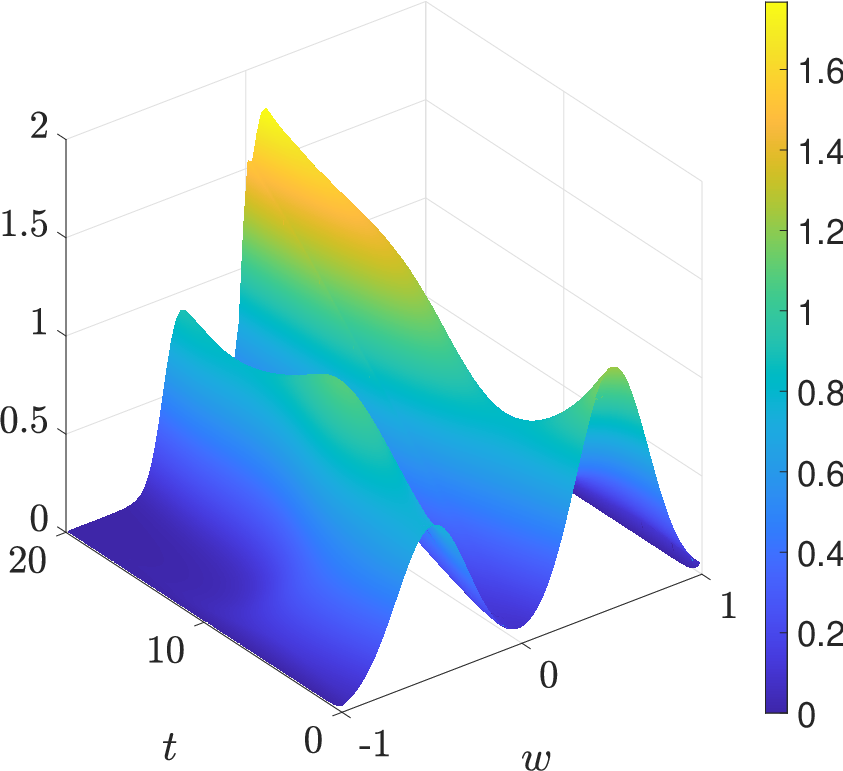}&\includegraphics[height=4cm]{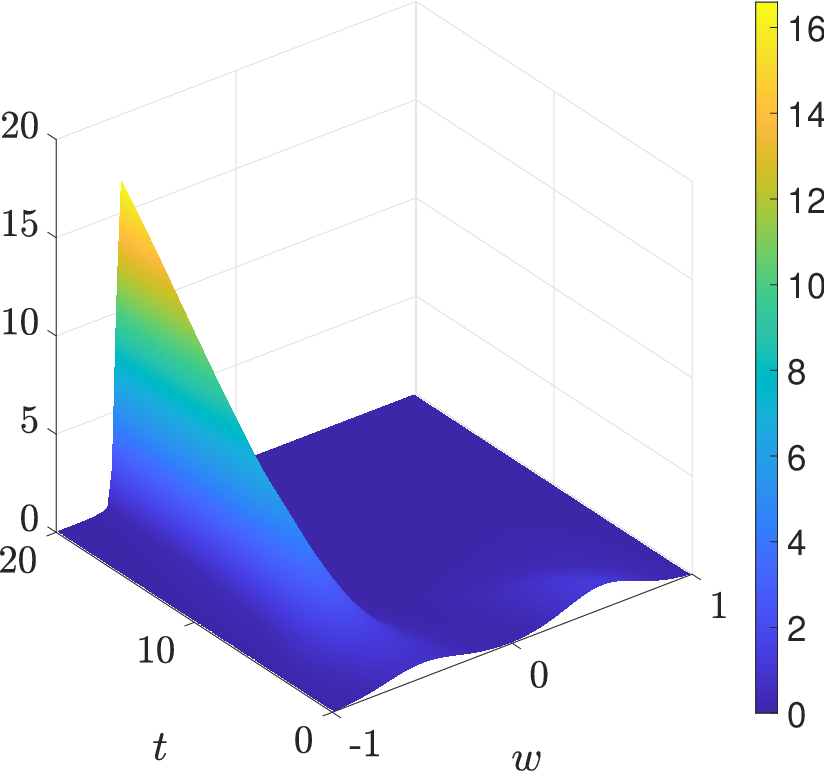}&\includegraphics[height=4cm]{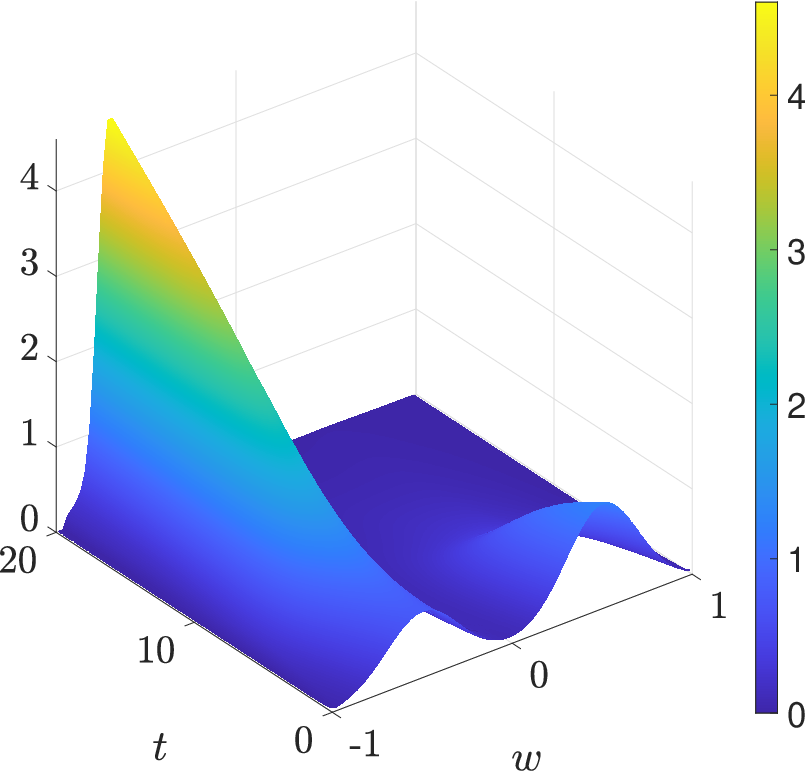}\\ \includegraphics[height=4cm]{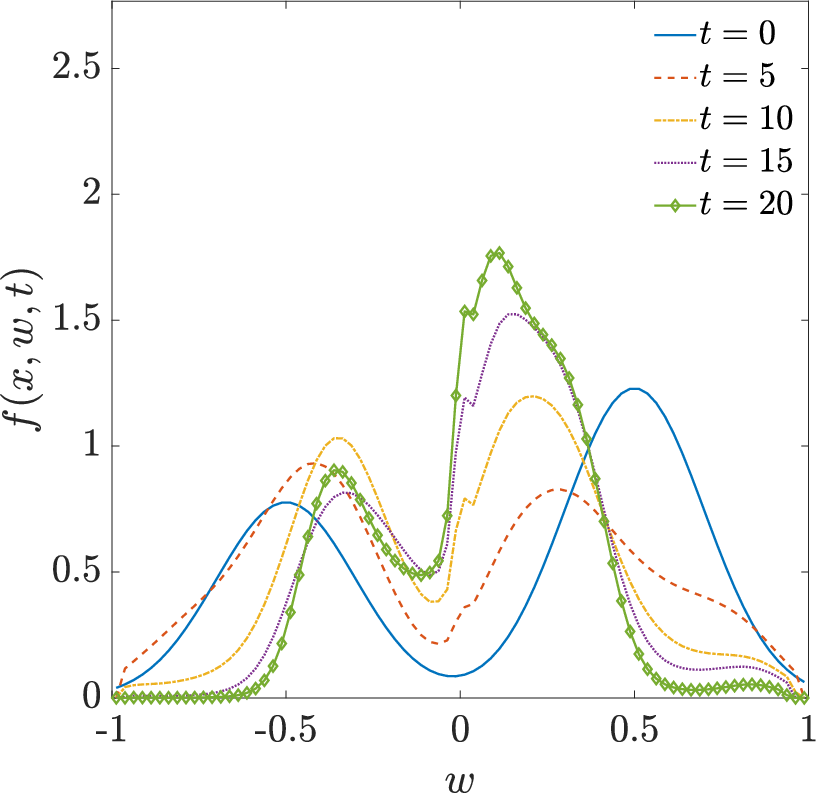}& \includegraphics[height=4cm]{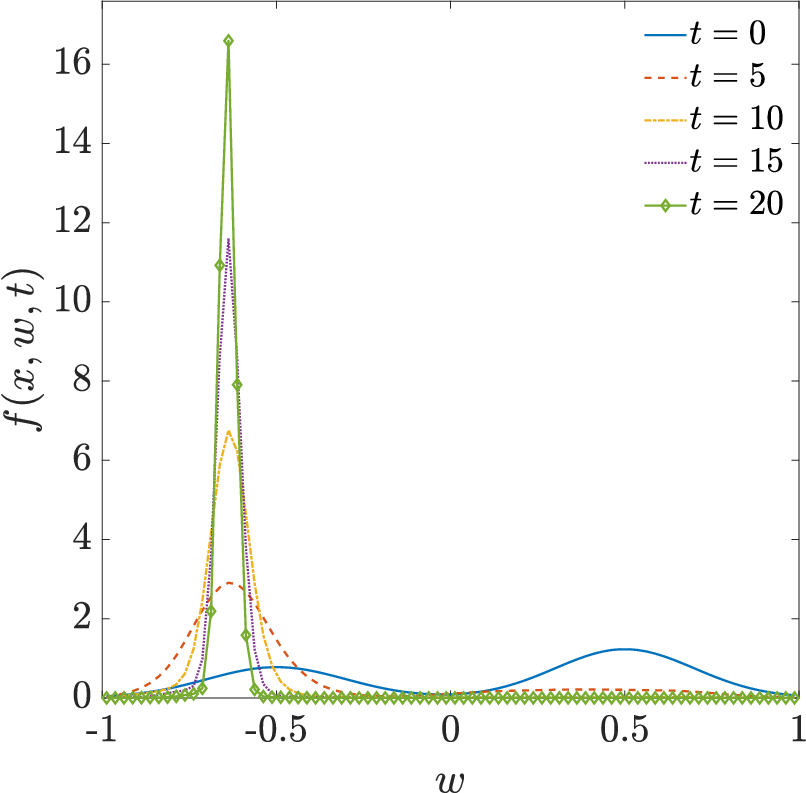} & \includegraphics[height=4cm]{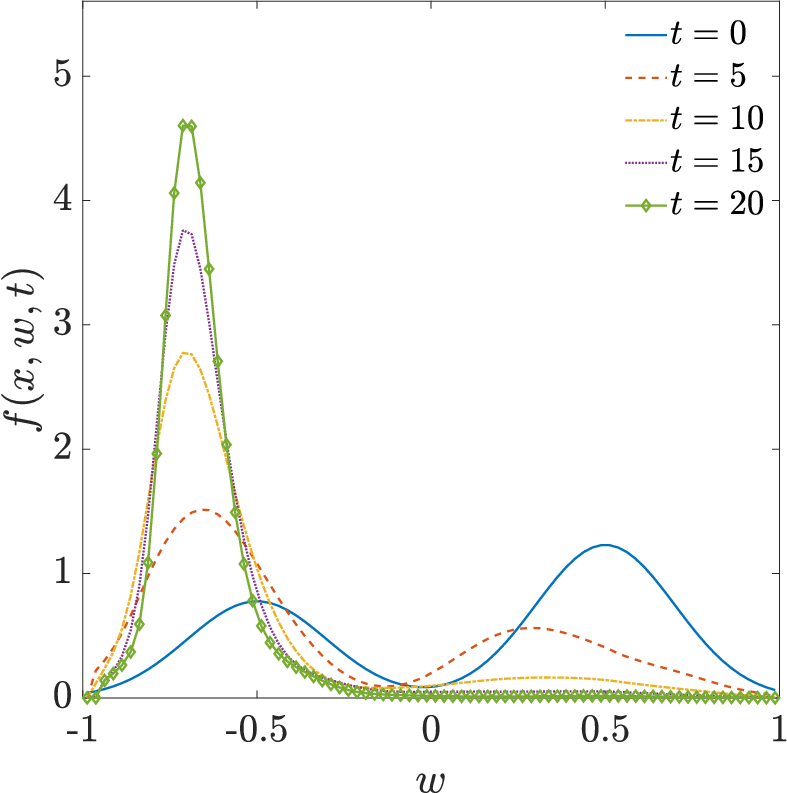} 
	\end{tabular}
	\caption{Real graphon from the arXiv astroPh collaboration network ($\mathcal{B}_A(x,y)$). Surface plot (top), different time frames (bottom) for the evolution of the opinions in time. First column: $P_1(w,w_*,x,y)$ in \eqref{eq:kernels}. Second column: $P_2(w,w_*,x,y)$ in \eqref{eq:kernels}. Third column: $P_3(w,w_*,x,y)$ in \eqref{eq:kernels}} \label{fig:astro} 
\end{figure}

\begin{figure}[h] 
	\centering  
	\begin{tabular}{ccc}
		\includegraphics[height=4cm]{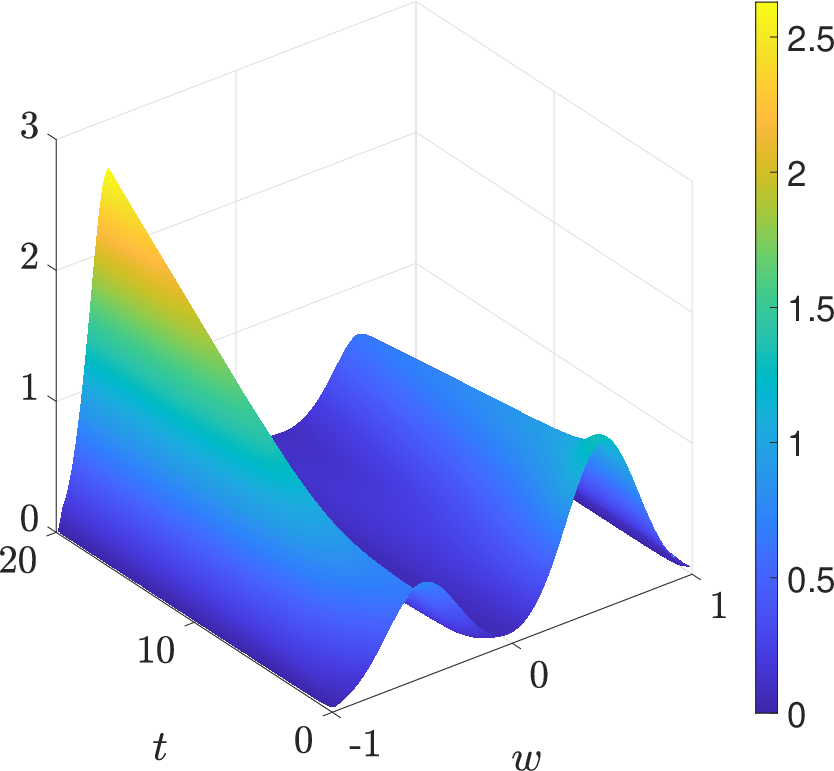}& \includegraphics[height=4cm]{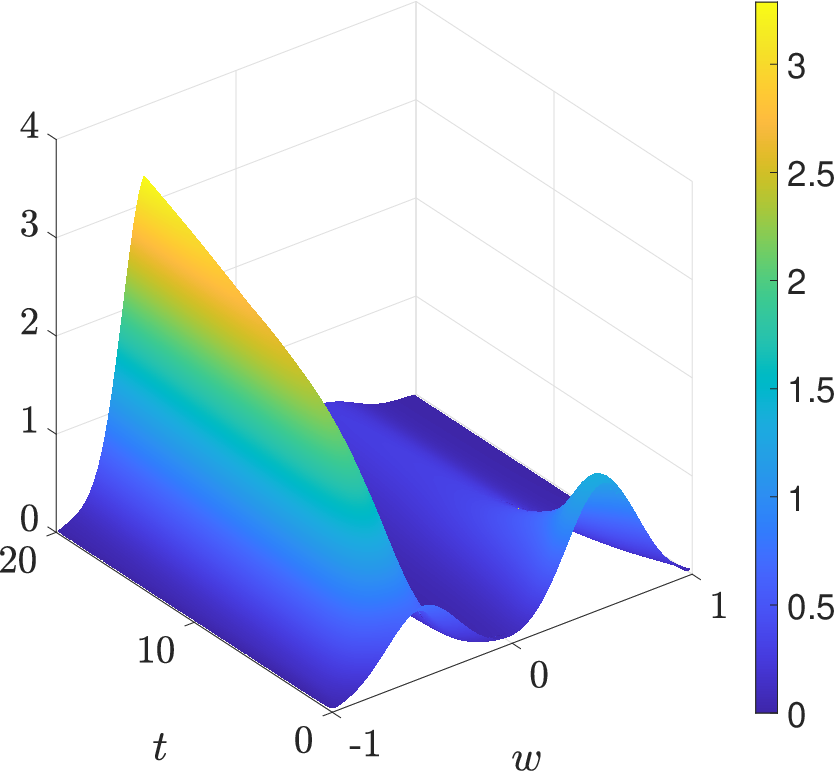} &
		\includegraphics[height=4cm]{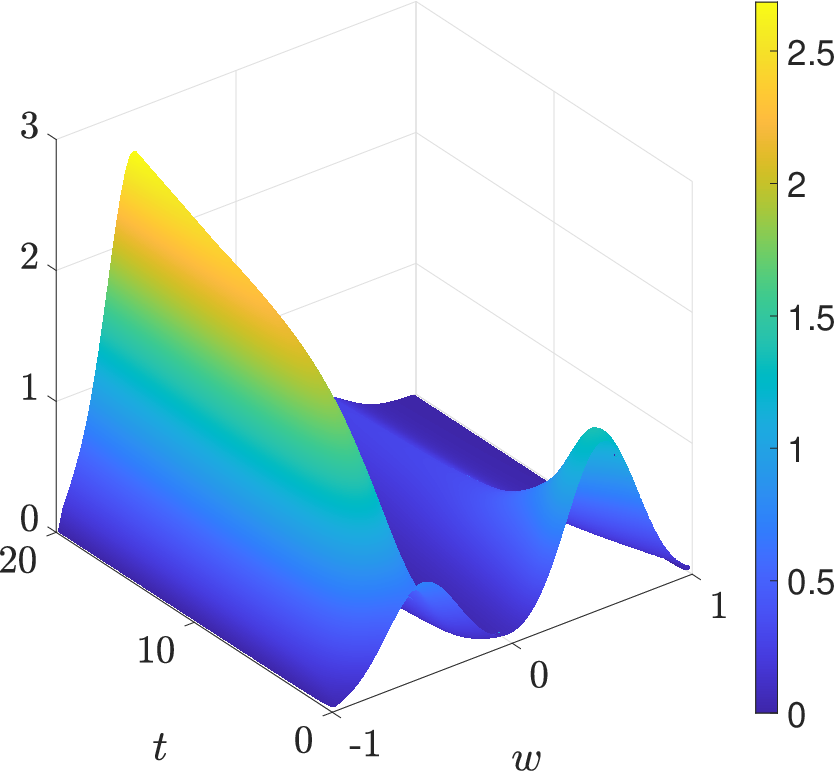} \\
		\includegraphics[height=4cm]{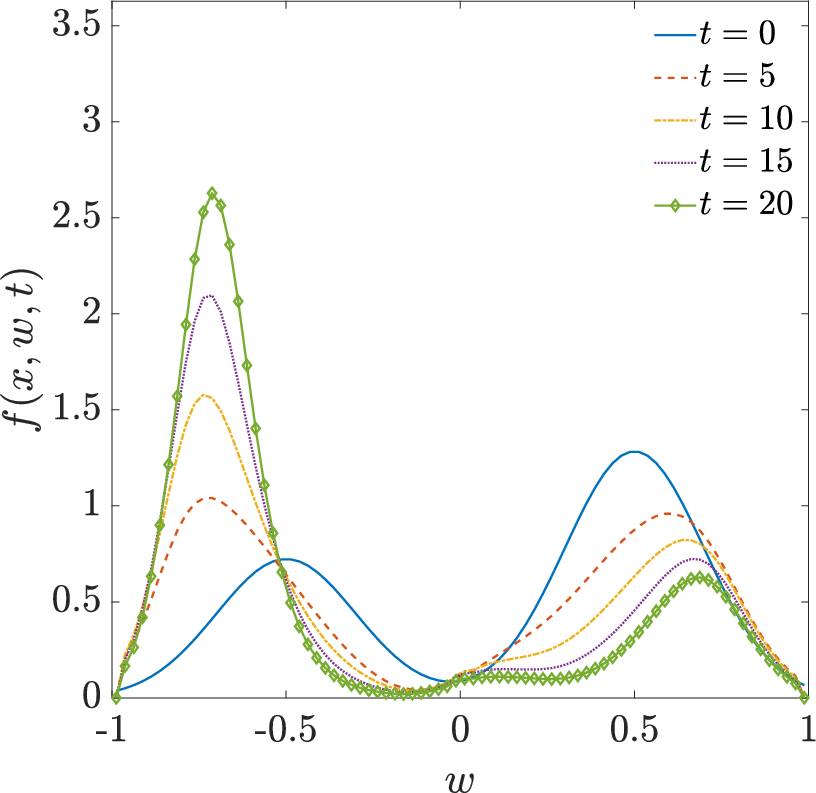}& \includegraphics[height=4cm]{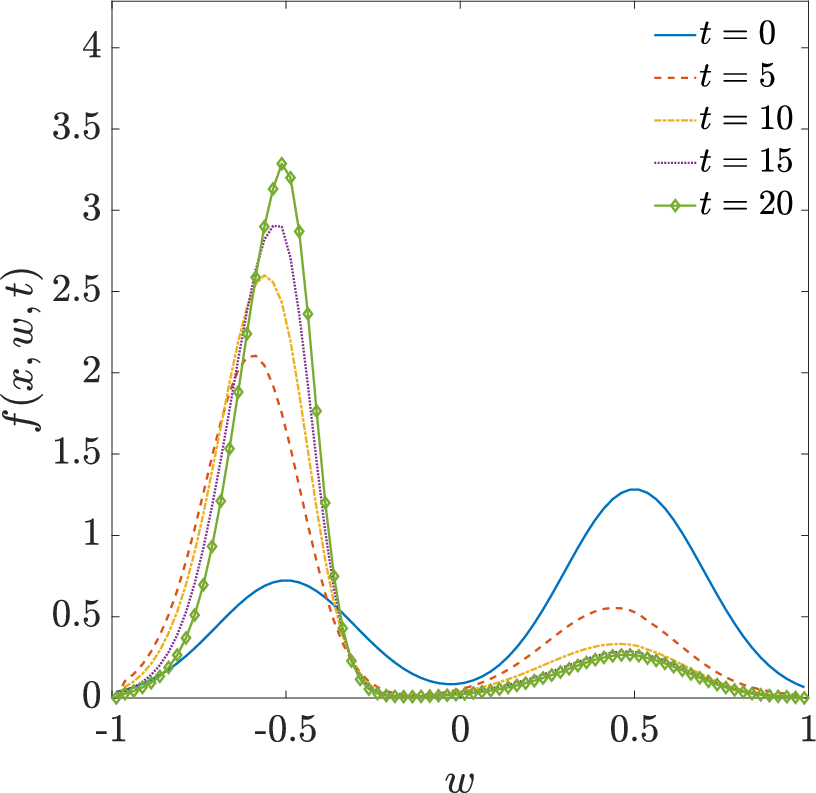} & \includegraphics[height=4cm]{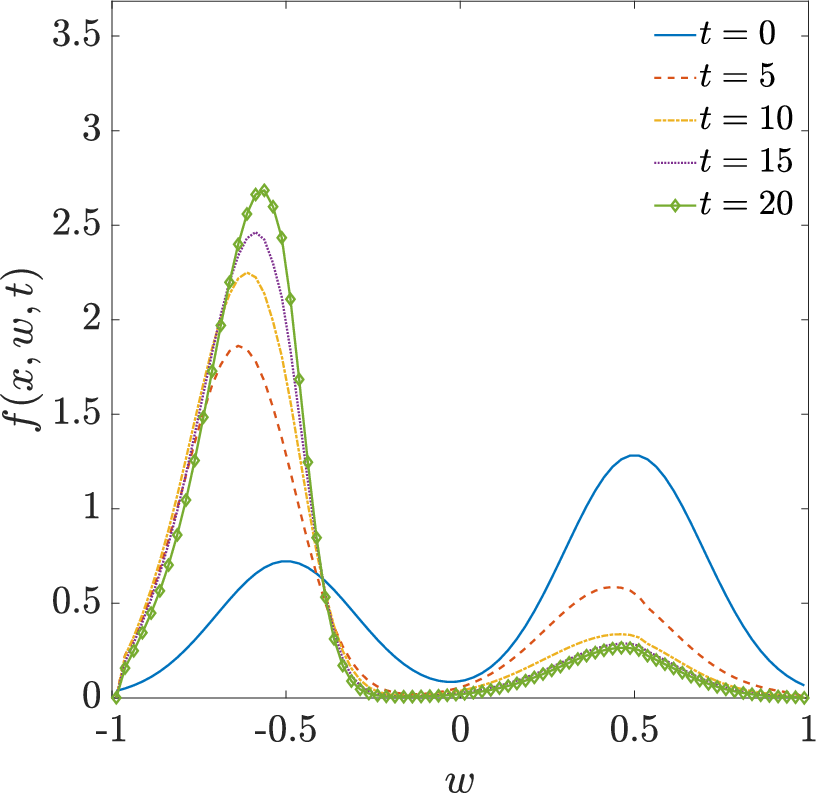} 
	\end{tabular}
	\caption{Real graphon from the who-trusts-whom network ($\mathcal{B}_E(x,y)$). Surface plot (top), different time frames (bottom) for the evolution of the opinions in time. First column: $P_1(w,w_*,x,y)$ in \eqref{eq:kernels}. Second column: $P_2(w,w_*,x,y)$ in \eqref{eq:kernels}. Third column: $P_3(w,w_*,x,y)$ in \eqref{eq:kernels}} \label{fig:epi} 
\end{figure}

Figures~\ref{fig:separable}, \ref{fig:astro}, and \ref{fig:epi} correspond to the simulations of opinion formation using respectively  the separable graphon, the arXiv astroPh network, and the who-trusts-whom network. In each figure, the first, second, and third columns refer to the kernels $P_1$, $P_2$, and $P_3$, respectively.

The kernel $P_1$ is such that the activation of the interaction depends only on the proximity of opinions, namely agents interact whenever $|w-w_*|<1$. The strength of the interaction, however, is modulated by the network structure through $H(x,y)$: interactions are stronger when an agent with lower in-degree $d_i(x)$ interacts with a more connected agent with larger out-degree $d_o(y)$. The kernel $P_2$ models a situation in which the activation of interactions depends on both opinion proximity and network structure. In particular, the confidence threshold is modulated by $H(x,y)$, meaning that agents are more willing to interact when facing more influential (i.e., highly connected) individuals. In this case, the magnitude of the interaction does not depend on the network structure. Finally, the kernel $P_3$ combines both mechanisms: both the interaction threshold and its strength depend on the connectivity of the agents involved, leading to a fully coupled opinion--network interaction.

A comparison of Figures~\ref{fig:separable}--\ref{fig:epi} reveals that the qualitative behavior of the opinion dynamics is remarkably consistent across the different graphon structures. At the same time, the choice of the interaction kernel plays a significant role. In particular, moving from $P_1$ to $P_2$ and $P_3$ leads to progressively stronger distortions of the opinion distribution, highlighting how the inclusion of network-dependent interaction mechanisms enhances the influence of highly connected individuals. The main differences are instead quantitative, affecting the rate of convergence and the sharpness of the resulting opinion clusters. This suggests that the macroscopic features of the opinion dynamics are primarily driven by the interaction rules, while the network structure acts as a modulating factor provided it exhibits sufficient heterogeneity.

In particular, these results highlight the prominent role of highly connected individuals, whose influence on the global opinion distribution becomes increasingly evident when interaction mechanisms depend on network connectivity. This aspect will be further investigated in the following section.

\subsection{On the physical contact influence on the epidemic spread}\label{sec:epi_ph}
\begin{figure}[h] 
	\centering  
	\includegraphics[height=4cm]{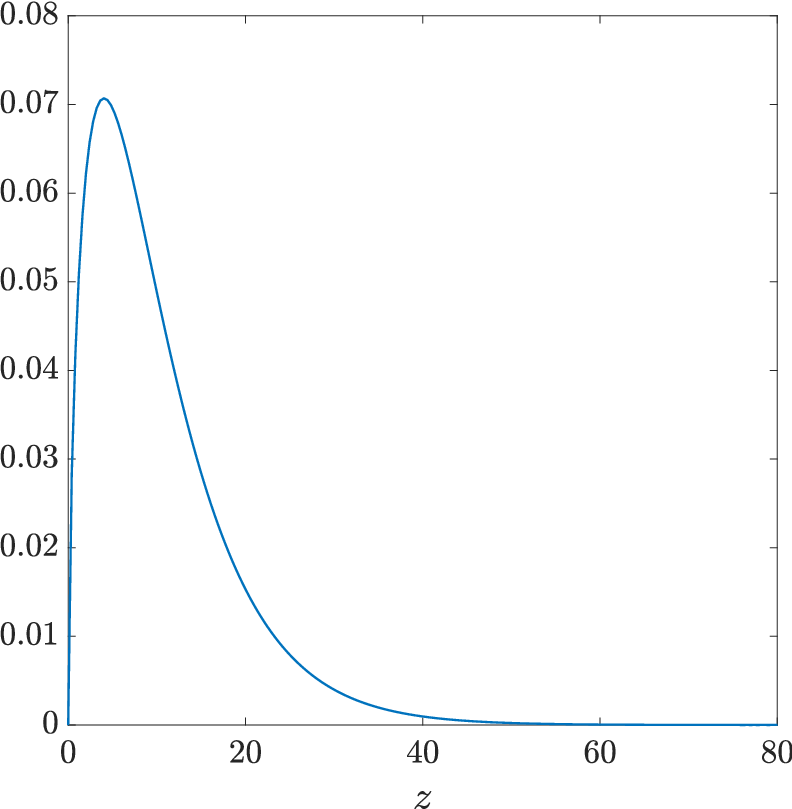}\quad\quad
	\includegraphics[height=4cm]{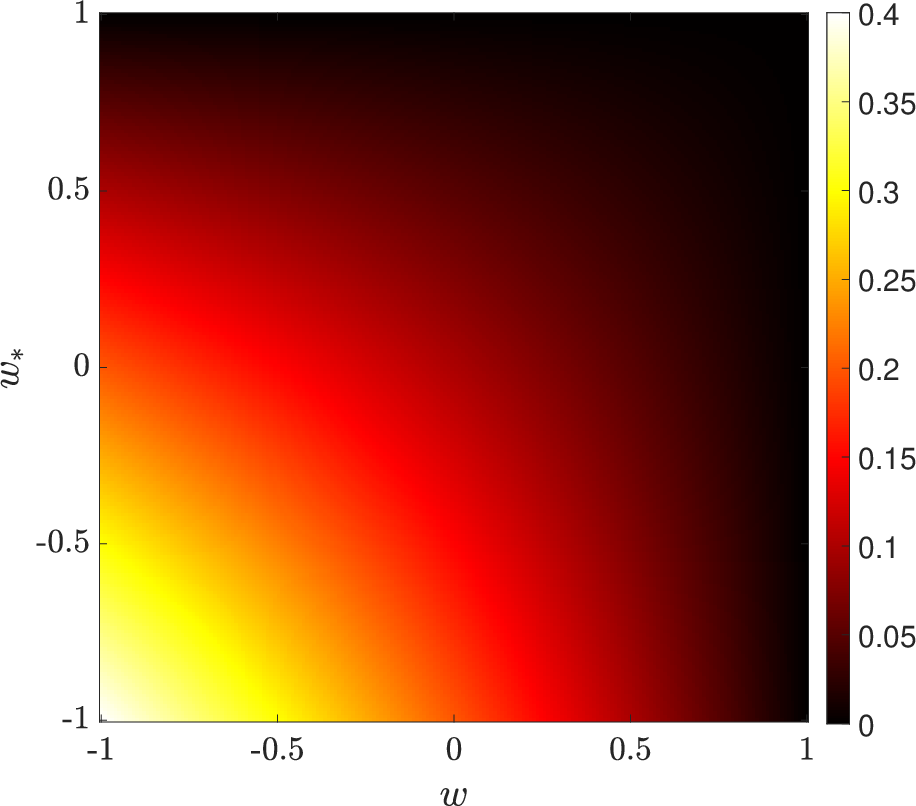} 
	\caption{Distribution of physical contacts (left) and of $\beta(1-w)(1-w_*)/4$ in the case $z=z_*=1$, $\eta = 1$, $\beta=0.4$ (right).} 
	\label{fig:plot_seir_param} 
\end{figure}

In this section, we investigate how heterogeneity in the number of physical contacts influences the spread of the epidemic within the population. We consider the case in which the social connectivity is described by the empirical graphon $\mathcal{B}_A(x,y)$. For the distribution of physical contacts, we adopt the stationary distribution introduced in \cite{Perthame}. In particular, we assume that the marginal distribution of physical contacts is given by the generalized Gamma probability density in \eqref{equili}, which we recall here for convenience:
\[
f_z(z) = C_z \, z^{\nu/\delta + \delta -2} \exp\left\{-\frac{\nu}{\delta^2}\left(\frac{z}{\overline z}\right)^\delta\right\},
\]
with parameters $\nu = 1.65$, $\delta = 1$, and $\overline z = 10.25$, in agreement with the empirical data reported in \cite{French}.
For the epidemic transmission, we assume that the interaction kernel $\kappa(w,w_*;z,z_*)$ is given by \eqref{eq:kappa}, with parameters $\beta = 0.4$, $\eta = 1$, and $\alpha = \alpha_* = 0.7$. In the SEIR system \eqref{eq:SEIR_updated}, we fix the transition rate $\sigma_E = 0.5$ and the recovery rate $\gamma = 1/12$. A graphical representation of the distribution of physical contacts and of the factor $\beta(1-w)(1-w_*)/4$ is shown in Figure~\ref{fig:plot_seir_param}. The latter term highlights how the transmission rate is modulated by the opinions of the interacting individuals: in particular, the infection risk is maximized when both agents exhibit low adherence to protective measures (i.e., $w,w_* \approx -1$), while it is significantly reduced when at least one of them adopts protective behaviors (i.e., $w$ or $w_* \approx 1$). This reflects the behavioral feedback mechanism embedded in the model, where individual attitudes directly impact the epidemic transmission dynamics.
In what follows, we consider two different scenarios. In the first one, individuals with a large number of physical contacts are assumed to be opposed to non-pharmaceutical protective measures. In the second scenario, highly connected individuals are instead in favor of such measures. In both cases, the initial distribution of the compartments is chosen such that $\rho_E(0)=\rho_I(0)=\rho_R(0)=0.1$, while $\rho_S(0) = 1 - (\rho_E(0)+\rho_I(0)+\rho_R(0))$.

\paragraph{Setting $1a$.}
As introduced above, we split the population into two groups: individuals with a high number of physical contacts, characterized by $z \geq \kappa_z$, and individuals with fewer contacts, for which $z < \kappa_z$. At the initial time, the opinions of the first group (high-contact individuals) are centered around the \emph{negative} value $-0.7$. The opinions of the second group instead depend on the epidemiological compartment. More precisely, the opinions of Susceptible and Exposed individuals are uniformly distributed in $[-1,0]$, while those of Infected and Recovered individuals are uniformly distributed in $[0,1]$.
To formalize this setting, we introduce the auxiliary function
\begin{equation}\label{eq:aux_setting1a}
	\bar g_{1,\mathcal J}(w,x) = \bar K_{g_1}
	\begin{cases}
		\chi(w \leq 0)\, d_i(x), & \text{if } z < \kappa_z \text{ and } \mathcal{J} \in \{S,E\},\\[4pt]
		\chi(w \geq 0)\, d_i(x), & \text{if } z < \kappa_z \text{ and } \mathcal{J} \in \{I,R\},\\[4pt]
		\exp\!\left\{-\dfrac{(w+0.7)^2}{0.005}\right\} d_i(x), 
		& \text{if } z \geq \kappa_z \text{ and } \mathcal{J} \in \mathcal{C},
	\end{cases}
\end{equation}
where $\bar K_{g_1}$ is a normalization constant.
We also define the normalized in-degree
\[
\overline d_i(x) = \frac{d_i(x)}{\int_0^1 d_i(\overline x)\,\mathrm{d}\overline x}.
\]
Using these definitions, the initial condition is given by
\begin{equation}\label{eq:init_set1a}
	\begin{aligned}
		f_S(z,x,w,0) &= \rho_S(0)\, f_z(z)\, \bar g_{1,S}(w,x)\, \overline d_i(x), 
		& \quad f_E(z,x,w,0) &= \rho_E(0)\, f_z(z)\, \bar g_{1,E}(w,x)\, \overline d_i(x), \\[4pt]
		f_I(z,x,w,0) &= \rho_I(0)\, f_z(z)\, \bar g_{1,I}(w,x)\, \overline d_i(x), 
		& \quad f_R(z,x,w,0) &= \rho_R(0)\, f_z(z)\, \bar g_{1,R}(w,x)\, \overline d_i(x).
	\end{aligned}
\end{equation}
The threshold parameter is set to $\kappa_z = 20$.

\paragraph{Setting $2a$.}
As in \emph{Setting $1a$}, we split the population into the same two groups according to their number of physical contacts. At the initial time, the opinions of the first group are now centered around the \emph{positive} value $0.7$, while the opinions of the second group are distributed as in \emph{Setting $1a$}.
To formalize this configuration, we define the auxiliary function
\begin{equation}\label{eq:aux_setting2a}
	\bar g_{2,\mathcal J}(w,x) = \bar K_{g_2}
	\begin{cases}
		\chi(w \leq 0)\, d_i(x), & \text{if } z < \kappa_z \text{ and } \mathcal{J} \in \{S,E\},\\[4pt]
		\chi(w \geq 0)\, d_i(x), & \text{if } z < \kappa_z \text{ and } \mathcal{J} \in \{I,R\},\\[4pt]
		\exp\!\left\{-\dfrac{(w-0.7)^2}{0.005}\right\} d_i(x), 
		& \text{if } z \geq \kappa_z \text{ and } \mathcal{J} \in \mathcal{C},
	\end{cases}
\end{equation}
where $\bar K_{g_2}$ is a normalization constant.
The initial condition for \emph{Setting $2a$} is then given by
\begin{equation}\label{eq:init_set2a}
	\begin{aligned}
		f_S(z,x,w,0) &= \rho_S(0)\, f_z(z)\, \bar g_{2,S}(w,x)\, \overline d_i(x), 
		& \quad f_E(z,x,w,0) &= \rho_E(0)\, f_z(z)\, \bar g_{2,E}(w,x)\, \overline d_i(x), \\[4pt]
		f_I(z,x,w,0) &= \rho_I(0)\, f_z(z)\, \bar g_{2,I}(w,x)\, \overline d_i(x), 
		& \quad f_R(z,x,w,0) &= \rho_R(0)\, f_z(z)\, \bar g_{2,R}(w,x)\, \overline d_i(x).
	\end{aligned}
\end{equation}
In this setting, the threshold parameter is set to $\kappa_z = 15$.

\begin{figure}[h] 
	\centering  
	\begin{tabular}{ccc}
		\includegraphics[height=4.2cm]{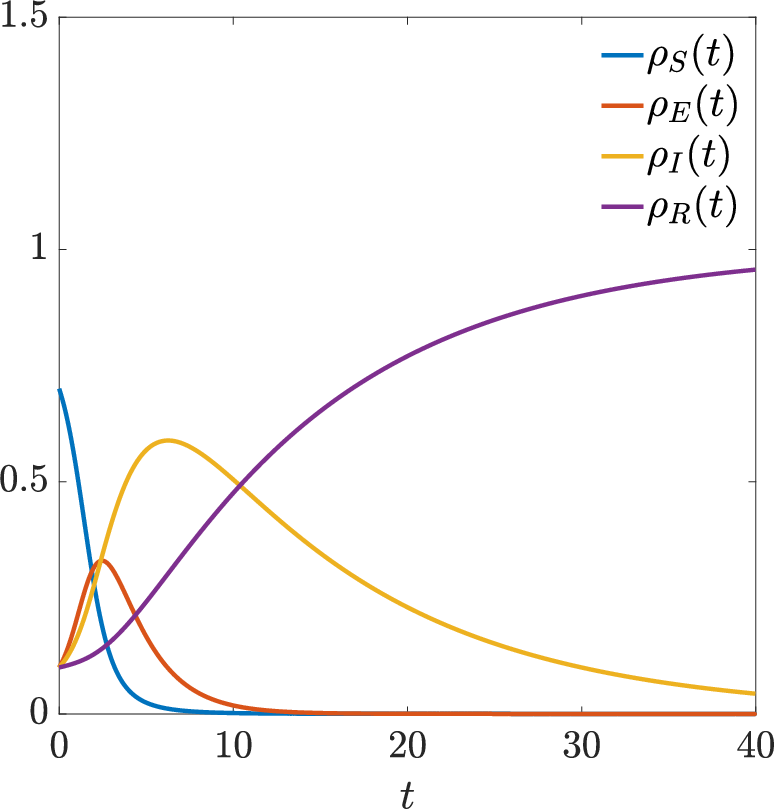}&\includegraphics[height=4.2cm]{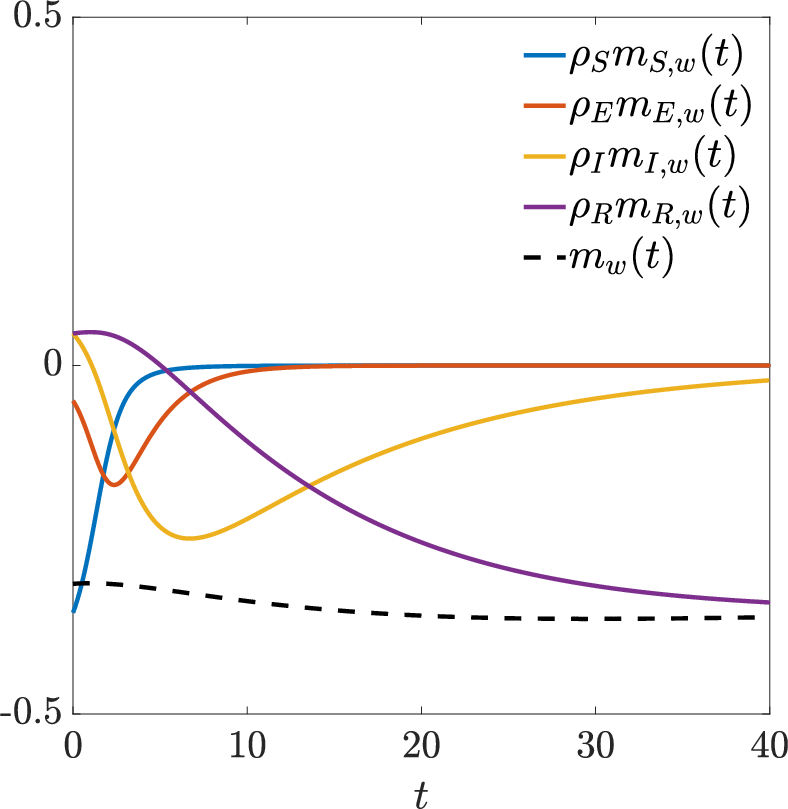} &\includegraphics[height=4.2cm]{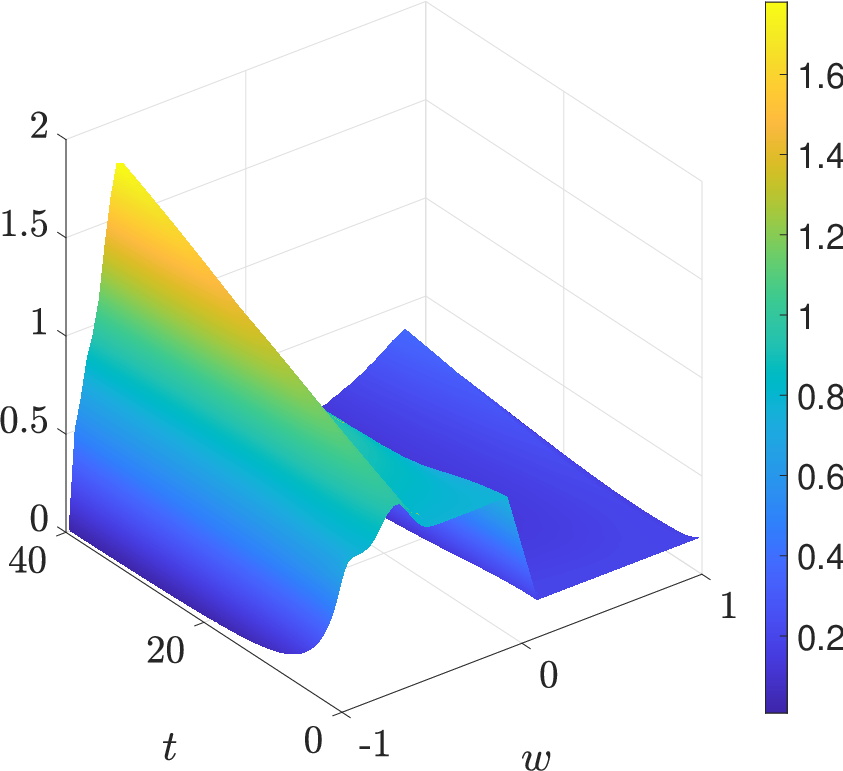} \\
		\includegraphics[height=4.2cm]{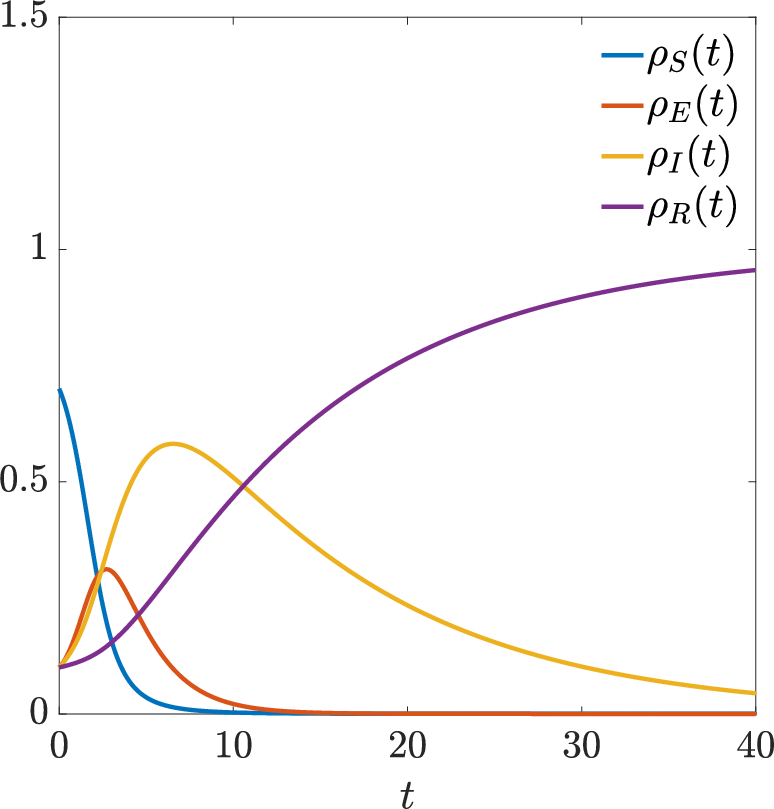}&\includegraphics[height=4.2cm]{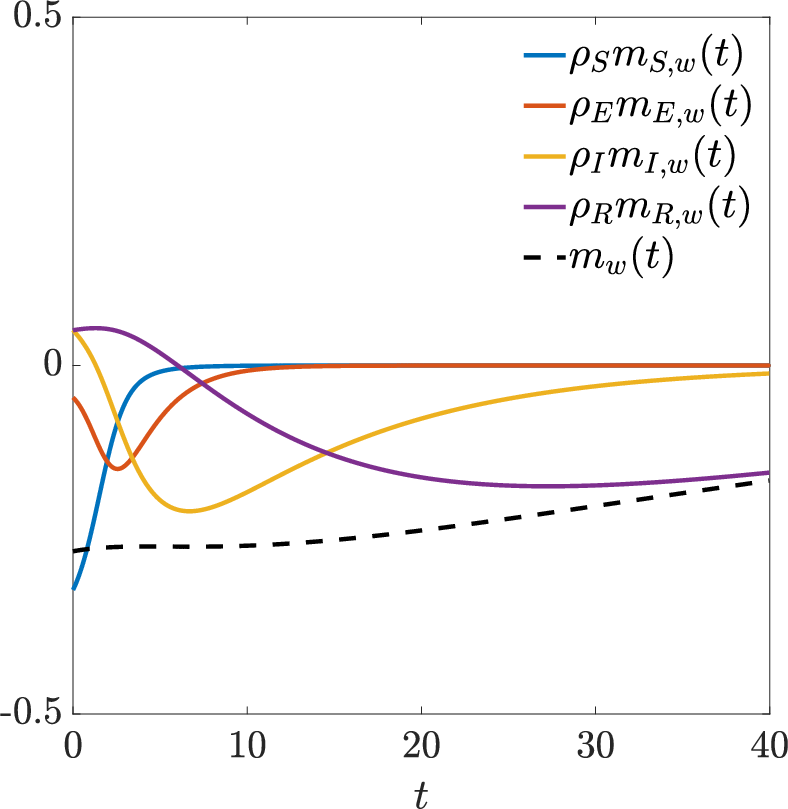} &\includegraphics[height=4.2cm]{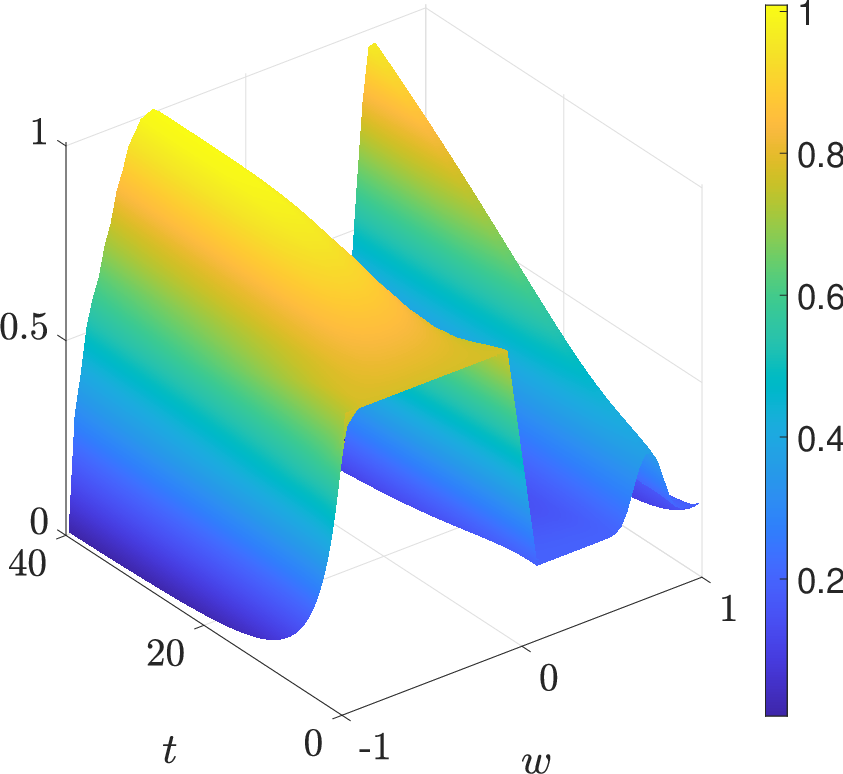} 
	\end{tabular}
	\caption{\emph{Setting $1a$} (top row): evolution of the compartment masses (left) and the marginal of opinions (right). \emph{Setting $2a$} (bottom row): evolution of the compartment masses (left) and the marginal of opinions (right).} \label{fig:plot_seir_contattifisici} 
\end{figure}

\medskip
We simulate the evolution of system \eqref{eq:SEIR_updated} up to the final time $T=40$ for both initial configurations. The results are reported in Figure~\ref{fig:plot_seir_contattifisici}, where the first row corresponds to \emph{Setting $1a$} and the second to \emph{Setting $2a$}. 
The left column displays the evolution of the compartment masses. We observe that the epidemic dynamics are qualitatively similar in the two settings, with no significant differences in the progression of the disease. This can be attributed to the fact that the fraction of individuals with a number of contacts above the threshold $\kappa_z$ is relatively small and therefore not sufficient to substantially alter the epidemic trajectory.
In contrast, the central column shows a marked difference in the evolution of the mean opinions, indicating that even a relatively small subgroup of highly connected individuals can significantly influence the global opinion dynamics.
Finally, the right column reports the time evolution of the full opinion distribution, further highlighting the impact of the initial opinion configuration of highly connected individuals on the overall behavioral dynamics of the population.

\subsection{On the joint influence of graphon connectivity and physical contacts on the epidemic spread}
This section is devoted to the study of the epidemic spread in the presence of both heterogeneous physical contacts and graphon-based social connectivity, accounting for the influence of opinion leaders with strong negative or positive attitudes toward protective measures.
As in Section~\ref{sec:epi_ph}, the distribution of physical contacts is described by the generalized Gamma law in \eqref{equili}, with the same parameters. We consider again the empirical graphon $\mathcal{B}_A(x,y)$. The epidemic parameters in \eqref{eq:kappa} are set to $\alpha = \alpha_* = 0.5$, $\eta = 1$, and $\beta = 0.4$. Moreover, we fix $\sigma_E = 0.5$ and the recovery rate $\gamma = 1/12$. 
For the opinion dynamics, we adopt the interaction kernel $P_2$ defined in \eqref{eq:kernels}, with $\alpha_{\mathcal J} = 1$ and $\sigma_{\mathcal J} = 0.1$ for all compartments. We consider two different initial configurations in order to highlight the impact of influential individuals on the epidemic evolution. In both cases, the initial masses are given by $\rho_E(0)=\rho_I(0)=\rho_R(0)=0.1$ and $\rho_S(0) = 1 - (\rho_E(0)+\rho_I(0)+\rho_R(0))$.

\paragraph{Setting $1b$.}
We split the population into two groups according to their in-degree: highly connected individuals satisfying
\[
d_i(x) \geq \kappa_x \max(d_i(x)),
\]
and the remaining population. At the initial time, the opinions of the first group are centered around the \emph{negative} value $-0.7$. For the second group, opinions depend on the epidemiological compartment. As in Section~\ref{sec:epi_ph}, we assume that Susceptible and Exposed individuals underestimate the severity of the epidemic and are therefore less inclined to adopt protective measures, so their opinions are uniformly distributed in $[-1,0]$. In contrast, Infected and Recovered individuals have already experienced the disease and thus have opinions uniformly distributed in $[0,1]$.
We define the auxiliary function
\begin{equation}\label{eq:aux_setting1}
	\widetilde{g}_{1,\mathcal J}(w,x) = \widetilde{K}_{g_1}
	\begin{cases}
		\chi(w \leq 0)\, d_i(x), & \text{if } d_i(x) < \kappa_x \max(d_i(x)) \text{ and } \mathcal{J} \in \{S,E\},\\[4pt]
		\chi(w \geq 0)\, d_i(x), & \text{if } d_i(x) < \kappa_x \max(d_i(x)) \text{ and } \mathcal{J} \in \{I,R\},\\[4pt]
		\exp\!\left\{-\dfrac{(w+0.7)^2}{0.005}\right\} d_i(x), 
		& \text{if } d_i(x) \geq \kappa_x \max(d_i(x)) \text{ and } \mathcal{J} \in \mathcal{C},
	\end{cases}
\end{equation}
where $\widetilde{K}_{g_1}$ is a normalization constant.
The initial condition is then given by
\begin{equation}\label{eq:init_set1}
	\begin{aligned}
		f_S(z,x,w,0) &= \rho_S(0)\, f_z(z)\, \widetilde{g}_{1,S}(w,x)\, \overline d_i(x), 
		& \quad f_E(z,x,w,0) &= \rho_E(0)\, f_z(z)\, \widetilde{g}_{1,E}(w,x)\, \overline d_i(x), \\[4pt]
		f_I(z,x,w,0) &= \rho_I(0)\, f_z(z)\, \widetilde{g}_{1,I}(w,x)\, \overline d_i(x), 
		& \quad f_R(z,x,w,0) &= \rho_R(0)\, f_z(z)\, \widetilde{g}_{1,R}(w,x)\, \overline d_i(x).
	\end{aligned}
\end{equation}
We set $\kappa_x = 0.8$.

\paragraph{Setting $2b$.}
As in \emph{Setting $1b$}, we split the population according to the in-degree. However, in this case the opinions of highly connected individuals are centered around the \emph{positive} value $0.7$, while the remaining population is initialized as before.
We define the auxiliary function
\begin{equation}\label{eq:aux_setting2}
	\widetilde{g}_{2,\mathcal J}(w,x) = \widetilde{K}_{g_2}
	\begin{cases}
		\chi(w \leq 0)\, d_i(x), & \text{if } d_i(x) < \kappa_x \max(d_i(x)) \text{ and } \mathcal{J} \in \{S,E\},\\[4pt]
		\chi(w \geq 0)\, d_i(x), & \text{if } d_i(x) < \kappa_x \max(d_i(x)) \text{ and } \mathcal{J} \in \{I,R\},\\[4pt]
		\exp\!\left\{-\dfrac{(w-0.7)^2}{0.005}\right\} d_i(x), 
		& \text{if } d_i(x) \geq \kappa_x \max(d_i(x)) \text{ and } \mathcal{J} \in \mathcal{C},
	\end{cases}
\end{equation}
where $\widetilde{K}_{g_2}$ is a normalization constant.
The corresponding initial condition reads
\begin{equation}\label{eq:init_set2}
	\begin{aligned}
		f_S(z,x,w,0) &= \rho_S(0)\, f_z(z)\, \widetilde{g}_{2,S}(w,x)\, \overline d_i(x), 
		& \quad f_E(z,x,w,0) &= \rho_E(0)\, f_z(z)\, \widetilde{g}_{2,E}(w,x)\, \overline d_i(x), \\[4pt]
		f_I(z,x,w,0) &= \rho_I(0)\, f_z(z)\, \widetilde{g}_{2,I}(w,x)\, \overline d_i(x), 
		& \quad f_R(z,x,w,0) &= \rho_R(0)\, f_z(z)\, \widetilde{g}_{2,R}(w,x)\, \overline d_i(x).
	\end{aligned}
\end{equation}

\begin{figure}[h] 
	\centering  
	\begin{tabular}{ccc}
		\includegraphics[height=4.2cm]{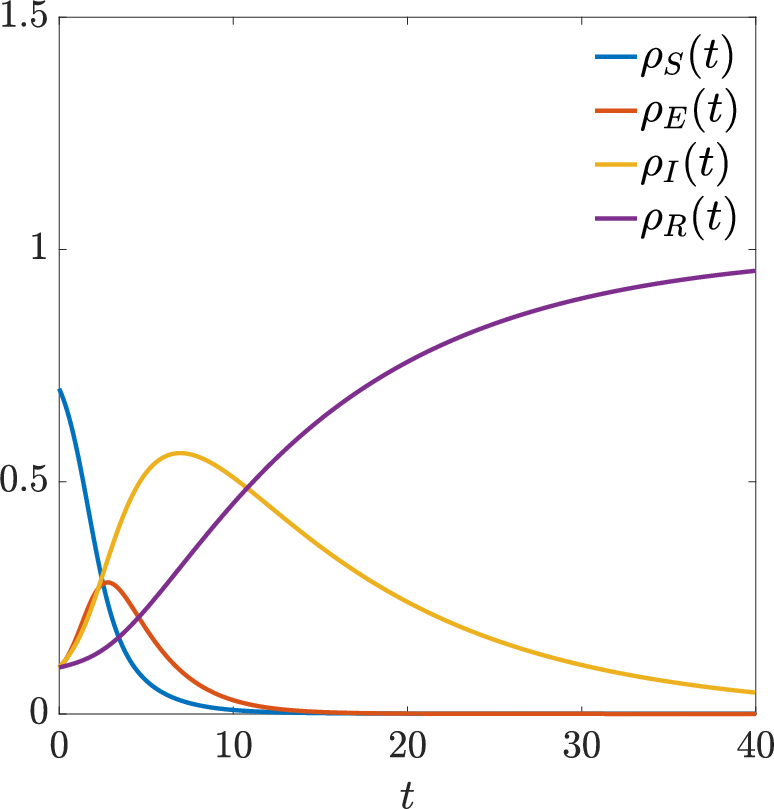}&\includegraphics[height=4.2cm]{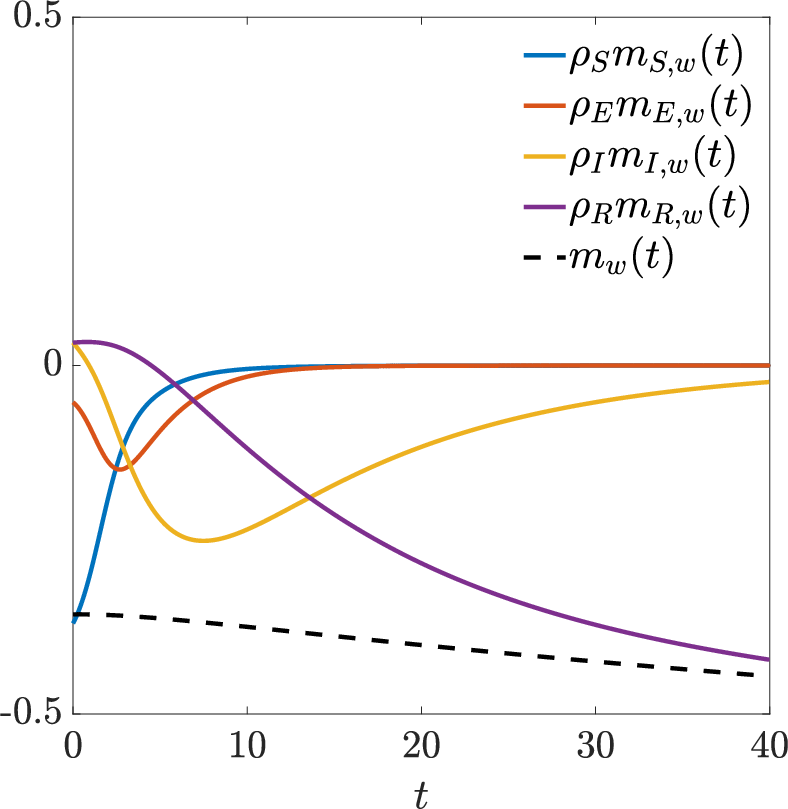}&\includegraphics[height=4.2cm]{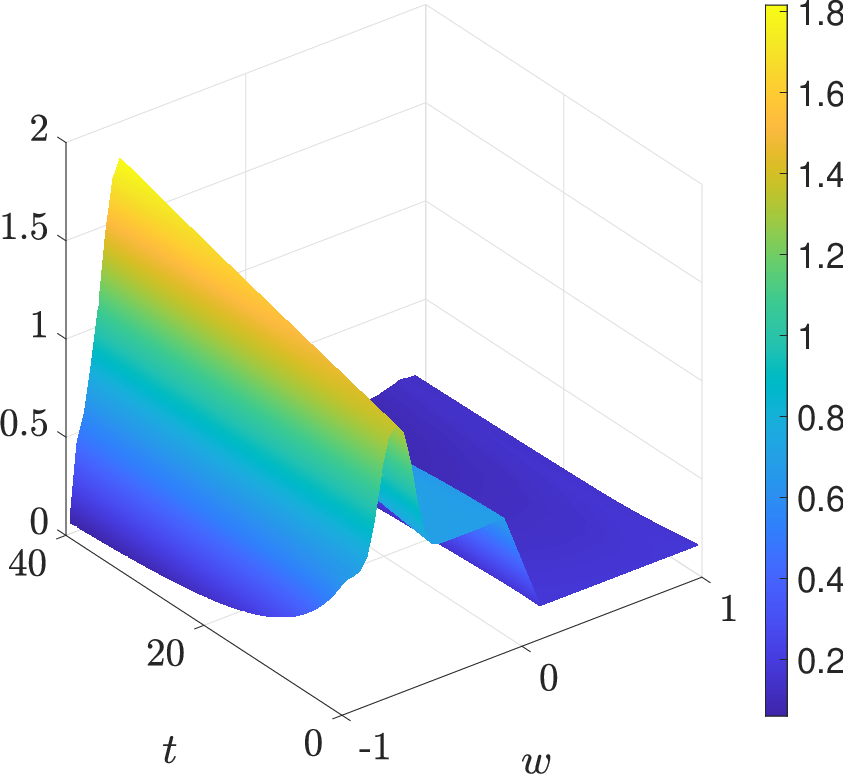} \\
		\includegraphics[height=4.2cm]{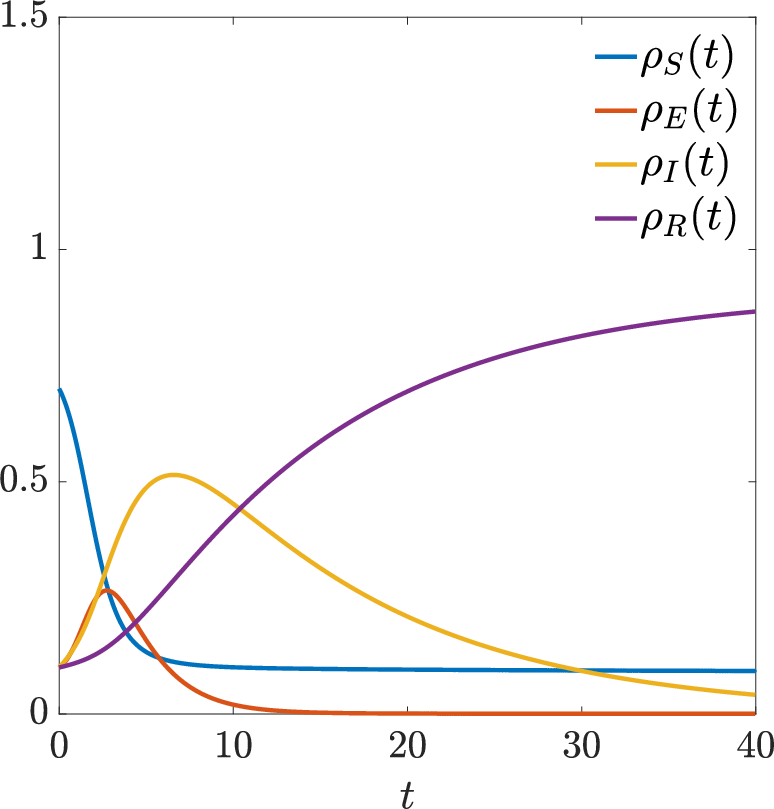}&\includegraphics[height=4.2cm]{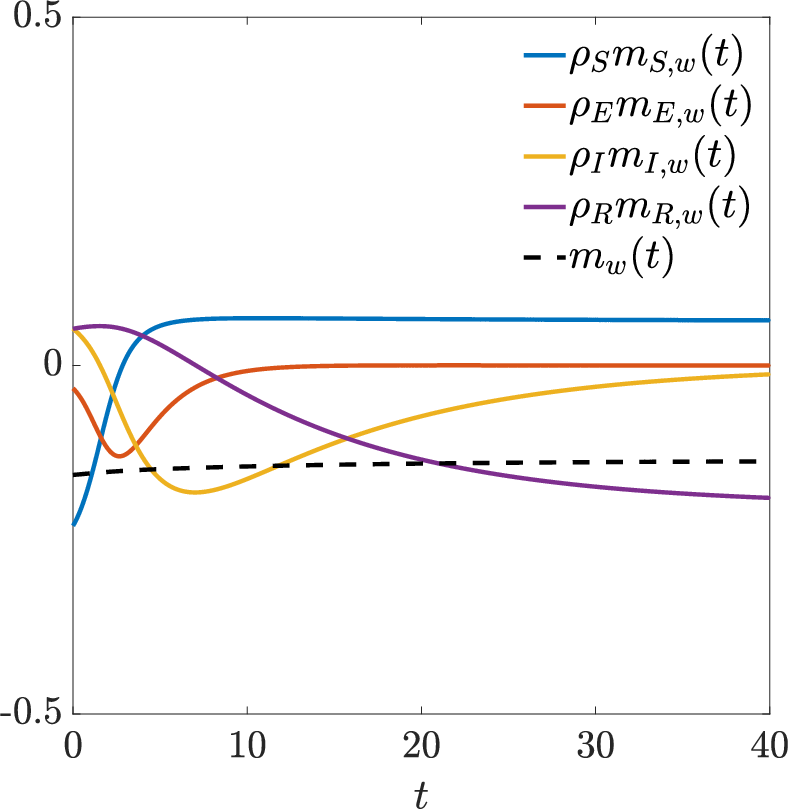}&\includegraphics[height=4.2cm]{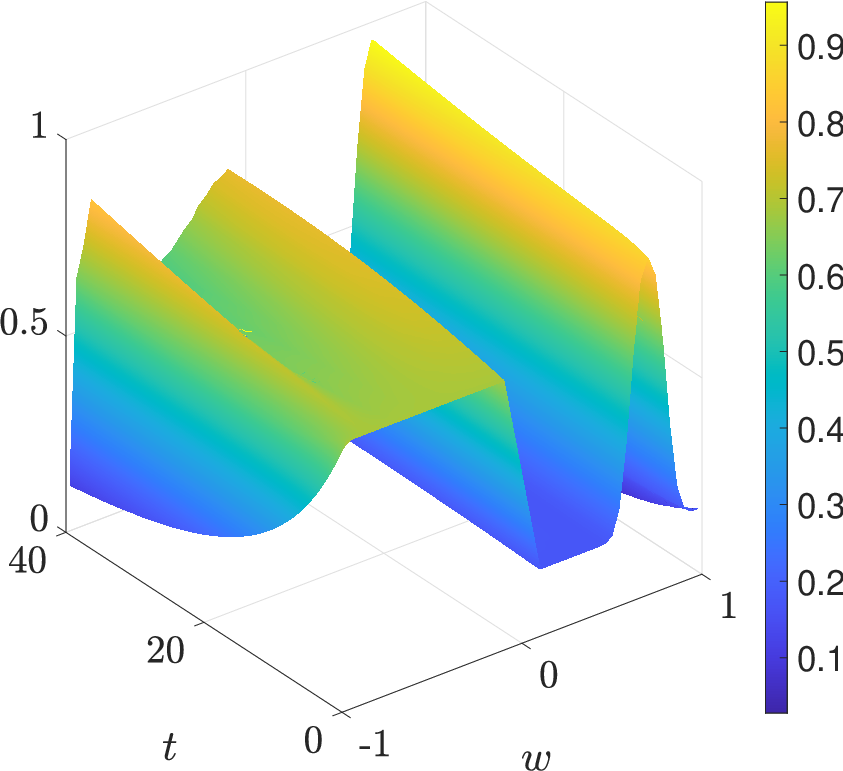} 
	\end{tabular}
	\caption{\emph{Setting $1b$} (top row): evolution of the compartment masses (left) and the marginal of opinions (right).
		\emph{Setting $2b$} (bottom row): evolution of the compartment masses (left) and the marginal of opinions (right).} \label{fig:plot_seir_contattisocial} 
\end{figure}

\begin{figure}[h!] 
	\centering  
	\includegraphics[height=5cm]{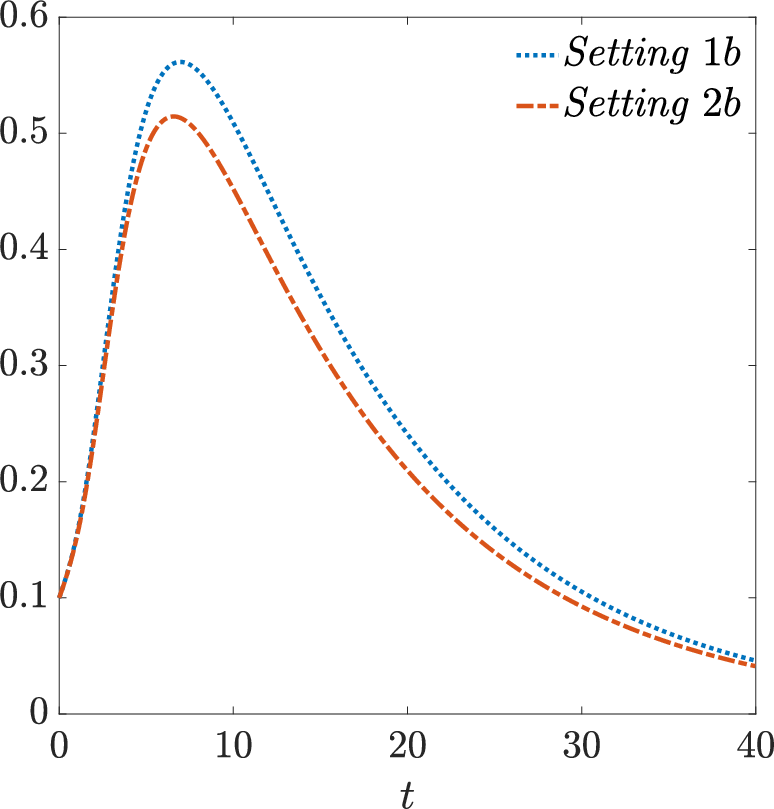}
	\caption{Comparison between the evolution of the mass of infected over time for \emph{Setting $1b$} and \emph{Setting $2b$}.} \label{fig:plot_contattisocial_comparison} 
\end{figure}

\medskip
We simulate the system \eqref{eq:SEIR_updated} up to time $T=40$ for both configurations. The results are shown in Figure~\ref{fig:plot_seir_contattisocial}, where the first row corresponds to \emph{Setting $1b$} and the second to \emph{Setting $2b$}. The left column displays the evolution of the compartment masses, the central column shows the evolution of the mean opinions, and the right column reports the full opinion distributions over time.
From the left column (see also Figure~\ref{fig:plot_contattisocial_comparison}), we observe that the epidemic peak in \emph{Setting $1b$} is significantly higher than in \emph{Setting $2b$}. This indicates that when highly connected individuals are opposed to protective measures, their influence leads to a more severe epidemic outbreak. In contrast, when these individuals favor protective behaviors (\emph{Setting $2b$}), the spread of the disease is mitigated.
Moreover, in \emph{Setting $1b$}, the entire susceptible population eventually becomes exposed and then infected, while in \emph{Setting $2b$} a non-negligible fraction of susceptible individuals remains unaffected over time. This clearly demonstrates that influential individuals can shape the collective behavior of the population, ultimately affecting the epidemic outcome.
These results highlight the ability of the model to capture the interplay between social influence, opinion formation, and epidemic spreading, showing how opinion leaders can significantly alter the course of an epidemic.

\section{Conclusion}\label{sec:conclusion}
In this work, we have proposed a novel modeling framework that couples classical epidemiological dynamics with social behavior and opinion formation processes on heterogeneous networks described by a graphon structure. This integrated approach provides a deeper understanding of how infectious diseases spread in modern societies, where both information diffusion and physical interactions play a crucial role.

By representing individual connectivity through a graphon, the model naturally accounts for heterogeneous social interaction patterns without relying on ad hoc or time-dependent structural assumptions. At the same time, the inclusion of a realistic distribution of physical contacts allows us to capture variability in exposure to infection. Within this setting, the classical susceptible-exposed-infected-recovered (SEIR) model is extended by incorporating a kinetic description of opinion dynamics on protective measures, evolving over the same underlying network. This results in a unified framework in which social influence, behavioral responses, and physical contacts jointly determine the epidemic evolution.

The numerical experiments, performed on real-world network data, highlight the significant impact of both opinion dynamics and contact heterogeneity on epidemic outcomes. In particular, influential individuals play a crucial role: positive opinion leaders promote the adoption of protective behaviors and mitigate disease spread, while negative influencers may limit such behaviors, leading to more severe or prolonged outbreaks. Moreover, the heterogeneity in physical contacts further amplifies these effects, showing that highly connected or highly exposed individuals can disproportionately affect the course of the epidemic.

Future research directions include the development of models where physical contact dynamics depend on the epidemic evolution, as well as the extension to multilayer and time-evolving networks. These developments would further enhance the realism and predictive capabilities of the proposed framework, providing a more comprehensive description of the interplay between social behavior, contact patterns, and epidemic spreading.

\appendix

\section*{Acknowledgments}

This work has been written within the activities of GNCS and GNFM groups of INdAM
(Italian National Institute of High Mathematics). EC and LP acknowledge the support by Fondo Italiano per la Scienza (FIS2023-01334) advanced grant "ADvanced numerical Approaches for MUltiscale Systems with uncertainties" - ADAMUS.

\section{Appendix: macroscopic contacts and opinion-based epidemic dynamics}\label{appendix}
In this appendix we derive a closed macroscopic system associated with the
kinetic SEIR model introduced in Section~2.3. The purpose is to show how,
under suitable assumptions, the kinetic description in the variables
\[
(z,x,w)\in \mathbb R_+\times \Psi\times \Phi,
\qquad 
\Psi=[0,1],\quad \Phi=[-1,1],
\]
leads to an ordinary differential system for the compartment masses and for the corresponding mean opinions. We denote by
$
\mathcal C=\{S,E,I,R\}
$
the set of epidemiological compartments. We also use the shorthand notation
$
d\varpi = dz\,dx\,dw.$ Let now \(f_J=f_J(z,x,w,t)\),  \(J\in\mathcal C\), be the kinetic density of
individuals in compartment \(J\). The total mass of each compartment is
defined by
\[
\rho_J(t)=\int_{\mathbb R_+\times\Psi\times\Phi}
f_J(z,x,w,t)\,d\varpi ,
\qquad J\in\mathcal C,
\]
where we assume the normalization
$
\sum_{J\in\mathcal C}\rho_J(t)=1,
\ t\geq 0.
$
The first moment in opinion and the corresponding mean opinion are defined by
\[
W_J(t)=\int_{\mathbb R_+\times\Psi\times\Phi}
w f_J(z,x,w,t)\,d\varpi,
\qquad
m_J(t)=\frac{W_J(t)}{\rho_J(t)},
\]
whenever \(\rho_J(t)>0\). Thus
$
W_J(t)=\rho_J(t)m_J(t).
$
In the following we also write \(m_J\) instead of \(m_{J,w}\) in order to simplify
the notation. We then introduce the second order moment in the opinion variable of the compartment,
\[
q_\mathcal{C}(t)=
\frac{1}{\rho_\mathcal{C}(t)}
\int_{\mathbb R_+\times\Psi\times\Phi}
w^2 f_\mathcal{C}(z,x,w,t)\,d\varpi .
\]
This moment will appear in the evolution of the mean opinion of susceptible individuals and will require a closure assumption. We assume that the contact variable is, at every time, at equilibrium. More precisely, we assume the factorization
\[
f_J(z,x,w,t)=f_z^\infty(z)g_J(x,w,t),
\qquad J\in\mathcal C,
\]
where
\[
\int_{\mathbb R_+} f_z^\infty(z)\,dz=1,
\qquad
\int_{\mathbb R_+} z f_z^\infty(z)\,dz=\bar z .
\]
In the case \(\delta=1\), the equilibrium contact distribution obtained from
the Fokker--Planck contact dynamics is the Gamma density
\[
f_z^\infty(z)
=
\left(\frac{\nu}{\bar z}\right)^\nu
\frac{1}{\Gamma(\nu)}
z^{\nu-1}
\exp\left\{-\frac{\nu}{\bar z}z\right\}.
\]
In the following, for the epidemic transmission kernel, we consider the linear case
$
\eta=1,
\
\alpha=\alpha_\ast=1,
$
so that
\[
\kappa(w,w_\ast;z,z_\ast)
=
\frac{\beta}{4}
(1-w)(1-w_\ast)zz_\ast ,
\]
and we define the shorthand notation
\[
\Lambda=\frac{\beta}{4}\bar z^2 .
\]
For the opinion dynamics we instead assume
\[
P(w,w_\ast,x,y)\equiv 1,
\qquad
\mathcal D(w)=\sqrt{1-w^2},
\]
and we replace the graphon \(B(x,y)\) by its average value
\[
\overline B
=
\int_0^1\int_0^1 B(x,y)\,dx\,dy .
\]
This corresponds to a mean-field closure of the graphon interactions. In
particular, the detailed dependence on the graphon position \(x\) is neglected
at the macroscopic level, while the average strength of the network
connectivity is retained through the scalar coefficient \(\overline B\) which will furnish a time scale for the opinion interaction.
The local incidence term is
\[
\mathcal K(f_I,f_S)(z,x,w,t)
=
f_S(z,x,w,t)
\int_{\mathbb R_+\times\Psi\times\Phi}
\kappa(w,w_\ast;z,z_\ast)
f_I(z_\ast,x_\ast,w_\ast,t)
\,dz_\ast dx_\ast dw_\ast .
\]
Under the assumptions above, its integral over
\(\mathbb R_+\times\Psi\times\Phi\) is
\[
\mathcal I(t)
:=
\int_{\mathbb R_+\times\Psi\times\Phi}
\mathcal K(f_I,f_S)(z,x,w,t)\,d\varpi,
\]
where, using the factorization in \(z\), we obtain
\[
\begin{aligned}
	\mathcal I(t)
	&=
	\frac{\beta}{4}
	\left(
	\int_{\mathbb R_+\times\Psi\times\Phi}
	z(1-w)f_S(z,x,w,t)\,d\varpi
	\right)
	\\
	&\hspace{2.5cm}\times
	\left(
	\int_{\mathbb R_+\times\Psi\times\Phi}
	z_\ast(1-w_\ast)f_I(z_\ast,x_\ast,w_\ast,t)\,
	dz_\ast dx_\ast dw_\ast
	\right)
	\\
	&=
	\frac{\beta}{4}\bar z^2
	\rho_S(t)\rho_I(t)
	\bigl(1-m_S(t)\bigr)
	\bigl(1-m_I(t)\bigr).
\end{aligned}
\]
Therefore,
\[
\mathcal I(t)
=
\Lambda
\rho_S(t)\rho_I(t)
\bigl(1-m_S(t)\bigr)
\bigl(1-m_I(t)\bigr).
\]
In the following, to get a macroscopic system, we will also need the opinion-weighted incidence term, defined as
\[
\mathcal I_w(t)
:=
\int_{\mathbb R_+\times\Psi\times\Phi}
w\,\mathcal K(f_I,f_S)(z,x,w,t)\,d\varpi,
\]
where a direct computation under the above hypothesis gives
\[
\begin{aligned}
	\mathcal I_w(t)
	&=
	\frac{\beta}{4}
	\left(
	\int_{\mathbb R_+\times\Psi\times\Phi}
	z\,w(1-w)f_S(z,x,w,t)\,d\varpi
	\right)
	\\
	&\hspace{2.5cm}\times
	\left(
	\int_{\mathbb R_+\times\Psi\times\Phi}
	z_\ast(1-w_\ast)f_I(z_\ast,x_\ast,w_\ast,t)\,
	dz_\ast dx_\ast dw_\ast
	\right)
	\\
	&=
	\frac{\beta}{4}\bar z^2
	\rho_S(t)\rho_I(t)
	\bigl(m_S(t)-q_S(t)\bigr)
	\bigl(1-m_I(t)\bigr).
\end{aligned}
\]
Hence
\[
\mathcal I_w(t)
=
\Lambda
\rho_S(t)\rho_I(t)
\bigl(m_S(t)-q_S(t)\bigr)
\bigl(1-m_I(t)\bigr).
\]

Now, integrating the kinetic SEIR system over
\(\mathbb R_+\times\Psi\times\Phi\), and using the zero-flux boundary
conditions in \(w\), the opinion operator does not contribute to the mass
balance. We therefore obtain
\[
\begin{aligned}
	\frac{d}{dt}\rho_S(t)
	&=
	-\Lambda
	\rho_S(t)\rho_I(t)
	\bigl(1-m_S(t)\bigr)
	\bigl(1-m_I(t)\bigr),
	\\
	\frac{d}{dt}\rho_E(t)
	&=
	\Lambda
	\rho_S(t)\rho_I(t)
	\bigl(1-m_S(t)\bigr)
	\bigl(1-m_I(t)\bigr)
	-\sigma_E\rho_E(t),
	\\
	\frac{d}{dt}\rho_I(t)
	&=
	\sigma_E\rho_E(t)-\gamma\rho_I(t),
	\\
	\frac{d}{dt}\rho_R(t)
	&=
	\gamma\rho_I(t).
\end{aligned}
\]
This system has the structure of a classical SEIR model, but with an effective
transmission rate depending on the mean opinions of susceptible and infected
individuals which need to be computed.

We now compute the contribution of the opinion operator to the evolution of the
first opinion moments. The Fokker--Planck opinion operator has the form
\[
\mathcal Q(f_J,f_H)
=
-\alpha_{JH}\partial_w\left(\mathcal P[f_H]f_J\right)
+
\frac{\sigma_{JH}^2}{2}
\partial_w^2\left(\mathcal L[f_H]\mathcal D^2(w)f_J\right).
\]
With \(P\equiv 1\), replacing the graphon by its average value
\(\overline B\), we have
\[
\mathcal P[f_H](w,t)
=
\overline B
\int_{\mathbb R_+\times\Psi\times\Phi}
(w_\ast-w)f_H(z_\ast,y,w_\ast,t)
\,dz_\ast dy dw_\ast .
\]
Therefore,
\[
\mathcal P[f_H](w,t)
=
\overline B\,\rho_H(t)\bigl(m_H(t)-w\bigr).
\]
Moreover,
\[
\mathcal L[f_H](t)
=
\overline B\rho_H(t).
\]

By multiplying \(\mathcal Q(f_J,f_H)\) by \(w\) and integrating over
\(\mathbb R_+\times\Psi\times\Phi\), the diffusion term gives no contribution
to the first moment, owing to the zero-flux boundary conditions at
\(w=\pm 1\). Hence, we get
\[
\begin{aligned}
	\int_{\mathbb R_+\times\Psi\times\Phi}
	w\,\mathcal Q(f_J,f_H)\,d\varpi
	&=
	\alpha_{JH}\overline B
	\rho_H(t)
	\int_{\mathbb R_+\times\Psi\times\Phi}
	\bigl(m_H(t)-w\bigr)f_J(z,x,w,t)\,d\varpi
	\\
	&=
	\alpha_{JH}\overline B
	\rho_J(t)\rho_H(t)
	\bigl(m_H(t)-m_J(t)\bigr).
\end{aligned}
\]
We then define the macroscopic opinion interaction term as
\[
\mathcal G_J(t)
=
\frac{\overline B}{\tau}
\sum_{H\in\mathcal C}
\alpha_{JH}
\rho_J(t)\rho_H(t)
\bigl(m_H(t)-m_J(t)\bigr).
\]
Multiplying the kinetic SEIR system by \(w\) and integrating over
\(\mathbb R_+\times\Psi\times\Phi\), we obtain consequently the following equations for the opinion moments \(W_J=\rho_Jm_J\):
\[
\begin{aligned}
	\frac{d}{dt}\bigl(\rho_S m_S\bigr)
	&=
	-\Lambda
	\rho_S\rho_I
	\bigl(m_S-q_S\bigr)
	\bigl(1-m_I\bigr)
	+\mathcal G_S,
	\\
	\frac{d}{dt}\bigl(\rho_E m_E\bigr)
	&=
	\Lambda
	\rho_S\rho_I
	\bigl(m_S-q_S\bigr)
	\bigl(1-m_I\bigr)
	-\sigma_E\rho_E m_E
	+\mathcal G_E,
	\\
	\frac{d}{dt}\bigl(\rho_I m_I\bigr)
	&=
	\sigma_E\rho_E m_E
	-\gamma\rho_I m_I
	+\mathcal G_I,
	\\
	\frac{d}{dt}\bigl(\rho_R m_R\bigr)
	&=
	\gamma\rho_I m_I
	+\mathcal G_R.
\end{aligned}
\]
Assume now that \(\rho_J(t)>0\). Since
\[
\frac{d}{dt}m_J
=
\frac{1}{\rho_J}
\frac{d}{dt}(\rho_Jm_J)
-
\frac{m_J}{\rho_J}
\frac{d}{dt}\rho_J,
\]
we obtain the time evolution for the mean opinions of the different compartments:
\[
\begin{aligned}
	\frac{d}{dt}m_S
	&=
	-\Lambda\rho_I
	\bigl(1-m_I\bigr)
	\bigl(m_S^2-q_S\bigr)
	+
	\frac{\overline B}{\tau}
	\sum_{H\in\mathcal C}
	\alpha_{SH}\rho_H
	\bigl(m_H-m_S\bigr),
	\\
	\frac{d}{dt}m_E
	&=
	\Lambda
	\frac{\rho_S\rho_I}{\rho_E}
	\bigl(1-m_I\bigr)
	\left[
	m_S-q_S
	-
	m_E\bigl(1-m_S\bigr)
	\right]
	+
	\frac{\overline B}{\tau}
	\sum_{H\in\mathcal C}
	\alpha_{EH}\rho_H
	\bigl(m_H-m_E\bigr),
	\\
	\frac{d}{dt}m_I
	&=
	\sigma_E
	\frac{\rho_E}{\rho_I}
	\bigl(m_E-m_I\bigr)
	+
	\frac{\overline B}{\tau}
	\sum_{H\in\mathcal C}
	\alpha_{IH}\rho_H
	\bigl(m_H-m_I\bigr),
	\\
	\frac{d}{dt}m_R
	&=
	\gamma
	\frac{\rho_I}{\rho_R}
	\bigl(m_I-m_R\bigr)
	+
	\frac{\overline B}{\tau}
	\sum_{H\in\mathcal C}
	\alpha_{RH}\rho_H
	\bigl(m_H-m_R\bigr).
\end{aligned}
\]
However, the system is not yet closed because of the second-order moment term \(q_S(t)\). To close the system, we approximate the susceptible opinion distribution by a
Beta-type local equilibrium with prescribed mean \(m_S(t)\). This closure is consistent with steady state solution of the underlying Fokker--Planck operator $\mathcal Q(f_J,f_H)$. Thus, for a given mean \(m_{\mathcal{J}}\in(-1,1)\) with \(\lambda>0\), this equilibrium distribution is given by
\[
g_\mathcal{J}(x,w,t)=g^\infty_\mathcal{J}(w)=\mathcal B_\lambda(w)
=
\frac{
	(1+w)^{\frac{1+m_\mathcal{J}}{\lambda}-1}
	(1-w)^{\frac{1-m_\mathcal{J}}{\lambda}-1}
}{
	2^{\frac{2}{\lambda}-1}
	\mathsf B\left(
	\frac{1+m_\mathcal{J}}{\lambda},
	\frac{1-m_\mathcal{J}}{\lambda}
	\right)
},
\qquad w\in[-1,1],
\]
where \(\mathsf B(\cdot,\cdot)\) denotes the Euler Beta function. Then
\[
\int_{-1}^{1}\mathcal B_\lambda(w)\,dw=1,
\qquad
\int_{-1}^{1}w\,\mathcal B_\lambda(w)\,dw=m_\mathcal{J},
\]
and
\[
\int_{-1}^{1}w^2\,\mathcal B_\lambda(w)\,dw
=
\frac{\lambda+2m_\mathcal{J}^2}{2+\lambda}.
\]
Thus, under the above hypothesis, we get
\[
q_S(t)
\simeq
\frac{\lambda_S+2m_S^2(t)}{2+\lambda_S},
\]
where \(\lambda_S>0\) measures the relative strength of self-thinking with
respect to compromise and in the case of class-dependent but partner-independent coefficients,
$
\alpha_{SH}=\alpha_S,
\
\sigma_{SH}=\sigma_S,
\ H\in\mathcal C,
$
one has
\[
\lambda_S=\frac{\sigma_S^2}{\alpha_S}.
\]
Finally, with this closure, one has
\[
m_S^2(t)-q_S(t)
=
\frac{\lambda_S}{2+\lambda_S}
\bigl(m_S^2(t)-1\bigr),
\]
and the macroscopic system reads
\[
\begin{aligned}
	\frac{d}{dt}\rho_S
	&=
	-\Lambda
	\rho_S\rho_I
	\bigl(1-m_S\bigr)
	\bigl(1-m_I\bigr),
	\\
	\frac{d}{dt}\rho_E
	&=
	\Lambda
	\rho_S\rho_I
	\bigl(1-m_S\bigr)
	\bigl(1-m_I\bigr)
	-\sigma_E\rho_E,
	\\
	\frac{d}{dt}\rho_I
	&=
	\sigma_E\rho_E-\gamma\rho_I,
	\\
	\frac{d}{dt}\rho_R
	&=
	\gamma\rho_I,
	\\[2mm]
	\frac{d}{dt}m_S
	&=
	-\Lambda\rho_I
	\bigl(1-m_I\bigr)
	\left(m_S^2-q_S\right)
	+
	\frac{\overline B}{\tau}
	\sum_{H\in\mathcal C}
	\alpha_{SH}\rho_H
	\bigl(m_H-m_S\bigr),
	\\
	\frac{d}{dt}m_E
	&=
	\Lambda
	\frac{\rho_S\rho_I}{\rho_E}
	\bigl(1-m_I\bigr)
	\left[
	m_S-q_S
	-
	m_E\bigl(1-m_S\bigr)
	\right]
	+
	\frac{\overline B}{\tau}
	\sum_{H\in\mathcal C}
	\alpha_{EH}\rho_H
	\bigl(m_H-m_E\bigr),
	\\
	\frac{d}{dt}m_I
	&=
	\sigma_E
	\frac{\rho_E}{\rho_I}
	\bigl(m_E-m_I\bigr)
	+
	\frac{\overline B}{\tau}
	\sum_{H\in\mathcal C}
	\alpha_{IH}\rho_H
	\bigl(m_H-m_I\bigr),
	\\
	\frac{d}{dt}m_R
	&=
	\gamma
	\frac{\rho_I}{\rho_R}
	\bigl(m_I-m_R\bigr)
	+
	\frac{\overline B}{\tau}
	\sum_{H\in\mathcal C}
	\alpha_{RH}\rho_H
	\bigl(m_H-m_R\bigr).
\end{aligned}
\]

The above resulting model is a closed approximation of the
kinetic SEIR--opinion system where the graphon structure enters only through the average connectivity \(\overline B\), while the dependence of epidemic transmission on behavior is retained through the compartmental mean opinions \(m_S\) and \(m_I\).

\bibliographystyle{abbrv}
\bibliography{biblio_SIS_model}

@article{bertaglia2021spatial,
  title={Spatial spread of {COVID-19} outbreak in {Italy} using multiscale kinetic transport equations with uncertainty},
  author={Bertaglia, Giulia and Boscheri, Walter and Dimarco, Giacomo and Pareschi, Lorenzo},
  journal={Mathematical Biosciences and Engineering},
  volume={18},
  number={5},
  pages={7028--7059},
  year={2021}
}

@article{zanella2021data,
  title={A data-driven epidemic model with social structure for understanding the {COVID-19} infection on a heavily affected {Italian Province}},
  author={Zanella, Mattia and Bardelli, Chiara and Dimarco, Giacomo and Deandrea, Silvia and Perotti, Pietro and Azzi, Mara and Figini, Silvia and Toscani, Giuseppe},
  journal={Mathematical Models and Methods in Applied Sciences},
  volume={31},
  number={12},
  pages={2533--2570},
  year={2021},
  publisher={World Scientific}
}

@book{bellomo2022predicting,
  title={Predicting pandemics in a globally connected world, Volume 1: Toward a Multiscale, Multidisciplinary Framework Through Modeling and Simulation},
  author={Bellomo, Nicola and Chaplain, Mark AJ},
  year={2022},
  publisher={Springer Nature}
}

@article{Durham2011, title = {Incorporating individual health-protective decisions into disease transmission models: a mathematical framework}, author = {Durham, D. P. and Casman, E. A.}, journal = {Journal of the Royal Society Interface}, year = {2011}, volume = {9}, issue = {68}, pages = {562-570}, doi = {10.1098/rsif.2011.0325} }

@article{Tunccgencc2021social,
  title={Social influence matters: We follow pandemic guidelines most when our close circle does},
  author={Tun{\c{c}}gen{\c{c}}, Bahar and El Zein, Marwa and Sulik, Justin and Newson, Martha and Zhao, Yi and Dezecache, Guillaume and Deroy, Ophelia},
  journal={British Journal of Psychology},
  volume={112},
  number={3},
  pages={763--780},
  year={2021},
  publisher={Wiley Online Library}
}

@article{Dezecache2020, author = {Dezecache, G. and Frith, C. and Deroy, O.}, title = {Pandemics and the great evolutionary mismatch}, journal = {Current Biology}, year = {2020}, volume = {30}, issue = {10}, pages = {R417-R419}, doi = {10.1016/j.cub.2020.04.010} }

@article{Barre2017, author = {Barré, J. and Degond, P. and Zatorska, E.}, title = {Kinetic theory of particle interactions mediated by dynamical networks}, journal = {Multiscale Modeling and Simulation}, year = {2017}, volume = {15}, issue = {3}, pages = {1294-1323}, doi = {10.1137/16m1085310} }

@article{Ciallella2021,
   author = {Ciallella, A. and  Pulvirenti, M. and Simonella,S.},
  title = {Kinetic {SIR} equations and particle limits},
  journal = {Proceedings of the Lincei National Academy of Sciences, Physics, Mat. Natur.},
  volume = {32},
  number = {2},
  pages = {295--315},
  year = {2021},
  doi = {10.4171/RLM/937}
}

@article{Motsch2014, author = {Motsch, S. and Tadmor, E.}, title = {Heterophilious dynamics enhances consensus}, journal = {SIAM Review}, year = {2014}, volume = {56}, issue = {4}, pages = {577-621}, doi = {10.1137/120901866} }

@article{Fornasier2011, author = {Fornasier, M. and Haškovec, J. and Toscani, G.}, title = {Fluid dynamic description of flocking via the {P}ovzner-{B}oltzmann equation}, journal = {Physica d Nonlinear Phenomena}, year = {2011}, volume = {240}, issue = {1}, pages = {21-31}, doi = {10.1016/j.physd.2010.08.003} }

@article{Degroot1974reaching,
  title={Reaching a consensus},
  author={DeGroot, Morris H},
  journal={Journal of the American Statistical Association},
  volume={69},
  number={345},
  pages={118--121},
  year={1974},
  publisher={Taylor \& Francis}
}

@article{Hegselmann2002opinion,
  title = {Opinion dynamics and bounded confidence: Models, analysis and simulation},
  author = {Hegselmann, R. and Krause, U.},
  journal = {Journal of Artificial Societies and Social Simulation},
  volume = {5},
  number = {3},
  year = {2002}
}

@article{Iacomini03042023,
author = {E. Iacomini and P. Vellucci},
title = {{Contrarian effect in opinion forming: Insights from Greta Thunberg phenomenon}},
journal = {The Journal of Mathematical Sociology},
volume = {47},
number = {2},
pages = {123--169},
year = {2023},
publisher = {Routledge},
doi = {10.1080/0022250X.2021.1981310}
}

@article{Goddard2021,
    author = {Goddard, B. D. and Gooding, B. and Short, H. and Pavliotis, G. A.},
    title = {Noisy bounded confidence models for opinion dynamics: the effect of boundary conditions on phase transitions},
    journal = {IMA Journal of Applied Mathematics},
    volume = {87},
    number = {1},
    pages = {80-110},
    year = {2021},
    month = {11},
    issn = {0272-4960},
    doi = {10.1093/imamat/hxab044}
}

@article{Jabin2014,
title = {Clustering and asymptotic behavior in opinion formation},
author = {Jabin, P.E. and Motsch, S.},
journal = {Journal of Differential Equations},
volume = {257},
number = {11},
pages = {4165-4187},
year = {2014},
issn = {0022-0396},
doi = {https://doi.org/10.1016/j.jde.2014.08.005}
}

@article{Toscani2006, author = {Toscani, G.}, title = {Kinetic models of opinion formation}, journal = {Communications in Mathematical Sciences}, year = {2006}, volume = {4}, issue = {3}, pages = {481-496}, doi = {10.4310/cms.2006.v4.n3.a1} }

@article{Albi2014, author = {Albi, G. and Pareschi, L. and Zanella, M.}, title = {Boltzmann-type control of opinion consensus through leaders}, journal = {Philosophical Transactions of the Royal Society a Mathematical Physical and Engineering Sciences}, year = {2014}, volume = {372}, issue = {2028}, pages = {20140138}, doi = {10.1098/rsta.2014.0138} }

@article{During2015, author = {Düring, B. and Wolfram, M.}, title = {Opinion dynamics: inhomogeneous {B}oltzmann-type equations modelling opinion leadership and political segregation}, journal = {Proceedings of the Royal Society a Mathematical Physical and Engineering Sciences}, year = {2015}, volume = {471}, issue = {2182}, pages = {20150345}, doi = {10.1098/rspa.2015.0345} }

@article{Pareschi2017, author = {Pareschi, L. and Vellucci, P. and Zanella, M.}, title = {Kinetic models of collective decision-making in the presence of equality bias}, journal = {Physica A: Statistical Mechanics and Its Applications}, year = {2017}, volume = {467}, pages = {201-217}, doi = {10.1016/j.physa.2016.10.003} }

@article{Pareschi2019, author = {Pareschi, L. and Toscani, G. and Tosin, A. and Zanella, M.}, title = {Hydrodynamic models of preference formation in multi-agent societies}, journal = {Journal of Nonlinear Science}, year = {2019}, volume = {29}, issue = {6}, pages = {2761-2796}, doi = {10.1007/s00332-019-09558-z} }

@article{Zanella2023, author = {Zanella, M.}, title = {Kinetic models for epidemic dynamics in the presence of opinion polarization}, journal = {Bulletin of Mathematical Biology}, year = {2023}, volume = {85}, issue = {5}, doi = {10.1007/s11538-023-01147-2} }

@article{Franceschi2023, author = {Franceschi, J. and Pareschi, L. and Bellodi, E. and Gavanelli, M. and Bresadola, M.}, title = {Modeling opinion polarization on social media: application to {COVID-19} vaccination hesitancy in {Italy}}, journal = {PLoS ONE}, year = {2023}, volume = {18}, issue = {10}, pages = {e0291993}, doi = {10.1371/journal.pone.0291993} }

@article{Bondesan2024, author = {Bondesan, A. and Toscani, G. and Zanella, M.}, title = {Kinetic compartmental models driven by opinion dynamics: Vaccine hesitancy and social influence}, journal = {Mathematical Models and Methods in Applied Sciences}, year = {2024}, volume = {34}, issue = {06}, pages = {1043-1076}, doi = {10.1142/s0218202524400062} }

@article{Danon2011, author = {Danon, L. and Ford, A. and House, T. and Jewell, C. and Keeling, M. and Roberts, G. and Ross, J. and Vernon, M.}, title = {Networks and the epidemiology of infectious disease}, journal = {Interdisciplinary Perspectives on Infectious Diseases}, year = {2011}, volume = {2011}, pages = {1-28}, doi = {10.1155/2011/284909} }

@article{lloyd2007network,
  title={Network models in epidemiology: an overview},
  author={Lloyd, Alun L and Valeika, Steve},
  journal={Complex Population Dynamics: Nonlinear Modeling in Ecology, Epidemiology and Genetics},
  pages={189--214},
  year={2007},
  publisher={World Scientific}
}

@article{Albi2017, author = {Albi, G. and Pareschi, L. and Zanella, M.}, title = {Opinion dynamics over complex networks: kinetic modelling and numerical methods}, journal = {Kinetic and Related Models}, year = {2017}, volume = {10}, issue = {1}, pages = {1-32}, doi = {10.3934/krm.2017001} }

@article{Fagioli2024, author = {Fagioli, S. and Favre, G.}, title = {Opinion formation on evolving network: the {DPA} method applied to a nonlocal cross-diffusion {PDE-ODE} system}, journal = {European Journal of Applied Mathematics}, year = {2024}, volume = {35}, issue = {6}, pages = {748-775}, doi = {10.1017/s0956792524000093} }

@article{Albi2024, author = {Albi, G. and Calzola, E. and Dimarco, G.}, title = {A data-driven kinetic model for opinion dynamics with social network contacts}, journal = {European Journal of Applied Mathematics}, year = {2024}, volume = {36}, issue = {2}, pages = {264-290}, doi = {10.1017/s0956792524000068} }

@inproceedings{Das2014modeling,
  title={Modeling opinion dynamics in social networks},
  author={Das, Abhimanyu and Gollapudi, Sreenivas and Munagala, Kamesh},
  booktitle={Proceedings of the 7th ACM international conference on Web search and data mining},
  pages={403--412},
  year={2014}
}

@article{li2020effect,
  title={Effect of the media on the opinion dynamics in online social networks},
  author={Li, Tingyu and Zhu, Hengmin},
  journal={Physica A: Statistical Mechanics and its Applications},
  volume={551},
  pages={124117},
  year={2020},
  publisher={Elsevier}
}

@InProceedings{Albi2016,
author={Albi, Giacomo and Pareschi, Lorenzo
and Zanella, Mattia},
editor={Bociu, Lorena
and D{\'e}sid{\'e}ri, Jean-Antoine
and Habbal, Abderrahmane},
title={On the Optimal Control of Opinion Dynamics on Evolving Networks},
booktitle={System Modeling and Optimization},
year={2016},
publisher={Springer International Publishing},
address={Cham},
pages={58--67},
isbn={978-3-319-55795-3}
}

@article{Lee2014,
 author = {Lee, J. and Choi, J. and Kim, C. and Kim, Y.}, title = {Social media, network heterogeneity, and opinion polarization}, journal = {Journal of Communication}, year = {2014}, volume = {64}, issue = {4}, pages = {702-722}, doi = {10.1111/jcom.12077} }

@article{Amelkin2017polar,
  title={Polar opinion dynamics in social networks},
  author={Amelkin, Victor and Bullo, Francesco and Singh, Ambuj K},
  journal={IEEE Transactions on Automatic Control},
  volume={62},
  number={11},
  pages={5650--5665},
  year={2017},
  publisher={IEEE}
}

@article{Toscani2018opinion,
  title={Opinion modeling on social media and marketing aspects},
  author={Toscani, Giuseppe and Tosin, Andrea and Zanella, Mattia},
  journal={Physical Review E},
  volume={98},
  number={2},
  pages={022315},
  year={2018},
  publisher={APS}
}

@article{Kermack1927contribution,
  title={A contribution to the mathematical theory of epidemics},
  author={Kermack, William Ogilvy and McKendrick, Anderson G},
  journal={Proceedings of the Royal Society of London. Series A, Containing Papers of a Mathematical and Physical Character},
  volume={115},
  number={772},
  pages={700--721},
  year={1927},
  publisher={The Royal Society London}
}

@inproceedings{Kendall1956deterministic,
  title={Deterministic and stochastic epidemics in closed populations},
  author={Kendall, David G},
  booktitle={Proceedings of the third Berkeley symposium on mathematical statistics and probability},
  volume={4},
  pages={149--165},
  year={1956},
  organization={University of California Press Berkeley, CA, USA}
}

@article{Beckley2013modeling,
  title={Modeling epidemics with differential equations},
  author={Beckley, Ross and Weatherspoon, Cametria and Alexander, Michael and Chandler, Marissa and Johnson, Anthony and Bhatt, Ghan S},
  journal={Tennessee State University Internal Report},
  volume={32},
  pages={33--34},
  year={2013},
  publisher={Tennessee State University}
}

@article{Loy2021viral,
  title={A viral load-based model for epidemic spread on spatial networks},
  author={Loy, Nadia and Tosin, Andrea and others},
  journal={Mathenatical Biosciences and Engineering},
  volume={18},
  number={5},
  pages={5635--5663},
  year={2021},
  publisher={AIMS}
}

@article{Marca2022sir,
  title={{An SIR--like kinetic model tracking individuals' viral load.}},
  author={Della Marca, Rossella and Loy, Nadia and Tosin, Andrea},
  journal={Networks \& Heterogeneous Media},
  volume={17},
  number={3},
  year={2022}
}

@article{Dimarco2022optimal,
  title={Optimal control of epidemic spreading in the presence of social heterogeneity},
  author={Dimarco, G and Toscani, G and Zanella, M},
  journal={Philosophical Transactions of The Royal Society A},
  volume={380},
  number={2224},
  pages={20210160},
  year={2022},
  publisher={The Royal Society}
}

@article{Pastor2002immunization,
  title={Immunization of complex networks},
  author={Pastor-Satorras, Romualdo and Vespignani, Alessandro},
  journal={Physical Review E},
  volume={65},
  number={3},
  pages={036104},
  year={2002},
  publisher={APS}
}

@article{Sun2017epidemic,
  title={Epidemic spreading on adaptively weighted scale-free networks},
  author={Sun, Mengfeng and Zhang, Haifeng and Kang, Huiyan and Zhu, Guanghu and Fu, Xinchu},
  journal={Journal of Mathematical Biology},
  volume={74},
  pages={1263--1298},
  year={2017},
  publisher={Springer}
}

@article{Salathe2010dynamics,
  title={Dynamics and control of diseases in networks with community structure},
  author={Salath{\'e}, Marcel and Jones, James H},
  journal={PLoS Computational Biology},
  volume={6},
  number={4},
  pages={e1000736},
  year={2010},
  publisher={Public Library of Science San Francisco, USA}
}

@article{Wang2003complex,
  title={Complex networks: small-world, scale-free and beyond},
  author={Wang, Xiao Fan and Chen, Guanrong},
  journal={IEEE Circuits and Systems Magazine},
  volume={3},
  number={1},
  pages={6--20},
  year={2003},
  publisher={IEEE}
}

@article{Iotti2017infection,
  title={Infection dynamics on spatial small-world network models},
  author={Iotti, Bryan and Antonioni, Alberto and Bullock, Seth and Darabos, Christian and Tomassini, Marco and Giacobini, Mario},
  journal={Physical Review E},
  volume={96},
  number={5},
  pages={052316},
  year={2017},
  publisher={APS}
}

@article{Saif2024sir,
  title={{SIR model on one dimensional small world networks}},
  author={Saif, M Ali and Shukri, MA and Al-makhedhi, FH},
  journal={Physica A: Statistical Mechanics and Its Applications},
  volume={633},
  pages={129430},
  year={2024},
  publisher={Elsevier}
}

@article{Caron2023, author = {Caron, F. and Panero, F. and Rousseau, J.}, title = {On sparsity, power-law, and clustering properties of graphex processes}, journal = {Advances in Applied Probability}, year = {2023}, volume = {55}, issue = {4}, pages = {1211-1253}, doi = {10.1017/apr.2022.75} }

@book{Van2024random,
  title={Random graphs and complex networks},
  author={Van Der Hofstad, Remco},
  volume={2},
  year={2024},
  publisher={Cambridge University Press}
}

@article{Bayraktar2023graphon,
  title={Graphon mean field systems},
  author={Bayraktar, Erhan and Chakraborty, Suman and Wu, Ruoyu},
  journal={The Annals of Applied Probability},
  volume={33},
  number={5},
  pages={3587--3619},
  year={2023},
  publisher={Institute of Mathematical Statistics}
}

@article{Nurisso2024network,
  title={Network-based kinetic models: Emergence of a statistical description of the graph topology},
  author={Nurisso, Marco and Raviola, Matteo and Tosin, Andrea},
  journal={European Journal of Applied Mathematics},
  pages={1--22},
  year={2024},
  publisher={Cambridge University Press}
}

@article{Coppini2022note,
  title={A note on {F}okker--{P}lanck equations and graphons},
  author={Coppini, Fabio},
  journal={Journal of Statistical Physics},
  volume={187},
  number={2},
  pages={15},
  year={2022},
  publisher={Springer}
}

@article{During2024breaking,
  title={Breaking consensus in kinetic opinion formation models on graphons},
  author={D{\"u}ring, Bertram and Franceschi, Jonathan and Wolfram, Marie-Therese and Zanella, Mattia},
  journal={Journal of Nonlinear Science},
  volume={34},
  number={4},
  pages={79},
  year={2024},
  publisher={Springer}
}

@article{glasscock2015graphon,
  title={What is... a graphon},
  author={Glasscock, Daniel},
  journal={Notices of the AMS},
  volume={62},
  number={1},
  pages={46--48},
  year={2015}
}

@article{lovasz2006limits,
  title={Limits of dense graph sequences},
  author={Lov{\'a}sz, L{\'a}szl{\'o} and Szegedy, Bal{\'a}zs},
  journal={Journal of Combinatorial Theory, Series B},
  volume={96},
  number={6},
  pages={933--957},
  year={2006},
  publisher={Elsevier}
}

@article{Pareschi2018structure,
  title={Structure preserving schemes for nonlinear {F}okker--{P}lanck equations and applications},
  author={Pareschi, Lorenzo and Zanella, Mattia},
  journal={Journal of Scientific Computing},
  volume={74},
  pages={1575--1600},
  year={2018},
  publisher={Springer}
}

@article{Bartel2024structure,
  title={Structure-preserving numerical methods for {F}okker--{P}lanck equations},
  author={Bartel, Hanna and Lampert, Joshua and Ranocha, Hendrik},
  journal={PAMM},
  volume={24},
  number={4},
  pages={e202400007},
  year={2024},
  publisher={Wiley Online Library}
}

@book{Wasserman2006all,
  title={All of nonparametric statistics},
  author={Wasserman, Larry},
  year={2006},
  publisher={Springer}
}

@article{Nowicki2001estimation,
  title={Estimation and prediction for stochastic blockstructures},
  author={Nowicki, Krzysztof and Snijders, Tom A B},
  journal={Journal of the American Statistical Association},
  volume={96},
  number={455},
  pages={1077--1087},
  year={2001},
  publisher={Taylor \& Francis}
}

@article{Airoldi2008mixed,
  title={Mixed membership stochastic blockmodels},
  author={Airoldi, Edo M and Blei, David and Fienberg, Stephen and Xing, Eric},
  journal={Advances in Neural Information Processing Systems},
  volume={21},
  year={2008}
}

@article{Hoff2002latent,
  title={Latent space approaches to social network analysis},
  author={Hoff, Peter D and Raftery, Adrian E and Handcock, Mark S},
  journal={Journal of the American Statistical Association},
  volume={97},
  number={460},
  pages={1090--1098},
  year={2002},
  publisher={Taylor \& Francis}
}

@inproceedings{Soufiani2012graphlet,
  title={Graphlet decomposition of a weighted network},
  author={Soufiani, Hossein Azari and Airoldi, Edo},
  booktitle={Artificial Intelligence and Statistics},
  pages={54--63},
  year={2012},
  organization={PMLR}
}

@inproceedings{Chan2014consistent,
  title={A consistent histogram estimator for exchangeable graph models},
  author={Chan, Stanley and Airoldi, Edoardo},
  booktitle={International Conference on Machine Learning},
  pages={208--216},
  year={2014},
  organization={PMLR}
}

@article{albi2025impact,
  title={Impact of opinion formation phenomena in epidemic dynamics: kinetic modeling on networks},
  author={Albi, Giacomo and Calzola, Elisa and Dimarco, Giacomo and Zanella, Mattia},
  journal={SIAM Journal on Applied Mathematics},
  volume={85},
  number={2},
  pages={779--805},
  year={2025},
  publisher={SIAM}
}

@article{hethcote2000mathematics,
  title={The mathematics of infectious diseases},
  author={Hethcote, Herbert W},
  journal={SIAM review},
  volume={42},
  number={4},
  pages={599--653},
  year={2000},
  publisher={SIAM}
}

@book{pareschi2013interacting,
  title={Interacting multiagent systems: kinetic equations and Monte Carlo methods},
  author={Pareschi, Lorenzo and Toscani, Giuseppe},
  year={2013},
  publisher={OUP Oxford}
}

@article {Perthame,
	AUTHOR = {Dimarco, G. and Perthame, B. and Toscani, G. and Zanella, M.},
	TITLE = {Kinetic models for epidemic dynamics with social
	heterogeneity},
	JOURNAL = {J. Math. Biol.},
	FJOURNAL = {Journal of Mathematical Biology},
	VOLUME = {83},
	YEAR = {2021},
	NUMBER = {1},
	PAGES = {Paper No. 4, 32},
}

@article {French,
	AUTHOR = {Béraud, G. and  Kazmercziak, S. and Beutels, P. and Levy-Bruhl, D. and Lenne, X. and Mielcarek, N. and Yazdanpanah, Y. and Boëlle, P.-Y. and Hens, N. and Dervaux, B.},
	TITLE = {{The French Connection: The First Large Population-Based Contact Survey in France Relevant for the Spread of Infectious Diseases.}},
	JOURNAL = {Plos one},
	FJOURNAL = {Plos one},
	VOLUME = {10},
	YEAR = {2015},
	NUMBER = {7},
	PAGES = {e0133203},
}

@article {Dol,
	AUTHOR = {Dolbeault, Jean and Turinici, Gabriel},
	TITLE = {Heterogeneous social interactions and the {COVID}-19 lockdown
	outcome in a multi-group {SEIR} model},
	JOURNAL = {Math. Model. Nat. Phenom.},
	FJOURNAL = {Mathematical Modelling of Natural Phenomena},
	VOLUME = {15},
	YEAR = {2020},
	PAGES = {Paper No. 36, 18},
	ISSN = {0973-5348,1760-6101},
	MRCLASS = {92D30 (34C60 92C60)},
	MRNUMBER = {4122507},
	DOI = {10.1051/mmnp/2020025},
	URL = {https://doi.org/10.1051/mmnp/2020025},
}

@article {TK,
author = {Amos Tversky AND Daniel Kahneman},
title = {Prospect Theory: An Analysis of Decision under Risk},
journal = {Econometrica},
volume = {47},
number = {2},
pages = {263-292},
year = {1979}
}

@Article{bonandin24,
title = {Effects of heterogeneous opinion interactions in many-agent systems for epidemic dynamics},
journal = {Networks and Heterogeneous Media},
volume = {19},
number = {1},
pages = {235-261},
year = {2024},
issn = {1556-1801},
doi = {10.3934/nhm.2024011},
url = {https://www.aimspress.com/article/doi/10.3934/nhm.2024011},
author = {Sabrina Bonandin and Mattia Zanella},
}

@article{bicego25,
    author = {Bicego, Sara and Kalise, Dante and Pavliotis, Grigorios A.},
    title = {Computation and control of unstable steady states for mean field multiagent systems},
    journal = {Proceedings of the Royal Society A: Mathematical, Physical and Engineering Sciences},
    volume = {481},
    number = {2311},
    pages = {20240476},
    year = {2025},
    month = {04},
    issn = {1364-5021},
    doi = {10.1098/rspa.2024.0476},
    url = {https://doi.org/10.1098/rspa.2024.0476},
    eprint = {https://royalsocietypublishing.org/rspa/article-pdf/doi/10.1098/rspa.2024.0476/2822707/rspa.2024.0476.pdf},
}

@misc{bondesan2026,
      title={Kinetic models of opinion-driven epidemic dynamics modulated by graphons}, 
      author={Andrea Bondesan and Jacopo Borsotti and Mattia Fontana},
      year={2026},
      eprint={2604.10614},
      archivePrefix={arXiv},
      primaryClass={math.AP},
      url={https://arxiv.org/abs/2604.10614}, 
}

@article{naldi26,
title = {On network-based epidemiological models: Analysis, simulations, and continuum limit},
journal = {Discrete and Continuous Dynamical Systems - B},
volume = {34},
number = {0},
pages = {243-274},
year = {2026},
issn = {1531-3492},
doi = {10.3934/dcdsb.2025171},
url = {https://www.aimsciences.org/article/id/690329226547bb6e5b3a4c41},
author = {Giovanni Naldi and Giuseppe Patanè}
}
\end{document}